\documentclass[header=ruled]{abnt}
\usepackage[brazil]{babel}
\usepackage[utf8]{inputenc}
\usepackage[num]{abntcite}
\usepackage{amsmath}
\usepackage{amssymb}
\usepackage{amsthm}
\usepackage{graphics}
\usepackage{graphicx}
\usepackage{float}
\usepackage{emanuel}

\usepackage{latexsym}
\usepackage{axodraw4j}
\usepackage{pstricks}
\usepackage{color}
\usepackage{feynmf}
\usepackage{axodraw}

\PassOptionsToPackage{caption=false}{subfig} 
\usepackage{subfig}
\usepackage{subeqnarray}
\citebrackets[]
\renewcommand{\thefootnote}{\alph{footnote}}

\makeglossary

\begin{document}

\autor{Emanuel Veras de Souza}

\titulo{Método de BPHZ para Pontos de Lifshitz $m$-Axiais Anisotrópicos}
\orientador{Prof.\ Dr.\ Paulo Renato Silva de Carvalho}
\coorientador{Prof.\ Dr.\ Marcelo de Moura Leite}

\comentario{Trabalho de dissertação de Mestrado apresentado ao Programa de Pós-Graduação em Física da Universidade Federal do Piauí
como requisito para obtenção do grau de Mestre em Física da Matéria Condensada.}

\instituicao{Universidade Federal do Piauí \par Centro de Ciências da Natureza \par Departamento de Física \par Curso de Pós-graduação em Física}

\local{Teresina}
\data{27 de Agosto de 2012}

\capaufpi

\folhaderostoufpi

\newpage
 \normalsize
\thispagestyle{empty}

\vspace{3.5cm}
\begin{center} \textbf{
\begin{tabular}{c}  FICHA CATALOGRÁFICA \\
 Universidade Federal do Piauí\\
Biblioteca Comunitária Jornalista Carlos Castelo Branco\\
Serviço de Processamento Técnico \\
\end{tabular}
}
\end{center}
\begin{footnotesize}
\begin{center}
\begin{tabular}{|cl|} \hline
\hspace{1cm} & \\
  S729m    & Souza, Emanuel Veras de \\
  & \hspace{0.6cm} Método de BPHZ para Pontos de Lifshitz m-Axiais 
 \\
      & Anisotrópicos./ Emanuel Veras de Souza -- Teresina: 2012, 95f. \\
      &  \\
            & \hspace{0.6cm} Dissertação (Mestrado em Física da Matéria Condensada) \\
      &  Teresina, 2012 \\
      & \hspace{0.6cm} Orientação: Prof. Dr. Paulo Renato Silva de Caravalho\\
       & \hspace{0.6cm} 1.Matéria Condensada. 2.Método de BPHZ.I.Título \\
       & CDD 530.41 \\
       & \\ \hline
 \end{tabular}
 \end{center}
 \setcounter{page}{1}
 \end{footnotesize}
 \hspace*{-1cm}

\newpage

\tesededicatoria{Dedico esta dissertação a todas as pessoas que contribuiram de forma direta ou indireta para a realização deste trabalho.}

\begin{agradecimentos}

Aos meus familiares, em especial a minha mãe (Telma), ao meu padrasto (Pantera), a Nádia (namorada) e ao meu filho (Dimitri), por terem me apoiado e incentivado por todos esses anos para a conclusão dos meus objetivos.

Ao meu orientador Paulo Renato Silva de Carvalho e o meu Co-orientador Marcelo de Moura Leite, por terem me orientado e contribuído para a minha formação profissional.

Aos professores e funcionários do Departamento de Física da UFPI.

A todos os meus amigos, colegas e irmãos escoteiros.

Ao grande amigo Moreira por ter me ajudado a trilhar essa jornada. Ao Rafael e Helder, pelas discursões sobre os trabalhos desenvolvidos pelo grupo, e pelos diagramas elaborados em \LaTeX.

A CAPES pelo apoio financeiro.

A UFPI pela estrutura física e a oportunidade de desenvolver este trabalho.

\end{agradecimentos}

\teseepigrafe{"A difusão do conhecimento é a chave para o progresso da humanidade."}{Emanuel Veras}

\begin{resumo}

Neste trabalho investigamos o comportamento crítico de sistemas físicos com interações competitivas que apresentam pontos de Lifshitz $m$-axiais. Para esse estudo usamos as técnicas de Teoria Quântica de Campos Escalares Massivos com interações do tipo $\lambda \phi^4$ para obtermos uma expansão perturbativa para as funções de vértice de dois pontos até a ordem de três {\it loops} e de quatro pontos até a ordem de dois {\it loops}. Essas funções de vértice foram regularizadas usando o método de regularização dimensional e renormalizadas usando o método de subtração mínima de pólos  dimensionais, onde foi adicionado contra-termos à Lagrangiana inicial, caracterizando o método BPHZ (Bogoliubov-Parasiuk-Hepp-Zimmermann). Através das ideias do Grupo de Renormalização, foram definidas as funções de Wilson que originam os pontos fixos, e a partir dessas funções e dos pontos fixos, calculamos os expoentes críticos anisotrópicos $\eta_{\tau}$, até a ordem três em número de {\it loops}, e $\nu_{\tau}$ até a ordem dois em número de {\it loops}, que caracterizam o comportamento crítico do tipo Lifshitz $m$-axial. Os expoentes calculados usando a presente técnica estão em perfeita concordância com os resultados obtidos usando outros métodos, confirmando assim a conhecida e importante hipótese de universalidade.

{\bf Palavras-chave}: Subtração mínima, expoentes críticos, ponto de Lifshitz, renormalização, sistemas competitivos.
\end{resumo}

\begin{abstract}

In this work we investigate the critical behavior of physical systems with competing interactions that present points Lifshitz $m$-axial. For this study we used the techniques of Quantum Field Theory with Massive Scalar interactions of type $\lambda\phi^4$ in order to obtain a perturbative expansion for the two-point vertex part up to the 3-loop order and four-point vertex function up to 2-loop level. These vertex functions were regularized using the method of dimensional regularization and renormalized using the minimal subtraction of dimensional poles method.In particular, counterterms have been added to the original Lagrangian, which is the main feature of the method BPHZ (Bogoliubov-Parasiuk-Hepp-Zimmermann). Through the renormalization group ideas, we defined the Wilson functions which shall produce nontrivial fixed points, and from these functions and fixed points, we calculate the anisotropic critical exponents $\eta_{\tau}$, to the order of three in number loops, and $\nu_{\tau}$ to the number of two in order loops, which characterize the critical behavior of the Lifshitz type $m$-axial. The exponents calculated using this technique are in perfect agreement with results obtained using other methods, thus confirming the known and important hypothesis of universality.


\end{abstract}

\listoffigures  

\tableofcontents

\chapter{Introdução}

 Transições de Fase e Fenômenos Críticos sempre ocuparam bastante espaço na literatura científica por desempenhar um papel muito importante na história da humanidade. Em particular, estas transições podem ocorrer em vários contextos desde a formação do universo, passando pela estrutura microscópica das interações fundamentais, e mais importante, na descrição de sistemas em física da matéria condensada que através das informações adquiridas podemos usá-las no desenvolvimento e aprimoramento de novas tecnologias. Este tema representa uma grande importância tanto para aplicações tecnológicas como para os fundamentos de física. Dentre as aplicações, encontramos facilmente uma extensa lista do uso da física de transições de fase e fenômenos críticos, como um exemplo temos o uso do conhecimento deste ramo da física na confecção e funcionamento de aparelhos eletroeletrônicos que inclusive fazem parte do nosso cotidiano.

Ao desenvolvermos aplicações nos vários setores da nossa sociedade moderna, usando sistemas físicos exibindo transições de fase, necessitamos de um conhecimento preciso das propriedades físicas desses sistemas. Para tal, temos que usar um formalismo que nos dê a quantidade de informação necessária sobre essas propriedades, com base em seus aspectos fundamentais. Isto é feito usando modelos microscópicos que representam os sistemas físicos estudados. Com esses modelos podemos extrair informações que nos permitem entender as transições de fase que ocorrem em tais sistemas em seu nível macroscópico.

As Transições de Fase mencionadas anteriormente, possuem uma classificação. De acordo com tal classificação \cite{fisher:1974}, uma transição de fase é dita de primeira ordem se a primeira derivada da energia livre é descontínua, ou seja, quando há descontinuidades, por exemplo, na energia interna e/ou na mag\-ne\-ti\-za\-ção do sistema. Talvez o melhor exemplo para uma transição deste tipo seja a transição da água entre os estados sólido e líquido, onde percebe-se a coexistência de fases e uma descontinuidade na energia interna do sistema, cujo parâmetro de ordem do sitema é dado pela densidade. Ainda de acordo com a classificação de Fisher \cite{fisher:1974}, transições contínuas ou críticas (muitas vezes chamadas de segunda ordem) são aquelas que apresentam a primeira derivada da energia livre contínua e sua segunda derivada descontínua ou infinita.

\begin{figure}[htb]
\begin{center}
\includegraphics[scale=0.4]{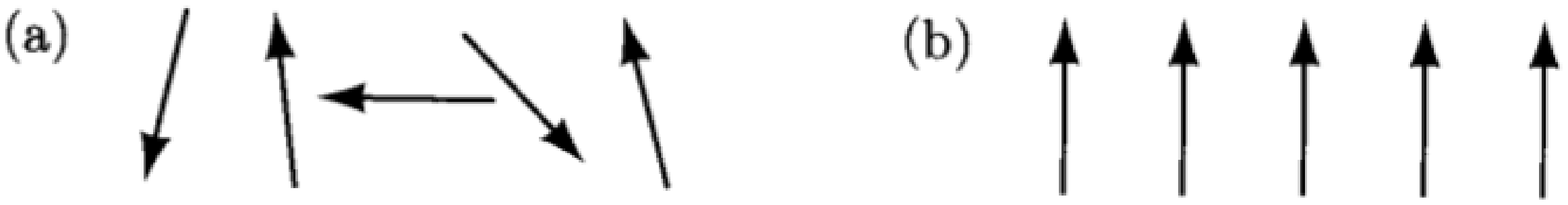}
\caption{Representação esquemática dos spins em (a) na fase paramagnética, (b) na fase ferromagnética.}
\label{fig:spins}
\end{center}
\end{figure}

Um exemplo deste tipo de transição é a sofrida por alguns materiais magnéticos. Abaixo de uma determinada temperatura crítica $T_c$, também conhecida como temperatura de Curie, o sistema encontra-se magnetizado. À medida que a temperatura aumenta, percebe-se que a magnetização diminui continuamente até atingir um valor nulo na temperatura crítica. Ao mesmo tempo observa-se uma divergência na susceptibilidade magnética do sistema. As formações ferromagnéticas (ou domínios magnéticos) mergulhadas na fase paramagnética e vice-versa ocorrem em todas as escalas de distância, conforme aumentamos ou diminuímos a temperatura em torno da temperatura crítica. Na fase ferromagnética, a interação entre os spins é mais forte que os efeitos térmicos e os deixam alinhados com uma ordenação uniforme dentro de um domínio. Ao passo que na fase paramagnética, os efeitos térmicos vencem a interação entre os spins e os desalinham desordenadamente, como mostra a Figura \ref{fig:spins} .

O modelo mais simples utilizado no estudo de transições de fase em sistemas magnéticos é o modelo de Ising \cite{ising:1925}. Esse modelo foi proposto para tentar entender o comportamento do tipo imã (ferromagnético) de sistemas magnéticos, e teve uma solução exata, em sua versão unidimensional, obtida pela primeira vez por E. Ising. O modelo de Ising bidimensional foi resolvido exatamente, sem um campo magnético externamente aplicado, onde o pioneiro nessa solução foi L. Onsager \cite{onsager:1944}. Na presença de um campo magnético, o caso bidimensional ainda não tem solução analítica, assim como o caso tridimensional, mesmo na ausência de um campo externo.

Uma característica comum em transições de fases é a existência de parâmetros que têm um valor não nulo em uma temperatura abaixo da temperatura crítica e que se anulam continuamente na temperatura crítica. Tal parâmetro é conhecido como parâmetro de ordem, e foi inicialmente introduzido por Lev Landau. Em alguns casos, a identificação do parâmetro de ordem não é trivial, podendo ocorrer deste não ser um escalar e/ou ter que ser descrito por números complexos. No contexto de sistemas magnéticos, temos que o momento magnético por unidade de volume define a magnetização espontânea $M$ do material. Assim uma magnetização $M=0$ define o estado desordenado e $M\not=0$ o estado ordenado. Uma vez que, nesse contexto, as transições estão associadas a uma mudança de um estado desordenado de alta temperatura para um estado ordenado de baixa temperatura. Dizemos então que a magnetização $M$ é o parâmetro de ordem do sistema, pois dá informação de quando se tem ordem ou não no sistema. Transições de fase de segunda ordem são caracterizadas por um parâmetro de ordem que vai continuamente a zero na temperatura de transição.

No estudo de transições de fases estamos interessados em estudar o comportamento do sistema em regiões próximas ao ponto crítico. Resultados experimentais e teóricos indicam que o comportamento das propriedades do sistema podem, em geral, ser descritas por leis de potência
simples, caracterizando um conjunto de  expoentes críticos. Através da introdução do parâmetro denominado {\bf temperatura reduzida $t$}, onde

\begin{equation}
t=\frac{T-T_c}{T_c} , \label{tepreduzida}
\end{equation}
que mede a distância em relação ao ponto crítico e nos dá a informação sobre em qual lado da transição estamos, podemos expressar o comportamento assintótico das funções termodinâmicas, definindo com isso os expoentes críticos da seguinte forma:

\begin{equation}
C\sim|t|^{-\alpha} , \label{alfa}
\end{equation}
onde $\alpha$ é o expoente crítico associado à singularidade do calor específico $C$ quando $t \rightarrow 0$;

\begin{equation}
M\sim |t|^{\beta} , \label{beta}
\end{equation}
onde $\beta$ é o expoente crítico que caracteriza o parâmetro de ordem $M$ (magnetização espontânea no caso de sistemas magnéticos, ou a diferença de densidade no caso líquido-gás), mostrando como $M \rightarrow 0$, para $T \rightarrow 0$. Ainda temos

\begin{equation}
\chi \sim |t|^{-\gamma} , \label{gama}
\end{equation}
onde o expoente crítico $\gamma$ está associado à divergência da susceptibilidade/compressibilidade i\-so\-tér\-mi\-ca $\chi$ quando $t \rightarrow 0$, e

\begin{equation}
M \sim H^{1/\delta} . \label{delta}
\end{equation}
Essa última expressão (equação de estado), que define o expoente crítico $\delta$, nos dá a relação entre o parâmetro de ordem na isoterma crítica $(t=0)$ e o campo externo/pressão aplicada $H$.

Olhando agora para as correlações, no qual significa uma dependência de termos entre si, existentes nos sistemas que exibem o comportamento crítico, por exemplo, as correlações entre os spins no caso magnético, podemos considerar ainda mais dois expoentes críticos. As funções de correlação são de grande utilidade no estudo das flutuações que ocorrem em tais sistemas e definem os dois últimos expoentes. Consideremos a função de correlação $g(R)$ entre dois spins, indicados genericamente por $s_i$ e $s_j$, localizados nas posições $\vec{r}_i$ e $\vec{r}_j$, respectivamente, participantes de um sistema magnético que pode ser tratado pelo modelo Ising, situados a uma distância $R=|\vec{r}_i-\vec{r}_j|$, onde

\begin{equation}
g(R)=\langle s_i s_j \rangle - \langle s_i \rangle \langle s_j \rangle , \label{corre}
\end{equation}
e $\langle \cdots \rangle$ indica a média termodinâmica do argumento em questão. A equação (\ref{corre}) mede a probabilidade condicional de que o spin em um dado sítio aponte em uma certa direção, dado que o spin numa origem definida também aponte na mesma direção. O segundo termo à direita na equação (\ref{corre}) garante que se está descontando a possibilidade de os spins serem paralelos não devido à correlação direta entre eles, mas por se estar numa fase de baixa temperatura, onde a magnetização espontânea tende a alinhar todos os spins na mesma direção \cite{queiroz:2000}. Podemos escrever ainda a função de correlação da seguinte forma

\begin{equation}
g(R)=\langle (s_i-\langle s_i\rangle)(s_j-\langle s_j\rangle)\rangle = \langle (s_i-M)(s_j-M)\rangle .
\end{equation}
Essa função é denominada função de correlação spin-spin. De acordo com a definição \cite{queiroz:2000}, espera-se que $g(R)$ se comporte da seguinte forma 

\begin{equation}
g(R) \sim \frac{exp(-R/\xi)}{R^{d-2+ \eta}} , \label{corre2}
\end{equation}
onde $d$ é o número de dimensões do sistema, $\xi$ é o comprimento de correlação, o qual nos dá a escala do alcance das correlações entre as flutuações, neste caso específico, flutuações da magnetização. Em uma transição de primeira ordem, o comprimento de correlação é sempre finito, o que impede a ocorrência, de invariância de escala, para esse tipo de transição. Porém, próximo a uma transição de segunda ordem, o comprimento de correlação, ou seja, o alcance das correlações entre as flutuações, diverge e sua forma assintótica para $t \rightarrow 0$ é dada por

\begin{equation}
\xi \sim |t|^{-\nu} , \label{nu}
\end{equation}
onde $\nu$ é o expoente crítico associado à divergência do comprimento de correlação numa transição de segunda ordem. Em $T_c$, a equação (\ref{corre2}) mostra que as correlações tomam a forma de uma lei de potência, $g(R) \sim \frac{a^{\eta}}{R^{d-2+\eta}}$, onde $\eta$ é um dos expoentes críticos, recebendo o nome de {\bf dimensão anômala}, pois aparentemente há uma violação da análise dimensional quando comparado com o resultado fornecido pela teoria de Landau. A explicação desse paradoxo está em que outra escala de comprimento deve necessariamente estar envolvida. Como o comprimento de correlação é infinito no ponto crítico, ele não pode fornecer a escala procurada \cite{dudumiranda:2005}.

Foi mostrado pela primeira vez por Rushbrooke \cite{rushbrooke:1963}, Griffiths \cite{griffiths:1965,griffiths2:1965}, Josephson \cite{josephson:1967,josephson2:1967} e Fisher \cite{fisher:1969}, que usaram apenas termodinâmica básica e algumas suposições razoáveis, que os seis expoentes críticos deveriam satisfazer a quatro desigualdades. Posteriormente, experimentos indicaram que essas desigualdades deveriam ser, na verdade, igualdades. Enfatizando assim, a importância de se obter experimentalmente esses expoentes críticos. 

Antes do desenvolvimento da teoria do grupo de renormalização, alguns pesquisadores perceberam que na região crítica as quantidades físicas tinham um comportamento mais simples como função dos parâmetros externos, como temperatura e campo magnético no caso dos ferromagnetos. Um dos primeiros a notar essa simplificação foi Ben Widom \cite{widom:1965}. Essa simplificação leva o nome de Teoria de Escala (ou de escalamento, "{\it scaling theory}"). Uma das consequências mais importantes da teoria de escala é mostrar que os expoentes críticos não são independentes, mas satisfazem algumas relações que formam as leis de escala. Através da hipótese de escala de Widom \cite{widom:1965}, ele supôs que a energia livre de Helmotz $f(T,H)$ poderia ser escrita como

\begin{equation}
f(T,H)=t^{1/y}\psi \biggl(\frac{H}{t^{x/y}}\biggr) ,
\end{equation} 
que leva às relações de escala de Rushbrooke (\ref{Rushbrooke}) e a de Widom (\ref{Widon}) dadas por

\begin{eqnarray}
\alpha + 2\beta + \gamma =2 , \label{Rushbrooke} \\
\gamma=\beta(\delta-1) . \label{Widon}
\end{eqnarray}

 As outras duas relações de escalas entre os expoentes críticos relacionados às funções de correlação, são as relações de escala de Fisher (\ref{Fisher}) e Josephson (\ref{Josephson}), dadas por

\begin{eqnarray}
\gamma= \nu(2-\eta) , \label{Fisher} \\
\nu d=2-\alpha . \label{Josephson}
\end{eqnarray}

Destas relações vemos que, obtendo independentemente apenas dois expoentes dentre os seis existentes, pode-se obter os quatro expoentes restantes através dessas quatro relações de escala.

\begin{figure}[htb]
\begin{center}
\includegraphics[scale=0.5]{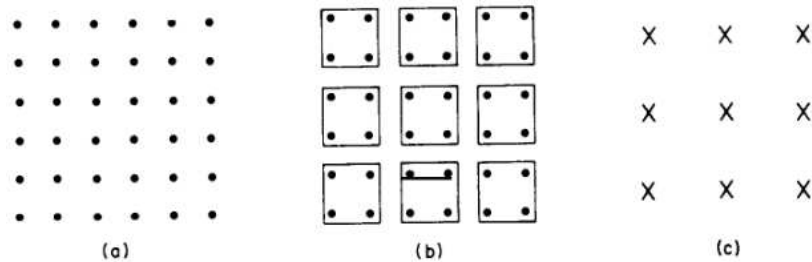}
\caption{Visualização da construção dos blocos de spins de Kadanoff. A rede original (a) de spins de cada sítio é dividido em blocos (b) de 4 spins por bloco que é substituído por uma nova rede (c) de spins "efetivos".}
\label{fig:kadanoff}
\end{center}
\end{figure}

Utilizando a teoria do grupo de renormalização, pode-se obter essas relações de escalas por primeiros princípios e não meramente através de hipóteses apresentada acima. A ideia pioneira foi devida a Kadanoff \cite{kadanoff:1966}, com sua técnica de dizimação ou grupo de renormalização no espaço real. A brilhante ideia de Kadanoff permitiu integrar as flutuações da magnetização com pequenos comprimentos de onda de maneira a incluir apenas as flutuações com grandes comprimentos de onda responsável pelo comportamento coletivo em grandes distâncias. Por simplicidade, podemos visualizar a ideia considerando o modelo Ising bidimensional em uma rede quadrada que possui uma determinada constante de rede (distância entre dois vizinhos mais próximos). Considerando uma pequena região contendo quatro spins, temos que o comprimento de correlação é muito grande perto da temperatura crítica, então todos os spins de tal bloco estão fortemente correlacionados, então esses spins estão em apenas dois estados possíveis: todos para cima ou todos para baixo. Assim, uma dizimação corresponde a transformar quatro spins de um bloco em um único spin efetivo, como mostra a figura \ref{fig:kadanoff}. Aplicando esta técnica a todos os pontos da rede, obtemos uma rede efetiva cujo parâmetro de rede é agora maior que o original. Este processo pode ser realizado um grande número de vezes, sendo eficiente por exemplo em cálculos numéricos. Esta técnica originou a uma linha de pesquisa conhecida como grupo de renormalização no espaço real. Do ponto de vista quantitativo, esta abordagem não rendeu grandes resultados, pois o método permite obter apenas pequenas correções às soluções de campo médio para as grandezas desejadas.

Entretanto, esta ideia foi adaptada ao espaço dos momentos levando a resultados espetaculares: as transformações de dizimação sobre blocos de spin ganhavam agora um significado mais preciso em termos matemáticos, permitindo pela primeira vez obter resultados analíticos muito além da teoria de campo médio para as grandezas de interesse. Estas transformações de grupo de renormalização no espaço dos momentos correspondem a resolver o problema não-trivial da interação entre todas as escalas de comprimento envolvidas. De acordo com Wilson \cite{wilson:1971,wilson2:1971}, para examinarmos o problema no regime IR (para grandes escalas $\gg 1$, que caracteriza a região infravermelho) tudo o que temos que fazer é integrar os modos com pequenos comprimentos de onda (dizimação) sucessivamente. Quando as grandezas desejadas não mudam mais depois de um número de interações de grupo de renormalização, dizemos que o sistema físico está em um ponto fixo, que nos leva aos pontos críticos do sistema, e são caracterizados pela invariância de escala. Estes conceitos juntamente com a expansão perturbativa em um parâmetro pequeno ($\varepsilon = 4-d$, onde $d$ é a dimensão espacial do sistema) fornece a chave para obter os resultados analíticos mais interessantes.

Percebe-se que existem grupos de sistemas caracterizados pelo mesmo conjunto de expoentes críticos, apesar das temperaturas críticas serem, em geral, diferentes. Desta forma, podemos definir classes de universalidades, que são conjuntos de sistemas que possuem os mesmos expoentes críticos \cite{stanley:1971}. Os expoentes críticos dependem apenas da dimensão espacial do sistema ($d$), da simetria e dimensionalidade do parâmetro de ordem ($n$) e do alcance das interações \cite{hohenberg:1977} definindo um par ($N,d$), e não de detalhes microscópicos do sistema. As classes de universalidade permitem que através do estudo de sistemas mais simples possamos obter importantes propriedades de sistemas mais complexos. Em geral, as classes de universalidade recebem o nome do modelo mais simples para o qual um determinado conjunto de expoentes críticos é observado. Outros exemplos de grandezas físicas universais são razões entre amplitudes de potenciais termodinâmicos acima e abaixo da transição, como o calor específico e a susceptibilidade.

\begin{figure}[htb]
\begin{center}
\includegraphics[scale=0.5]{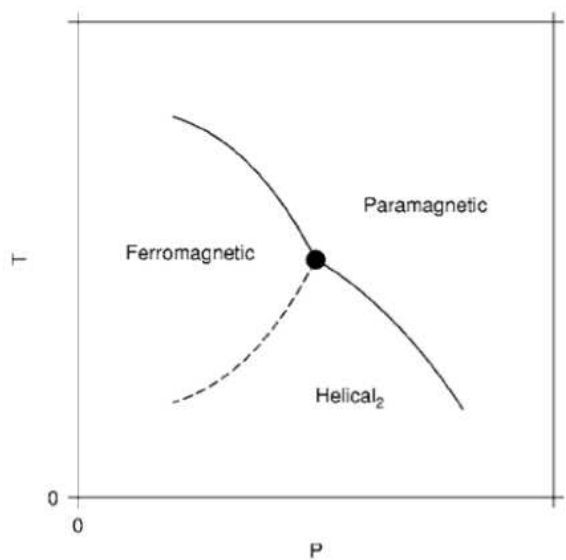}
\caption{Diagrama de fases para um comportamento crítico de Lifshitz uniaxial de segundo caráter, onde o termo segundo caráter refere-se ao acoplamento até segundos vizinhos, exibido pelo modelo $ANNNI$. A linha tracejada indica a transição de primeira ordem entre duas fases ordenadas. A linha contínua corresponde a transição de segunda ordem entre fases desordenadas e ordenadas. A intersecção das linhas das transições é o denominado de ponto multicrítico de Lifshitz (ponto de Lifshitz). O parâmetro $p$ é definido por $p=J_2/J_1$.}
\label{fig:lifshitz}
\end{center}
\end{figure}

Sistemas competitivos, dentre eles os que exibem competição do tipo Lifshitz, têm atraído grande atenção em diferentes contextos nos últimos anos. No contexto de sistemas magnéticos \cite{stanley:1971,ma:1976,amit:1978,parisi:1988,itzykson:1989,Le:1991,Zinn:1989,binney:1993}, a competição  entre as interações ferromagnéticas entre os spins de uma rede, que são primeiros vizinhos, e a temperatura do sistema de spins, provoca o surgimento de duas fases, uma desordenada associada à temperatura, que tende a desordenar as orientações relativas dos spins e uma ordenada, onde os spins tendem a se alinhar em uma dada direção preferencial. Quando há a competição entre as interações ferromagnéticas entre os spins de uma rede que são primeiros vizinhos, as interações antiferromagnéticas entre spins que são segundos vizinhos e a temperatura, originam-se três fases distintas: uma desordenada, uma ordenada ferromagnética e uma ordenada modulada. Em cristais, a fase modulada no modelo $ANNNI$ ({\it axial next-nearest-neighbor Ising}, \cite{selke:1988,selke:1998}) descreve uma organização quasi-periódica unidimensional desenvolvida ao longo da direção da magnetização. Nesta fase o spin é representado por $s(x)=\cos(k_0 z)$, onde $k_0 = 2 \pi / \lambda$ é o vetor de onda da modulação. A fase incomensurada tem a razão $\lambda / a$ irracional, em que $a$ é o espaçamento da rede. O diagrama de fase do modelo é  representado na Figura \ref{fig:lifshitz}.

Ainda no contexto de sistemas magnéticos que exibem competição, vamos considerar $J_1$ como sendo a constante de acoplamento da interação de troca entre primeiros vizinhos e $J_2$ a constante de acoplamento análoga associadada à interação entre segundos vizinhos. Se para um determinado valor das interações de troca entre primeiros vizinhos for ferromagnético, isto é, $J_1>0$  e aquela interação de troca entre segundos vizinhos também for ferromagnético, isto é, $J_2>0$, é obtido os mesmos expoentes críticos de antes com interações entre apenas primeiros vizinhos, pois a interação favorece o alinhamento dos spins. Mas se variarmos $J_1$, $J_2$, obtemos que para alguns valores $J_1>0$ e $J_2<0$, o sinal negativo de $J_2$ favorece o estado onde os spins são antiparalelos. Assim variando a razão $J_2/J_1$ é equivalente a introduzirmos competição entre ferromagnetismo e antiferromagnetismo nesses modelos. 

Para esse último caso, temos para uma particular temperatura $T_L$ (Temperatura de Lifshitz) um ponto onde as três fases coexistem, chamado de ponto de Lifshitz. Esses pontos de Lifshitz foram introduzidos teoricamente na literatura pela primeira vez em 1975 \cite{hornreich:1975}. Desde então houve um grande estudo experimental desses pontos críticos em sistemas físicos de natureza completamente diferentes como supercondutores de alta temperatura \cite{sachdev:1992}, crístais líquidos \cite{huang:1981}, cristais líquidos ferroelétricos \cite{skarabot:2000}, ferroelétricos uniaxiais\cite{vysochanskii:1992}, alguns tipos de polímeros \cite{fredrickson:1991}, materiais magnéticos e ligas \cite{becerra:1980}. Há também trabalhos sobre esses pontos críticos envolvendo transições de fase quânticas \cite{ardonne:2004}.

\begin{figure}[htb]
\begin{center}
\includegraphics[scale=0.45]{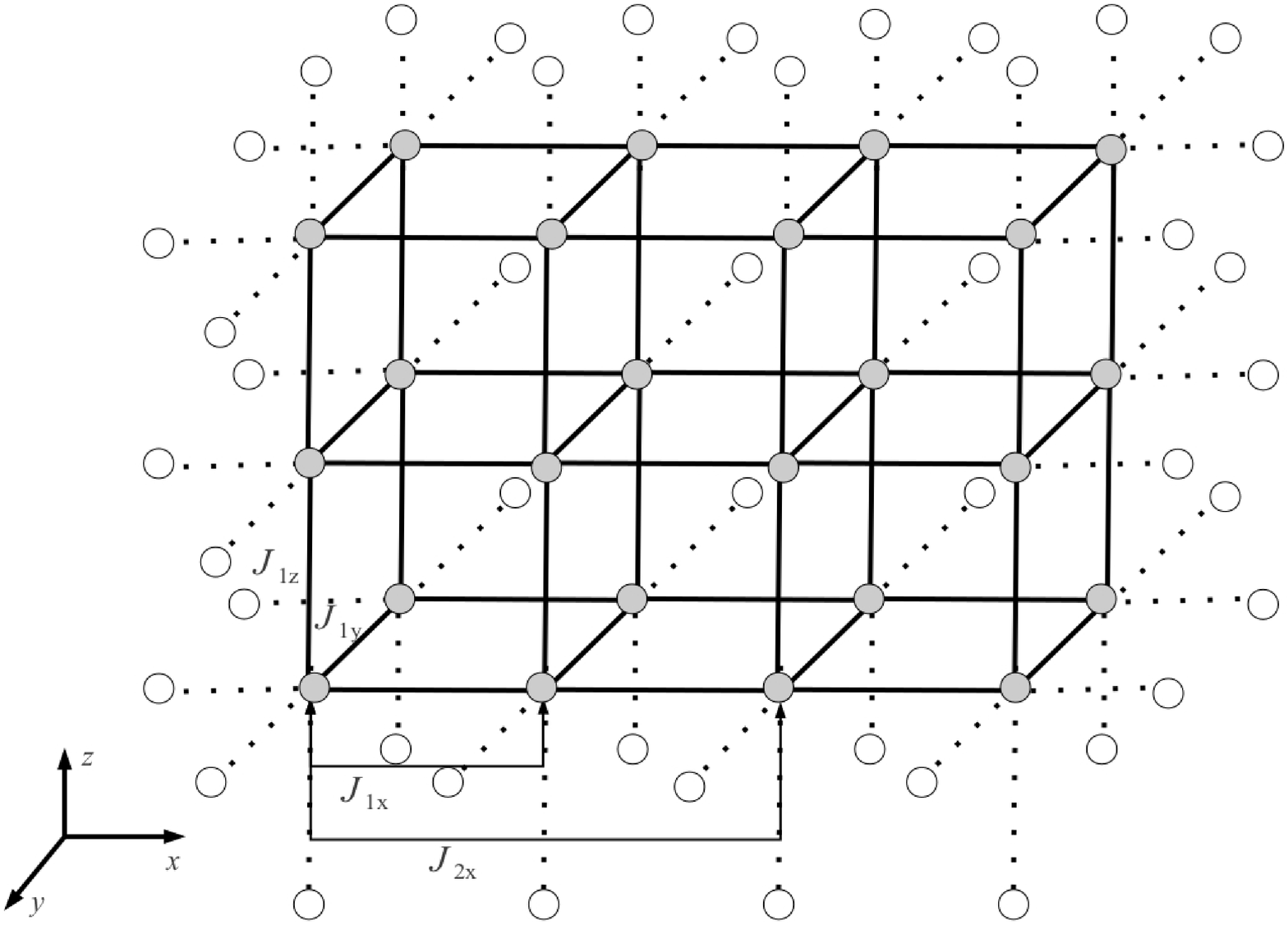}
\caption{Representação do modelo $ANNNI$ em $d=3$, que exibe o comportamento crítico de Lifshitz uniaxial de segundo caráter.}
\label{fig:uniaxial}
\end{center}
\end{figure}

A competição do tipo Lifshitz pode ser introduzida de duas maneiras, anisotrópica e isotrópica. O caso anisotrópico, o mais simples, ocorre quando existem interações ferromagnéticas ($J_1>0$) entre primeiros vizinhos ao longo de todas as direções espaciais e interações antiferromagnéticas ($J_2<0$) entre segundos vizinhos ao longo de uma determinada direção espacial, como mostra a Figura \ref{fig:uniaxial}. Essa anisotropia é chamada uniaxial e o comportamento crítico em tal situação é chamado de comportamento crítico de Lifshitz uniaxial. Esse comportamento crítico ocorre quando a razão $J_2/J_1$ tem um determinado valor na temperatura de Lifshitz. Para esse caso, existem sistemas reais com medidas experimentais descritos pelo modelo {\it ANNNI} \cite{leite:2003}. 

\begin{figure}[htb]
\begin{center}
\includegraphics[scale=0.5]{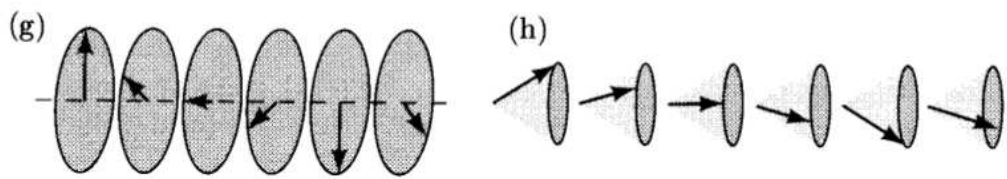}
\caption{Representação esquemática dos spins na fase modulada em (g) para o modelo $ANNNXY$ e (h) para o modelo $ANNNH$. Ambos são chamados de fases hélicas.}
\label{fig:fasemodulada}
\end{center}
\end{figure}

\begin{figure}[htb]
\begin{center}
\includegraphics[scale=0.45]{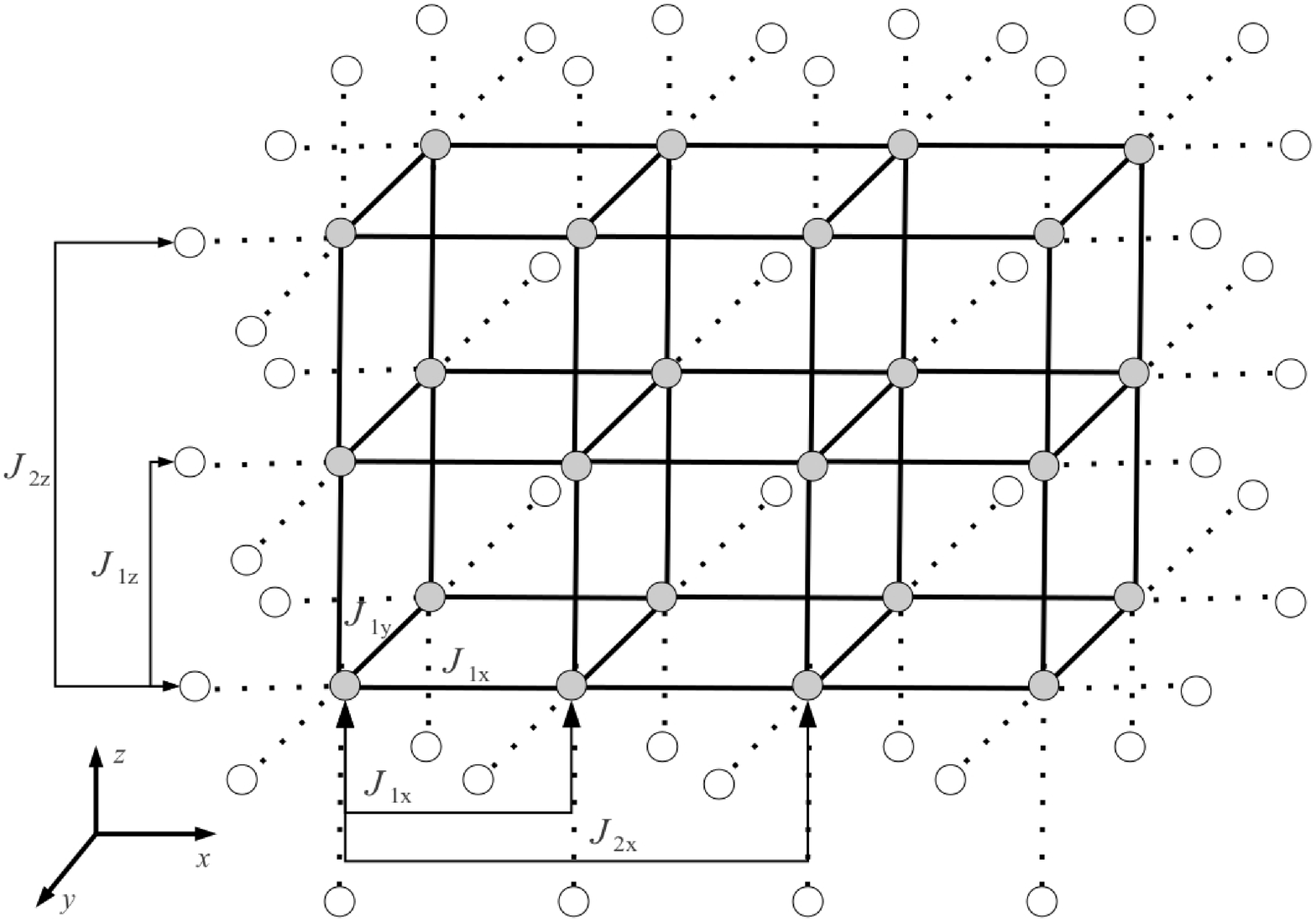}
\caption{Representação do modelo $ANNNI$ em $d=3$, que exibe o comportamento crítico de Lifshitz m-axial de segundo caráter, onde $m=2$.}
\label{fig:maxial}
\end{center}
\end{figure}

Várias propriedades críticas de pontos de Lifshitz têm sido estudadas recentemente \cite{leite:2000,albuquerque:2001,leite:2003}. Posteriormente, um modelo mais geral foi proposto \cite{leite2:2003}, no qual o sistema estudado agora possui uma dimensão arbitrária $d$ e parâmetro de ordem com um número de componentes $N$ qualquer ($N=1$ no caso de sistemas de spins de Ising representado pelo modelo $ANNNI$, $N=2$ para o modelo $XY$ representado pelo modelo $ANNNXY$ \footnote{Axial-Next-Nearest-Neighbor XY} e $N=3$ para o modelo de Heinsenberg representado pelo modelo $ANNNH$ \footnote{Axial-Next-Nearest-Neighbor Heisenberg}, onde esses modelos possuem diagramas de fases similares. Nesses últimos a fase modulada apresenta vetores de spin cujos valores médios precessionam em torno do eixo de modulação como mostra a Figura \ref{fig:fasemodulada} \cite{messias:2010,junior:2010}) e as interações antiferromagnética $J_2$ ocorrem entre spins que são segundos vizinhos e ao longo de $m$ direções espaciais, conhecidas como eixos de competição, definindo assim pontos de Lifshitz de segundo caráter $m$-axial, como mostra a Figura \ref{fig:maxial}. Esses sistemas pertencem a uma classe de universalidade caracterizada por ($N,d,m$) e exibem um comportamento crítico do tipo Lifshitz $m$-axial. Podemos ter competição isotrópica quando $J_1$ e $J_2$ têm componentes em todas as direções espaciais $d=m$. As classes de universalidade para esses sistemas são definidas por ($N,m$).

Em 2005 \cite{leite:2005}, o modelo anterior foi extendido mais ainda, onde as interações antiferromagnéticas ocorrem entre vizinhos mais distantes, até o L-ésimo vizinho. Esse é o caso de pontos de Lifshitz de caráter geral. Em ambos os casos, de segundo caráter e de caráter geral, a teoria estudada foi uma teoria quântica de campo escalar com interações do tipo $\lambda \phi^4$ em $d$ dimensões com massa nula, regularizada usando o método de regularização dimensional e renormalizada usando condições de normalização, fixando os momentos externos dos diagramas de Feynman. Também foi usado o método de subtração mínima de pólos dimensionais, onde os momentos externos são mantidos arbitrários em uma teoria sem massa ( o resultado final para grandezas físicas universais, usando métodos diferentes, são os mesmos e só dependem de $d$, $N$ e $m_2,m_3,...,m_L$). Posteriormente, o mesmo modelo foi usado para o cálculo dos mesmos expoentes críticos, mas agora em uma teoria massiva usando regularização dimensional e condições de normalização, tanto para pontos de Lifshitz de segundo caráter como para os de caráter geral \cite{carvalho:2008,carvalho:2009,carvalho2:2009}.

Nesta dissertação, investigamos o comportamento crítico de sistemas físicos com interações competitivas que apresentam pontos de Lifshitz $m$-axiais. Para esse estudo, resumimos os aspectos fundamentais no capítulo 2, onde abordamos as técnicas de Teoria Quântica de Campos Escalares Massivos com interações do tipo $\lambda \phi^4$ para obtermos uma expansão perturbativa para as funções de vértice de 2-pontos e de 4-pontos até a ordem de 2-{\it loops}. Essas funções de vértice foram regularizadas usando o método de regularização dimensional e renormalizadas usando o método de subtração mímina de pólos  dimensionais, onde foi adicionado contra-termos à Lagrangiana inicial, caracterizando o método BPHZ (Bogoliubov-Parasiuk-Hepp-Zimmermann) \cite{kleinert:2000}. Através das ideias do Grupo de Renormalização, foram definidas as funções de Wilson que originam os pontos fixos, e a partir dessas funções e dos pontos fixos, podemos calcular os expoentes críticos do sistema.

No capítulo 3, calculamos os expoentes críticos anisotrópicos $\eta_{\tau}$, até a ordem três em número de {\it loops}, e $\nu_{\tau}$ até a ordem dois em número de {\it loops}, que caracterizam o comportamento crítico do tipo Lifshitz $m$-axial. Todos os expoentes calculados estão em perfeita concordância com os correspondentes expoentes calculados anteriormente usando outros métodos, confirmando assim a conhecida e importante hipótese de universalidade.

No capítulo 4 discutimos os nossos resultados e concluímos com algumas perspectivas futuras resultantes dessa discussão.

\chapter{Método de BPHZ para Teoria $\lambda \phi^4$}

A fim de definir adequadamente uma teoria de campo, é necessário regularizar o comportamento de curtas distâncias dos sistemas. Isso pode ser feito com a ajuda da estrutura original do sistema ou por uma técnica matemática. Outra possiblidadae é de trabalhar no espaço-tempo contínuo, mas assumindo que todo o momento está confinado em uma esfera de raio $\Lambda$ grande, chamado { \bf espaço de momento de corte} ( {\it momentum space cutoff}). Tal corte, fornece um menor comprimento de escala sobre o qual o parâmetro de ordem pode variar. Temos que distâncias muito curtas só se tornam visíveis sob luz ultravioleta, neste caso $\Lambda$ é chamado de {\bf corte ultravioleta} ({\it ultraviolet cutoff}) ou {\bf corte UV} ({\it UV cutoff}), e uma teoria de campo contínuo com corte ultravioleta $\Lambda$ não possue divergências ultravioleta.

Na teoria quântica de campos, um sistema descrito pelo funcional de Ginzburg-Landau

\begin{equation}
E[\phi]=\int d^Dx\left\{\frac{A_1}{2}\partial_{i} \phi(\vec{x}) \partial_{i} \phi(\vec{x}) + \frac{A_2}{2}\phi^2(\vec{x}) + \frac{A_4}{4!}[\phi^2(\vec{x})]^2 \right\} ,
\end{equation}
é chamado de {\bf teoria $\phi^4$} \cite{kleinert:2000}. Equipado com diferentes simetrias, a teoria $\phi^4$ tem se tornado a mais atraente ferramenta teórica para o estudo de fenômenos críticos em uma grande variedade de sistemas estatísticos, onde sua relevância para a compreensão desses fenômenos foi enfatizado por Wilson e  posteriormenre por Fisher \cite{kleinert:2000}. Usando técnicas de teoria de campos, tem sido possível compreender de forma muito satisfatória todas as transições de fase de segunda-ordem de sistemas magnéticos e suas generalizações. Estes explicam outras transições de fase de muita importância. Por exemplo, em uma teoria $\phi^4$ de simetria $O(2)$, tem as mesmas propriedades críticas do superfluido $^{4}He$ na transição para a fase normal. Se os expoentes críticos de modelos $O(N)$ são contínuos analiticamente para $N=0$, obtém-se valores obervados em soluções diluidas de polímeros. 

Neste capítulo, iremos descrever de forma sucinta a ferramenta necessária para o cálculo dos expoentes críticos de sistemas do tipo Lifshitz. O leitor interessado, pode consultar a referência \cite{kleinert:2000} para um estudo mais detalhado da teoria abordada neste capítulo.

\section{Definição da Teoria $\lambda\phi^4$}

Temos que uma teoria de campo térmico flutuante é definida por meio da representação de integrais funcionais de uma função de partição, em que os campos são acoplados linearmente às fontes externas. Isso constitui um funcional gerador, a partir do qual, todas as quantidades termodinâmicas e funções de correlação do sistema podem ser obtidos por uma diferenciação funcional.

Todos os objetos de estudo abordados nesse capítulo, são descritos em termos de $N$ componentes do campo flutuante $\phi=(\phi_1(\vec{x}) \cdots \phi_N(\vec{x}))$ no espaço euclidiano $D$-dimensional. As componentes do campo interagem entre si, via um termo de quarta ondem nos campos: $\lambda T_{\alpha \beta \gamma \delta}\phi_{\alpha}\phi_{\beta}\phi_{\gamma} \phi_{\delta}$, $(\alpha, \beta, \gamma, \delta =1,\ldots,N)$, onde o parâmetro $\lambda > 0$ caracteriza a força de interação e é chamado de constante de acoplamento. A quantidade $T_{\alpha \beta \gamma \delta} $ é o tensor de acoplamento. 

O campo $\phi(\vec{x})$ realiza flutuações do parâmetro de ordem e $m^2$ é o termo que realiza as flutuações térmicas do sistema. Logo podemos investigar as propriedades do sistema, principalmente na fase normal, onde o valor esperado do campo $\phi$ é zero, que acontece no regime da temperatura em que a simetria do sistema é quebrada. Seu tamanho é controlado por uma energia funcional local, dado por

\begin{equation}
E[\phi]=E_0[\phi]+E_{int}[\phi] \label{energiafuncional} ,
\end{equation}
onde

\begin{equation}
E_0[\phi]=\int d^Dx \frac{1}{2} \left\{ [\partial_x \phi(\vec{x})]^2 + m^2\phi^2(\vec{x})  \right\} ,
\end{equation}
é o termo quadrático do campo chamado de energia do campo livre, e

\begin{equation}
E_{int}[\phi]=\int d^Dx \frac{\lambda}{4!} \phi^4(\vec{x}) ,
\end{equation}
é o termo de quarta ordem do campo chamado de energia de interação.

Assim, a energia funcional pode escrita na forma

\begin{equation}
E[\phi]=\int d^Dx \left\{\frac{1}{2}[(\partial_x \phi(\vec{x}))^2 + m^2\phi^2(\vec{x})] + \frac{\lambda}{4!} \phi^4(\vec{x})   \right\} . \label{energiafuncional2}
\end{equation}


Na expressão anterior, temos que o termo quadrático no campo é o termo de massa em Teoria Quântica de Campos. E temos também que, em vez de flutuações térmicas, o campo $\phi(\vec{x})$ realiza flutuações quânticas que podem equivalentemente ser descrita por um operador quântico de campos $\hat{\phi}(\vec{x})$, que é capaz de criar e aniquilar partículas de massa $m$ em um espaço de Fock.

Todas as propriedades termodinâmicas do sistema são descritas pela função de partição do campo flutuante, que é dado pelo funcional

\begin{equation}
Z^{phys}=\int D\phi(\vec{x})e^{-E[\phi]/k_bT} , \label{fparti}
\end{equation}
no qual $k_b$ é a constante de Boltzmann's e $T$ é a temperatura. A expressão da função de partição pode ser abreviada renormalizando o campo, a massa e a constante de acoplamento $\lambda$ de tal forma que $k_bT=1$.

Temos que a função de partição (\ref{fparti}) descreve todas as propriedades termodinâmicas do sistema, mas ela não dá qualquer informação das propriedades locais do sistema que são observadas em experimentos de espalhamento. Essa informação é dada pela função de correlação do campo $\phi(\vec{x})$, através

\begin{equation}
G^{(n)}(\vec{x_1}, \ldots , \vec{x_n})= \langle \phi(\vec{x_1}) \cdots \phi(\vec{x_n}) \rangle , \label{greenfunci1}
\end{equation}
com a notação

\begin{equation}
\langle \phi(\vec{x_1}) \cdots \phi(\vec{x_n}) \rangle = \frac{\int D\phi(\vec{x}) \phi(\vec{x_1}) \cdots \phi(\vec{x_n}) e^{-E[\phi]}}{\int D\phi(\vec{x})e^{-E[\phi]}} .
\end{equation}

A função de correlação pode ser escrita na forma

\begin{equation}
G^{(n)}(\vec{x_1}, \ldots , \vec{x_n})=Z^{-1}\int D\phi(\vec{x}) \phi(\vec{x_1}) \cdots \phi(\vec{x_n}) e^{-E[\phi]} ,
\end{equation}
que também é chamada de { \bf Funções de Green} ou {\bf Funções de $n$-pontos}.

Podemos descrever todas as funções de correlação do sistema de maneira compacta, através da introdução de um campo auxiliar externo $j(\vec{x})$, chamado de {\bf fonte} ({\it current}),  adicionado à energia funcional (\ref{energiafuncional}), temos uma interação linear de energia dessa corrente com o campo $\phi(\vec{x})$, onde

\begin{equation}
E_{source}[\phi ,j]=-\int d^Dx \phi(\vec{x}) j(\vec{x}) .
\end{equation}

A energia total é dada por

\begin{equation}
E[\phi , j]=E[\phi]+E_{source}[\phi ,j] .
\end{equation}

A função de partição formada com essa energia pode ser escrita como

\begin{equation}
Z[j]=\left[Z_0^{phys} \right]^{-1}\int D\phi(\vec{x})e^{-E[\phi ,j]} ,
\end{equation}
onde $Z_0^{phys}$ corresponde a função de partição do campo livre quando $\lambda = 0$.

A derivada funcional de $Z[j]$ com respeito a $j(\vec{x})$ calculada no ponto $j \equiv 0$, nos dá a função de correlação do sistema

\begin{equation}
G^{(n)}(\vec{x_1}, \ldots , \vec{x_n})=Z^{-1}\left[\frac{\delta^n Z[j]}{\delta j(\vec{x_1}) \cdots \delta j(\vec{x_n})}  \right]_{j \equiv 0} , \label{green}
\end{equation}
assim, $Z[j]$ é referido como {\bf funcional gerador} da teoria interagente, pois é utilizado para gerar as funções de Green de $n$-pontos (ou pernas externas).

Temos que as funções de correlação da teoria do campo livre são obtidas a partir da derivada funcional do funcional gerador da teoria, ou seja,

\begin{equation}
G_0^{(n)}(\vec{x_1}, \ldots , \vec{x_n}) = \left[\frac{\delta^n Z_0[j]}{\delta j(\vec{x_1}) \cdots \delta j(\vec{x_n})}  \right]_{j \equiv 0} .
\end{equation}

O funcional gerador da teoria do campo livre pode ser escrito na forma

\begin{equation}
Z_0[j]=e^{\frac{1}{2}\int d^Dx_1 d^Dx_2 j(\vec{x_1})D^{-1}(\vec{x_1},\vec{x_2})j(\vec{x_2})} ,
\end{equation}
onde $D$ é uma matriz funcional simétrica definida por $D=D(\vec{x_1}, \vec{x_2})=\delta^{(D)}(\vec{x_1} - \vec{x_2})(-\partial_{x_2}^2 + m^2)$.

Assim, obtemos as funções de correlação da teoria livre dadas por

\begin{eqnarray}
G_0^{(n)}(\vec{x_1}, \ldots , \vec{x_n}) = \frac{1}{2^{n/2}(n/2)!}\left\{\frac{\delta}{\delta j(\vec{x_1})} \cdots \frac{\delta}{\delta j(\vec{x_n})} \left[\int d^Dx_1 d^Dx_2 j(\vec{x_1})D^{-1}(\vec{x_1},\vec{x_2})j(\vec{x_2})   \right]^{n/2}  \right\}_{j \equiv 0} . \label{green2}
\end{eqnarray}

Fazendo agora uma expansão perturbativa do funcional gerador $Z[j]$ em série de potência em termos da constante de acoplamento $\lambda$ na forma

\begin{equation}
Z[j]=Z_0[j] + \sum_{p=1}^{\infty}\frac{1}{p!}\left(\frac{-\lambda}{4!} \right)^p \int D'\phi(\vec{x}) \int d^Dz_1 \cdots d^Dz_p \phi^4(z_1)\cdots \phi^4(z_p) e^{-(E_0[\phi] + E_{source}[\phi, j])} ,
\end{equation}
junto com a definição

\begin{equation}
\int D'\phi(\vec{x}) \equiv e^{\frac{1}{2} Tr log D} \int D \phi(\vec{x}) ,
\end{equation}
e o uso da expressão (\ref{green}), obtemos as funções de Green de $n$-pontos expressas em termos de uma série perturbativa como

\begin{eqnarray}
G^{(n)}(\vec{x_1}, \ldots , \vec{x_n})=Z^{-1}\left[ G_0^{(n)}(\vec{x_1}, \ldots , \vec{x_n}) + \sum_{p=1}^{\infty} G_p^{(n)}(\vec{x_1}, \ldots , \vec{x_n})  \right] , \label{correlacao}
\end{eqnarray}
onde $G_0^{(n)}(\vec{x_1}, \cdots , \vec{x_n})$ são as funções de $n$-pontos do campo livre, e as contribuições de $p$-ésima ordem com $p>1$ são dadas pelas integrais

\begin{eqnarray}
G_p^{(n)}(\vec{x_1}, \ldots , \vec{x_n})&=&\frac{1}{p!} \left(\frac{-\lambda}{4!} \right)^p \int d^Dz_1 \cdots d^Dz_p \times \nonumber \\
&& G_0^{(n+4p)}(\vec{z_1},\vec{z_1},\vec{z_1},\vec{z_1}, \ldots , \vec{z_p},\vec{z_p},\vec{z_p},\vec{z_p}, \vec{x_1}, \ldots ,\vec{x_n}) .
\end{eqnarray}

Através dos resultados (\ref{green2}) e (\ref{correlacao}), as funções de correlação $G^{(n)}$ podem ser expressas na forma

\begin{equation}
G^{(n)}(\vec{x_1}, \ldots , \vec{x_n})=Z^{-1}\sum_{p=0}^{\infty}G_p^{(n)}(\vec{x_1}, \ldots , \vec{x_n}) ,
\end{equation}
onde

\begin{eqnarray}
G_p^{(n)}(\vec{x_1}, \ldots , \vec{x_n})&=&\left(\frac{-\lambda}{4!} \right)^p \frac{1}{p!} \frac{1}{(2p+n/2)!2^{2p+n/2}}\frac{\delta}{\delta j(\vec{x_1})} \cdots \frac{\delta}{\delta j(\vec{x_n})} \times \nonumber \\
&& \left[\prod_{i=1}^{p} \int d^Dz_i \left(\frac{\delta}{\delta j(\vec{z_i})} \right)^4  \right] \left[\int d^Dx d^Dy j(\vec{x})G_0(\vec{x}, \vec{y})j(\vec{y})  \right]^{2p+n/2} . \label{green3}
\end{eqnarray}

Escrevendo cada termo de tal forma que coincida com argumentos $\vec{z}$ dos propagadores livres que são inicialmente distintos e aplicada em uma integral estendida por meio de funções $\delta$, podemos reescrever a expressão (\ref{green3}) na forma 

\begin{eqnarray}
G_p^{(n)}(\vec{x_1}, \ldots , \vec{x_n}) &\equiv& \frac{1}{p!}\left(\frac{-\lambda}{4!} \right)^p\int d^Dz_1 \cdots d^Dz_p \int d^Dy_1 \cdots d^Dy_{4p+n} \prod_{l=1}^{n}\left[\delta^{(D)}(\vec{y}_{4p+l}-\vec{x}_l) \right] \times \nonumber \\
&& \prod_{k=1}^{p}\left[\delta^{D}(\vec{y}_{4k-3}-\vec{z}_k) \delta^{D}(\vec{y}_{4k-2}-\vec{z}_k)     \delta^{D}(\vec{y}_{4k-1}-\vec{z}_k) \delta^{D}(\vec{y}_{4k}-\vec{z}_k) \right] \times \nonumber \\
&& \sum_{i=1}^{(4p+n-1)!!} \prod_{j=1}^{\frac{4p+n}{2}}G_0(\vec{y}_{\pi_{i}^{(4p+n)}(2j-1)},\vec{y}_{\pi_{i}^{(4p+n)}(2j)}) , \label{diagrama}
\end{eqnarray}
onde os pares $\pi_{i}(2j-1)$, $\pi_{i}(2j)$ indicam o número total de índices a partir do qual os pares foram selecionados. A soma inclui todas as possíveis permutações das variáveis $\vec{y}_i$, exceto para aqueles que correspondem apenas a uma troca dos argumentos espaciais num propagador, ou para uma troca de propagadores idênticos no conjunto. Cada termo do produto total da soma (\ref{diagrama}) é chamado de {\bf diagrama de Feynman} ou {\bf grafo de Feynman}.

Ao fazermos a expansão perturbativa em potências de $\lambda$, obtemos diagramas que são desconectados e diagramas que são conectados. Os diagramas desconectados são diagramas que podem ser escritos como um produto disjunto de dois diagramas. Já os Diagramas conectados não podem ser escritos como o produto disjunto de dois diagramas. Assim, qualquer termo da expansão perturbativa pode, em geral, ser escrito como uma soma de diagramas conectados e desconectados.

Iremos considerar apenas diagramas conectados, para uma dada ordem em $\lambda$, com o sentido de trabalharmos com um número menor de diagramas. De fato, tal expansão pode ser obtida, considerando o funcional gerador

\begin{equation}
Z[j]=e^{W[j]} ,
\end{equation}
onde obtemos

\begin{equation}
G_c^{(n)}(\vec{x}_1, \ldots ,\vec{x}_n)=Z^{-1}\left[\frac{\delta}{\delta j(\vec{x}_1)}\cdots  \frac{\delta}{\delta j(\vec{x}_n)} W[j] \right]_{j \equiv 0} ,
\end{equation}
que são as funções de correlação conectadas no espaço das coordenadas.

A equação (\ref{green}) é uma representação das funções de Green de $n$-pontos no espaço das coordenadas. Para o problema que abordaremos, com simetria translacional, é conveniente trabalharmos em uma representação das funções de Green de $n$-pontos no espaço dos momentos. Aplicando a transformada de Fourier em todos os argumentos das funções de Green de $n$-pontos $G^{(n)}(\vec{x_1}, \ldots , \vec{x_n})$, obtemos as funções de $n$-pontos no espaço dos momentos definida por

\begin{equation}
G^{(n)}(\vec{k_1}, \ldots , \vec{k_n})= \int d^Dx_1 \cdots d^Dx_n e^{-i(\vec{k_1} \cdot \vec{x_1} + \cdots + \vec{k_n} \cdot \vec{x_n})} G^{(n)}(\vec{x_1}, \ldots , \vec{x_n}) , \label{fourier}
\end{equation}
que nos leva a obter 

\begin{equation}
G^{(n)}(\vec{k_1}, \ldots , \vec{k_n})= Z^{-1}\left[\frac{\delta^n Z[j]}{\delta j(-\vec{k_1}) \cdots \delta j(-\vec{k_n})}  \right]_{j\equiv 0} .
\end{equation}

Uma vez que as funções de correlação estão associadas com os diagramas de Feynman desconectados, que podem ser fatorados das partes conectadas, esta associação é também estendida para as suas transformadas de Fourier. Assim, é suficiente a criação de regras de Feynman para os diagramas conectados no espaço dos momentos. Logo, considerando $\vec{x}_k (k=1,\ldots,n)$ como pontos externos e $\vec{z}_i (i=1,\ldots,p)$ como pontos internos dos diagramas, podemos representar cada momento como sendo uma linha. Dependendo dos pontos finais, pode-se distinguir as linhas externas $(k=1,\ldots,n)$ e as linhas internas $\vec{p}_i (i=1, \ldots,I)$, onde o número de linhas internas $I$ é determinado por $n$ e o número de vértices $p$, através da expressão \cite{kleinert:2000}

\begin{equation}
I=\frac{(4p-n)}{2} .
\end{equation}

Assim, omitindo o fator de acoplamento $(-\lambda)^p$ e o fator de peso $W_G$\footnote{O fator de peso é dado pela expressão $W_G=\frac{M_G}{4!^pp!}$, onde $M_G$ caracteriza a multiplicidade da integral de Feynman.\cite{kleinert:2000}}, temos que as funções de correlação conectadas de $G_p^{(n)}(\vec{x}_1, \ldots, \vec{x}_n)$ podem ser dadas pela expressão

\begin{equation}
G_c^{(n)}(\vec{x}_1,\ldots,\vec{x}_n)=\int d^Dz_1 \cdots d^Dz_p \prod_{i=1}^{n}G_0(\vec{x}_i - \vec{z}_i) \times \prod_{j=1}^{I}G_0(\vec{z}_i - \vec{z}_j) , \label{greenconect}
\end{equation}
onde a transformada de Fourier das funções de correlação conectadas é definida como na equação (\ref{fourier})

\begin{equation}
G_c^{(n)}(\vec{k_1}, \ldots , \vec{k_n})= \int d^Dx_1 \cdots d^Dx_n e^{-i(\vec{k_1} \cdot \vec{x_1} + \cdots + \vec{k_n} \cdot \vec{x_n})} G_c^{(n)}(\vec{x_1}, \ldots , \vec{x_n}) . \label{fourierconect}
\end{equation}

Através da expressão (\ref{fourierconect}), devemos observar que ela contém $n$ fatores $e^{-i\vec{k}_k\cdot\vec{x}_k}$, cujos momentos $\vec{k}_k$ são representados por linhas externas. E cada ponto externo $\vec{x}_k$ que aparece na função de Green, é representado pela transformada de Fourier

\begin{equation}
G_0(\vec{x}_k - \vec{z}_i)=\int \frac{d^Dk_k}{(2\pi)^D}e^{i\vec{k}_k(\vec{x}_k - \vec{z}_i)}G_0(\vec{k}_k) ,
\end{equation}
que contribui para a integral (\ref{greenconect}) com um fator exponencial $e^{i\vec{k}_k(\vec{x}_k - \vec{z}_i)}$. Temos também que cada par de pontos internos $\vec{z}_i$ que aparece na função de Green

\begin{equation}
G_0(\vec{z}_i - \vec{z}_j)=\int \frac{d^Dp}{(2\pi)^D} e^{i\vec{p}(\vec{z}_i - \vec{z}_j)}G_0(\vec{p}) ,
\end{equation}
contribui com um fator exponencial $e^{i\vec{p}(\vec{z}_i - \vec{z}_j)}$.

Como cada um dos pontos externos $\vec{x}_i$ aparece duas vezes nos fatores de fase, as integrais sobre $\vec{x}_i$ produzem um fator $(2\pi)^D\delta^{(D)}(\vec{k} - \vec{p})$ para cada linha externa. Subsequentemente, as $n$ integrais sobre o correspondente momento $\vec{p}$, pode todos serem feitos, levando as $\frac{(4p-n)}{2}$ integrais do momento não triviais, para cada linha interna. Temos também que cada vértice aparece em quatro fatores exponenciais $e^{i\vec{p}\vec{x}}$. As integrais sobre os vértices internos de posições $\vec{z}_i$, produz $p$ distribuições $\delta$ que expressam a conservação do momento em cada vértice. Um desses podem ser escolhidos para conter a soma sobre todos os momentos externos. Isso garante totalmente a conservação do momento. As demais integrais, podem simplesmente serem feitas, por meio da remoção de $p-1$ integrações. Assim, é possível acabar com

\begin{equation}
L=I-p+1 ,
\end{equation}
integrais não triviais sobre as variáveis do momento $I_i$, que é o {\bf {\it loop} do momento}, que está associado com {\it loops} independentes nos diagramas.

Logo, a transformada de Fourier de $G_c^{(n)}(\vec{x}_1, \ldots, \vec{x}_n)$ é dado na forma

\begin{eqnarray}
G_c^{(n)}(\vec{k}_1, \ldots, \vec{k}_n)&=&G_0(\vec{k}_1)\cdots G_0(\vec{k}_n)(2\pi)^D\delta^{(D)} \left(\sum_{i=1}^{n}\vec{k}_i  \right) \times \nonumber \\
&& \int \frac{d^Dl_1}{(2\pi)^D}\cdots \frac{d^Dl_L}{(2\pi)^D}G_0(\vec{p}_1(\vec{l},\vec{k}))\cdots G_0(\vec{p}_I(\vec{l},\vec{k})) , \label{greenconet2}
\end{eqnarray}
onde cada linha do momento é expressado pela combinação do {\it loop} do momento $\vec{l}_i(i=1, \ldots, L)$ e do momento externo $\vec{k}_i(i=1, \ldots, n)$, abreviado por $(\vec{l},\vec{k})$.

Com isso, reintroduzindo os fatores omitidos anteriormente, vemos que os diagramas de Feynman no espaço dos momentos contém:

1. Um fator $G_0(\vec{k}_i)$ para cada linha externa;

2. Um fator $G_0(\vec{p}_j)$ com $j=1, \ldots, I$ para cada linha interna, onde cada momento $\vec{p_j}$ tem uma orientação e é expresso por uma combinação de $L=I-p+1$ loops de momento e $n$ momentos externos;

3. Uma integração sobre cada loop de momento independente $\int \frac{d^Dl_i}{(2\pi)^D}(i=1, \ldots, L)$;

4. Um fator $(2\pi)^D\delta^{(D)}(\vec{k}_1 + \cdots + \vec{k}_n)$, para garantir totalmente a conservação do momento;

5. Um fator $-\frac{\lambda}{4!}$ para cada vértice; e

6. Um fator de peso $W_G$ do diagrama.

Após estabelecido tais regras dos diagramas de Feynman no espaço dos momentos, temos que é conveniente trabalhar com uma expansão perturbativa ainda mais simplificada onde temos apenas diagramas cujos propagadores associados às pernas externas são omitidos, pois estes propagadores das pernas externas entram como fatores multiplicativos que não envolvem integrais. Desses diagramas, os que não podem ser separados em dois cortando-se apenas uma linha são chamados de $1PI$ (irredutíveis a uma partícula). Eles representam o menor diagrama de Feynman não trivial que forma blocos de construção básica para todos os diagramas.

As partes de vértice 1PI são obtidas, formalmente, das funções de Green conectadas através de uma transformação de Legendre

\begin{equation}
\Gamma\{\overline{\phi}\}=\sum_{i}\overline{\phi}(i)J(i)-F\{J\} \label{gerador1pi}
\end{equation}

de onde podemos obter

\begin{equation}
\langle \phi(i) \rangle \equiv \langle \phi(k_i) \rangle \equiv \overline{\phi}_i = \frac{\delta F\{J\}}{\delta J(i)}
\end{equation}

e 

\begin{equation}
\frac{\delta \Gamma\{\overline{\phi}\}}{\delta \overline{\phi}(i)}=J(i)
\end{equation}

Da mesma forma que $Z\{J\}$ foi expandido nos campos externos para defini-lo como funcional gerador das funções de Green, podemos expandir $\Gamma\{\overline{\phi}\}$ em termos das médias dos campos $\overline{\phi}$ e então identificarmos os coeficientes como sendo as partes de vértice 1PI da seguinte maneira

\begin{equation}
\Gamma^{(N)}(1,\ldots, N) \equiv \frac{\delta^{N}\Gamma\{\overline{\phi}\}}{\delta\overline{\phi}(1)\cdots \delta \overline{\phi}(N)}
\end{equation}

Assim, para cada diagrama $1PI$ com $n>2$, é considerado o produto de {\it loops} de integrais na equação (\ref{greenconet2}), como sendo {\bf Funções de Vértice}, que são definidas a menos de um sinal negativo por convenção, dado por

\begin{equation}
\overline{\Gamma}^{(n)}(\vec{k}_1, \ldots, \vec{k}_n) \equiv - \int \frac{d^Dl_1}{(2\pi)^D} \cdots \frac{d^Dl_L}{(2\pi)^D} G_0(\vec{p}_1(\vec{l},\vec{k}))\cdots G_0(\vec{p}_I(\vec{l},\vec{k})), \hspace*{0.4cm} n>2.
\end{equation}

A parte $1PI$ da função de correlação $G_c^{(n)}(\vec{k}_1, \ldots, \vec{k}_n)$ pode ser expressa em termos das funções de vértice na forma

\begin{eqnarray}
G_c^{(n)}(\vec{k}_1, \ldots, \vec{k}_n)|_{1PI}=-G_0(\vec{k_1})\cdots G_0(\vec{k_n})(2\pi)^D\delta^{(D)}\left(\sum_{i=1}^{n}\vec{k}_i \right)\overline{\Gamma}^{(n)}(\vec{k}_1, \ldots, \vec{k}_n), \hspace*{0.4cm} n>2.
\end{eqnarray}

Agora definindo a função de correlação, em que possamos remover a conservação do momento, na forma

\begin{equation}
G^{(n)}(\vec{k}_1, \ldots, \vec{k}_n) \equiv (2\pi)^D \delta^{(D)}\left(\sum_{i=1}^{n}\vec{k}_i  \right)\overline{G}^{(n)}(\vec{k}_1, \ldots, \vec{k}_n) ,
\end{equation}
onde as funções $\overline{G}^{(n)}(\vec{k}_1, \ldots, \vec{k}_n)$ são definidas somente para argumentos em que a soma é nula, e assim podemos escrever a função de vértice de 4-pontos em partes $1PI$ na forma

\begin{equation}
\overline{\Gamma}^{(4)}(\vec{k}_1,\vec{k}_2,\vec{k}_3,\vec{k}_4) \equiv -G^{-1}(\vec{k}_1)G^{-1}(\vec{k}_2)G^{-1}(\vec{k}_3)G^{-1}(\vec{k}_4)\overline{G}_c^{4}(\vec{k}_1,\vec{k}_2,\vec{k}_3,\vec{k}_4) , \label{relavergreen}
\end{equation}
onde a sua expansão diagramática até a ordem de 2-{\it loops} é dado por

\begin{eqnarray}
\overline{\Gamma}^{(4)}(\vec{k}_1,\vec{k}_2,\vec{k}_3,\vec{k}_4)= - \Biggl[
 \parbox{8mm}{ \begin{picture}(19,20) (324,-152)
    \SetWidth{1.0}
    \SetColor{Black}
    \Line(325,-134)(342,-151)
    \Line(325,-151)(342,-134)
    \Vertex(333.8,-142.6){1}
  \end{picture}} +
\frac{3}{2}\parbox{8mm}{ \begin{picture}(30,24) (218,-169)
    \SetWidth{1.0}
    \SetColor{Black}
    \Arc(233,-157)(8,252,612)
    \Line(241.5,-157)(246,-151)
    \Line(241.5,-157)(246,-163)
    \Line(224.5,-157)(220,-151)
    \Line(224.5,-157)(220,-163)
    \Vertex(224.5,-157){1.5}
    \Vertex(241.5,-157){1.5}
  \end{picture}} \hspace*{0.2cm} +
3 \parbox{8mm}{\begin{picture}(39,29) (156,-201)
    \SetWidth{1.0}
    \SetColor{Black}
    \Arc(179,-186)(11,270,630)
    \Line(168,-186)(157,-175)
    \Line(168,-186)(157,-197)
    \Arc(193.5,-186.5)(13.509,87.879,272.121)
    \Vertex(168,-186){1.5}
    \Vertex(184,-177){1.5}
    \Vertex(184,-196){1.5}
  \end{picture}} \hspace*{0.4cm}  +
\frac{3}{4}\parbox{8mm}{ \begin{picture}(44,16) (142,-142)
    \SetWidth{1.0}
    \SetColor{Black}
    \Arc(156,-134)(7.28,254,614)
    \Arc(171,-134)(7.28,254,614)
    \Vertex(163.5,-134){1.5}
    \Line(178,-134)(185,-128)
    \Line(178,-134)(185,-141)
    \Line(149,-134)(143,-128)
    \Line(149,-134)(143,-141)
    \Vertex(149,-134){1.5}
    \Vertex(178,-134){1.5}
  \end{picture}} \hspace*{0.7cm} +
  \frac{3}{2} \parbox{8mm}{ \begin{picture}(30,24) (218,-168)
    \SetWidth{1.0}
    \SetColor{Black}
    \Arc(233,-157)(8,252,612)
    \Arc(233,-145)(4.472,117,477)
    \Line(241,-157)(246,-151)
    \Line(241,-157)(246,-163)
    \Line(224,-157)(220,-151)
    \Line(224,-157)(220,-163)
    \Vertex(233,-149){1.5}
    \Vertex(224.7,-157){1.5}
    \Vertex(241.5,-157){1.5}
  \end{picture}} \hspace*{0.2cm}
  \Biggr]    .
\end{eqnarray}

De modo análogo, podemos obter a expansão diagramática da função de vértice de 2-pontos em partes $1PI$ até a ordem de 2-{\it loops}, que é dada por

\begin{eqnarray}
\overline{\Gamma}^{(2)}(\vec{k}_1,\vec{k}_2)=
\left(\begin{picture}(33,2) (113,-160)
    \SetWidth{1.0}
    \SetColor{Black}
    \Line(114,-159)(145,-159)
  \end{picture} \right)^{-1} - \left[
  \frac{1}{2}  \parbox{8mm}{\begin{picture}(30,17) (212,-157)
    \SetWidth{0.9}
    \SetColor{Black}
    \Arc(227,-144)(7.071,262,622)
    \Line(213,-152)(241,-152)
    \Vertex(227,-151.5){1.5}
  \end{picture}} \hspace*{0.2cm} +
 \frac{1}{4} \parbox{8mm}{ \begin{picture}(28,30) (204,-141)
    \SetWidth{0.9}
    \SetColor{Black}
    \Arc(218,-129)(6.083,261,621)
    \Line(205,-136)(231,-136)
    \Vertex(218,-135.5){1.5}
    \Arc(218,-116)(6.083,261,621)
    \Vertex(218,-122){1.5}
  \end{picture}} \hspace*{0.2cm} +
  \frac{1}{6} \parbox{8mm}{ \begin{picture}(29,18) (152,-182)
    \SetWidth{1.0}
    \SetColor{Black}
    \Arc(166,-173)(8.246,256,616)
    \Line(152,-173)(180,-173)
    \Vertex(158,-173){1.5}
    \Vertex(174,-173){1.5}
  \end{picture}} \hspace*{0.2cm} \right] .
\end{eqnarray}

Com isso, mostramos as expansões diagramáticas até a ordem de 2-loops das funções de vértices de 2 e 4-pontos que será o objeto de estudo das seções subsequentes.

\section{Divergências na Teoria $\phi^4$}

Ao fazermos a expansão diagramática das funções $1PI$, encontramos o fato de que as integrais de Feynman decorrentes da expansão são divergentes. Estas divergências dependem da dimensão do espaço onde a teoria de campo está definida. Considerando a integral associado ao diagrama da função de 4-pontos de um {\it loop} dada por

\begin{eqnarray}
\int \frac{d^Dp}{(2\pi)^D}\frac{1}{(p^2+m^2)[(p+k)^2+m^2]} ,
\end{eqnarray}
podemos observar através de uma simples análise da dimensionalidade dessa função de vértice, que ela possui $D-4$ potências de momento. Então, se $D=4$, teremos uma divergência logarítmica, que é a divergência mais fraca que pode ocorrer, pois para $D<4$ a integral será finita, e para $D>4$ teríamos divergências linear, quadrática e assim por diante, onde estamos considerando o limite superior da integração sendo $\zeta$, onde as divergências aparecem para $\zeta \longrightarrow \infty$. Podemos observar que quanto mais propagadores internos a função de vértice possuir, menor será o grau de divergência para uma determinada dimensão, sendo que cada propagador é responsável por uma diminuição quadrática na divergência total da função. As divergências citadas acima são conhecidas como divergências do regime ultravioleta $(\zeta \longrightarrow 0)$, ou seja, de pequeno comprimento de onda. Elas não estão associadas à física do sistema e por isso devemos utilizar algum método para extraí-las. 

Métodos de regularização são empregados com esse propósito. Dentre os vários procedimentos de regularização existentes, neste trabalho usamos o método da regularização dimensional de 't Hooft e Veltman. A ideia desse método é calcular a integral de Feynman para um número de valores contínuos de dimensões $D$ para o qual a convergência é assegurada. Para isso as integrais de Feynman para um inteiro $D$, obriga a ser extrapolada analiticamente para um complexo $D$. O conceito de dimensão contínua foi introduzida por Wilson e Fisher \cite{kleinert:2000} que inicialmente foi calculado quantidades físicas em $D=4-\varepsilon$ dimensões com $Re(\varepsilon)>0$, e expandida em potências de $\varepsilon$ para a dimensão $D=4$. Esse conceito foi subsequentemente incorporado na Teoria Quântica de Campos \cite{kleinert:2000}, dando origem a muitas aplicações na física estatística. A regularização dimensional de 't Hooft e Veltman foi a ferramenta matemática apropriada para tais expansões, onde uma explanação deste método é mostrado no apêndice B.2. Antes de fazermos o tratamento e remoção das divergências ultravioletas que damos o nome de { \bf renormalização}, onde abordaremos na próxima seção, iremos fazer algumas considerações sobre as divergências encontradas nos diagramas.

Para localizar as divergências UV dos diagramas, é usado uma simples contagem de potências. De acordo com as regras de Feynman estabelecidas na seção 1.1, uma integral de Feynman $I_G$ de um diagrama $G$ com $p$ vértices, podem conter {\it loops} de momento em geral. Assim, um diagrama com $I$ linhas internas contém $L=I-p+1$ {\it loops} de interação e assim $DL=D(I-p+1)$ potências de momento no numerador. Cada linha interna $I$ é associada com um propagador, assim contribuindo $2I$ potências no denominador. Assim, temos que

\begin{equation}
\omega(G)=DL-2I=(D-2)I+D-Dp ,
\end{equation}
são as potências de momento na integral de Feynman de um diagrama hipotético qualquer. O comportamento da integral em grandes momentos pode ser caracterizado pelo redimensionamento de todos os momentos internos como $\vec{p}\longrightarrow \lambda \vec{p}$ e pela observação do comportamento da potência 
 
\begin{equation}
I_G \propto \lambda^{\omega(G)},  \hspace{0.3cm} \lambda \longrightarrow \infty .
\end{equation}

A potência $\omega(G)$ é chamado de {\bf grau superficial de divergência} do diagrama $G$. Um diagrama $G$ com $\omega(G)\geq 0$ é {\bf superficialmente divergente}. Para $\omega(G)=0,2,\ldots$, a divergência superficial dos diagramas é {\bf logarítmica}, {\bf quadrática},$\ldots$, respectivamente. A divergência superficial surge a partir de regiões no espaço dos momentos das integrais de Feynman onde todos os {\it loops} dos momentos tornam-se simultaneamente grande. 

\begin{figure}[htb]
\begin{center}
\includegraphics[scale=0.6]{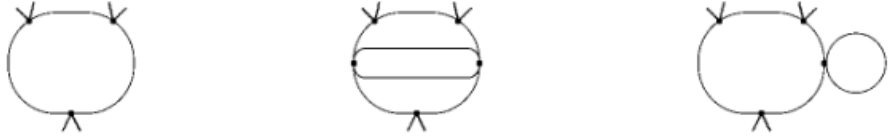}
\caption{Três diagramas superficialmente convergente com $\omega(G)=-2$. O primeiro diagrama não possui subdivergências. O segundo diagrama possui um subdiagrama $\iota$, do tipo (\ref{2klei}) com $\omega(\iota)=0$, logaritmicamente divergente e o terceiro diagrama possui um subdiagrama $\iota$, do tipo (\ref{tadklei}) com $\omega(\iota)=2$, quadraticamente divergente.}
\label{fig:klei}
\end{center}
\end{figure}

Um diagrama é dito ter {\bf subdivergências} se ele contém um {\bf subdiagrama} superficialmente divergente, isto é, um subdiagrama $\iota$ com $\omega(\iota) \geq 0$. Um subdiagrama é qualquer subconjunto de linhas e vértices de $G$ que forma um diagrama $\phi^4$ de ordem mais baixa na expansão perturbativa como podemos ver na figura (\ref{fig:klei}). Subdivergências vêm de regiões no espaço dos momentos onde o {\it loop} do momento do subdiagrama $\iota$ torna-se grande. Se um diagrama $G$ não possui subdivergências, mas se $\omega(G) \geqslant 0$, a divergência superficial é a única divergência da integral, e a integral de Feynman associada é convergente se um dos {\it loop} da integral é omitido, que corresponde a cortar uma das linhas. Logo podemos observar que um diagrama de Feynman $G$ é absolutamente convergente se o grau superficial de divergência $\omega(G)$ é negativo, e se o grau superficial de divergência $\omega(\iota)$ de todos os subdiagramas $\iota$ são também negativos. 

Para teorias $\phi^4$, o número de linhas internas $I$ no diagrama de Feynman pode ser expresso em termos do número de vértices $p$ e o número de linhas externas $n$ como

\begin{equation}
I=2p-\frac{n}{2} ,
\end{equation}
onde o grau superficial de divergência da integral associada torna-se portanto

\begin{equation}
\omega(G)=D + n(1-D/2) + p(D-4) , \label{graudivergencia}
\end{equation}
que em quatro dimensões fica na forma

\begin{equation}
\omega(G)=4-n ,
\end{equation}
implicando que em quatro dimensões, somente diagramas de 2-pontos e 4-pontos 1PI são superficialmente divergentes. Assim a única possibilidade de divergência das subintegrações são os subdiagramas de 2-pontos e 4-pontos. Se as integrais das funções de 2-pontos e 4-pontos são tornadas finitas por algum procedimento matemático, qualquer função de $n$-pontos serão finitos, de acordo com o teorema da divergência. Assim, uma teoria que possui essas propriedades é definida pelo termo "renormalizável". Daí a teoria $\phi^4$ é renormalizável em 4 dimensões.

Em 3 e 2 dimensões, temos que $\omega(G)=3-n/2 -p$ e $\omega(G)=2-2p$ respectivamente, implicando que somente alguns diagramas de baixa ordem possuem divergências. Uma teoria que possui essa propriedade é chamada {\bf super-renormalizável}. Acima de 4 dimensões, o último termo $p(D-4)$ na expressão (\ref{graudivergencia}) é positivo, implicando que novas divergências aparecerão em cada ordem mais alta na teoria de perturbação. Essa propriedade torna a teoria {\bf não-renormalizável}. 

Para uma teoria com uma potência arbitrária $r$ do campo na interação $\phi^r$, onde existe uma dimensão crítica dada por

\begin{equation}
D_c=\frac{r}{r/2-1} ,
\end{equation}
 para $D>D_c$, o grau superficial de divergência $\omega(G)$ torna-se independente do número $p$ de vértices. No caso da teoria $\phi^4$, a dimensão crítica é $D_c=4$.

\section{Renormalização}

Todas as integrais de Feynman associadas aos diagramas da expansão diagramática das funções de vértice de 2 e 4-pontos são infinitos em 4 dimensões. Após uma regularização dimensional em $4-\varepsilon$ dimensões, elas divergem especificamente para $\varepsilon \longrightarrow 0$. Para um dado número de {\it loops} $L$, as integrais de Feynman possuem singularidades do tipo $\frac{1}{\varepsilon^i}(i=1, \ldots, L)$ e essas divergências vem a conter todas as informações dos expoentes críticos da teoria na dimensão $4-\varepsilon$.

Ao expandirmos uma integral de Feynman de qualquer diagrama 1PI em uma série de potência no momento externo, podemos observar a seguinte propriedade: apartir de uma certa ordem do termo de expansão na função de 4-pontos e na função de 2-pontos, todos os maiores termos da expansão convergem par zero quando $\varepsilon \longrightarrow 0$. Precisamente esses termos são encontrados em uma expansão perturbativa da teoria em que a energia funcional (\ref{energiafuncional2}) é modificada, onde contém termos adicionais na forma $\int d^Dx\phi^4(x)$, $\int d^Dx \phi^2$ e $\int d^Dx(\partial \phi)^2$. Esses termos são da mesma forma que aqueles na energia funcional original, e esta propriedade faz com que a teoria seja {\bf renormalizável}. De fato, temos que observáveis finitos podem ser obtidos multiplicando, em cada função de correlação, campos, massa e constante de acoplamento por fatores de compensação, que são as constantes de renormalização $Z_{\phi}$, $Z_{m^2}$ e $Z_{g}$.

Usando a regularização dimensional para calcular as integrais de Feynman, é possível dar aos fatores a forma genérica

\begin{equation}
Z=1 + \sum_{k=1}^{L}g^k\sum_{i=1}^{k}\frac{c_i^k}{\varepsilon^i} , \label{constanteexpa}
\end{equation}
onde os coeficientes $c_i^k$ são c-números\footnote{ O termo c-número (ou número clássico) é uma antiga nomenclatura utilizada por Paul Dirac, que se refere aos números reais e complexos.} puros. As constantes de renormalização convertem objetos iniciais na energia funcional como campos, massa e constante de acoplamento não-renormalizados ({\it bare}), em campos, massa e constante de acoplamento renormalizados (finitos). Assim, os coeficientes $c_i$ da expansão são determinados para cada ordem de $g$\footnote{Temos que $g$ é a constante de acoplamento adimensional, dada através da relação de escala $g \equiv \mu^{D-4}\lambda = \lambda\mu^{-\varepsilon}$, onde $\mu$ é um parâmetro de massa arbitrário.} para cancelar as divergências existentes. Onde sendo feito de forma consistente, todos os observáveis tornam-se finitos em $\varepsilon \longrightarrow 0$.

Temos que o procedimento de renormalização pode ser realizado basicamente em duas maneiras diferentes, que diferem em suas ênfases entre as quantidades não-renormalizadas ({\it bare}) e quantidades renormalizadas na energia funcional. O método mais eficiente trabalha com quantidades renormalizadas que será adotado neste trabalho. Iniciando com a energia funcional contendo imediatamente o campo renormalizado, a massa renormalizada e a constante de acoplamento renormalizado determinado em cada ordem de $g$, certos contra-termos divergentes são adicionados no campo da energia para remover as divergências, caracterizando o método de contra-termos.

\subsection{Condições de Normalização}

As funções de vértice de 2-pontos e 4-pontos podem se tornar finitas através de uma re\-nor\-ma\-li\-za\-ção multiplicativa empregando as três constantes de renormalização $Z_{\phi}$, $Z_{m^2}$ e $Z_{g}$, que são as constantes de renormalização do campo, da massa e da constante de acoplamento  respectivamente. Através da renormalização multiplicativa são estabelecidas as seguintes relações entre quantidades renormalizadas e não-renormalizadas

\begin{equation}
\overline{\Gamma}_B^{(2)}(k)=Z_{\phi}^{-1}\overline{\Gamma}^{(2)}(k) ,
\end{equation}

\begin{equation}
m_B^2=Z_{m^2}Z_{\phi}^{-1}m^2 \label{massarenor} ,
\end{equation}

\begin{equation}
g_B=Z_{g}Z_{\phi}^{-2}g \label{acoplarenor} ,
\end{equation}

\begin{equation}
\phi_B = Z_{\phi}^{1/2}\phi  \label{camporenor} .
\end{equation}

Existem vários esquemas de renormalização. Em particular, estamos interessados em renormalizar a teoria quando os momentos externos são fixos, pois isto simplifica a nossa análise. Quando os momentos externos são fixos, utilizamos condições de normalização e no nosso caso, condições de normalização para teorias massivas. Os expoentes críticos podem ser calculados usando uma teoria massiva em que os valores dos momentos externos são nulos. Neste caso, considera-se que os parâmetros da energia funcional (\ref{energiafuncional2}), $m_B$ e $g_B$\footnote{Neste trabalho iremos considerar termos com índice $B$ como sendo termos não-renormalizados e sem índice, como sendo termos renormalizados.}, são infinitos, e portanto não representam os valores reais das grandezas físicas associadas. Os valores reais dessas grandezas são dadas pelos novos parâmetros $m$ e $g$ renormalizados, respectivamente. O mesmo ocorre com as funções de vértice $\overline{\Gamma}_B(k_i,m_B,g_B)$ e $\overline{\Gamma}(k_i,m,g)$. Assim, impomos, às duas partes de vértice 1PI, as seguintes {\bf condições de normalização} em momentos externos nulos

\begin{equation}
\overline{\Gamma}^{(2)}(0,m,g)=m^2 ,  \label{condinor1}
\end{equation}

\begin{equation}
\frac{\partial}{\partial k^2} \overline{\Gamma}^{(2)}(0,m,g)=1 , \label{condinor2}
\end{equation}

\begin{equation}
\overline{\Gamma}^{(4)}(0,m,g)=g  \label{condinor3} .
\end{equation}

Através da definição da função de Green (\ref{greenfunci1}), da relação (\ref{camporenor}) e da expressão (\ref{relavergreen}), pode-se estabelecer uma relação generalizada entre as funções de vértice renormalizada e não-re\-nor\-ma\-li\-za\-das, dada na forma

\begin{equation}
\overline{\Gamma}_{B}^{(n)}(\vec{k},m_B,g_B,\Lambda)=Z_{\phi}^{-n/2}\overline{\Gamma}^{(n)}(\vec{k},m,g) .
\end{equation}

As potências da constante de renormalização $Z_{\phi}$, determinado pela renormalização da função de 2-pontos, absorve todas as divergências que permanecem após ter renormalizado a massa e a constante de acoplamento.

Como veremos adiante, ao renormalizarmos a teoria, usaremos para o cálculo dos expoentes críticos as funções de vértice $\overline{\Gamma}^{(2)}$ e $\overline{\Gamma}^{(4)}$. Assim, para esse cálculo, usaremos a expansão perturbativa até a ordem de três {\it loops} para $\overline{\Gamma}^{(2)}$ o que nos habilita a calcular o expoente $\eta$ até a ordem de três {\it loops}, e até a ordem de dois {\it loops} para $\overline{\Gamma}^{(4)}$, tornando possível o cálculo do expoente $\nu$ até a ordem de dois {\it loops}.

\subsection{Método de Contra-Termos e Subtração Mínima}

Dependendo do esquema de regularização, as integrais de Feynman possuem divergências que se manifestarão como polos em $\varepsilon$ do tipo $\varepsilon^{-n}$ no limite $\varepsilon \longrightarrow 0$, logo essas divergências podem ser removidas empregando campos, massas e constantes de acoplamento renormalizados desde o início. Assim, as quantidades renormalizadas podem ser vistas como funções das quantidades não-renormalizadas de $\varepsilon$.

A teoria renormalizada é definida com a ajuda da energia funcional renormalizada, dada por

\begin{equation}
E[\phi]=E_0[\phi] + E_{int}[\phi] , \label{energiarenormalizada}
\end{equation}
onde a parte livre é dada por

\begin{equation}
E_0[\phi]=\int d^Dx \left[\frac{1}{2}(\partial \phi)^2 + \frac{1}{2}m^2 \phi^2 \right] ,
\end{equation}
enquanto que a parte de interação, inclui termos quadráticos adicionais nos campos, chamados {\bf contra-termos}, além dos termos polinomiais de ordem mais alta, a saber

\begin{equation}
E_{int}[\phi]= \int d^Dx \left[\frac{g\mu^{\varepsilon}}{4!}\phi^4 + c_{\phi}\frac{1}{2}(\partial \phi)^2 +c_{m^2}\frac{1}{2}m^2\phi^2 + c_g \frac{g\mu^{\varepsilon}}{4!}\phi^4 \right] . \label{eneginteadd}
\end{equation}

Como os termos adicionais são do mesmo tipo que os originais, podemos escrever

\begin{equation}
E[\phi]= \int d^Dx \left[(1+c_{\phi})\frac{1}{2}(\partial \phi)^2 + (1+c_{m^2})\frac{1}{2}m^2\phi^2 + (1+c_g)\frac{g\mu^{\varepsilon}}{4!}\phi^4 \right] .
\end{equation}

Os contra-termos $c_{\phi}$, $c_{m^2}$ e $c_g$ produzem vértices adicionais na expansão diagramática. Assim, no espaço dos momentos, eles têm a forma

\begin{equation}
\parbox{8mm} {\begin{picture}(51,13) (111,-204)
    \SetWidth{1.0}
    \SetColor{Black}
    \Line(118,-198)(157,-198)
    \SetWidth{1.0}
    \Line(133,-192)(142,-203)
    \Line(142,-192)(133,-203)
  \end{picture}} \hspace*{0.9cm} = (-c_{m^2})m^2 , \label{ct1} 
\end{equation}

\begin{equation}
\parbox{8mm} {\begin{picture}(68,14) (65,-167)
    \SetWidth{1.0}
    \SetColor{Black}
    \Line(84,-160)(115,-160)
    \Arc(100,-160)(3,270,630)
  \end{picture}} \hspace*{1cm} = (-c_{\phi})\vec{k}^2 , \label{ct2}
\end{equation}

\begin{equation}
\parbox{8mm} {\begin{picture}(19,20) (324,-152)
    \SetWidth{1.0}
    \SetColor{Black}
    \Line(325,-134)(342,-151)
    \Line(325,-151)(342,-134)
    \Vertex(333.8,-142.6){3}
  \end{picture}} = (-c_{g})g\mu^{\varepsilon} . \label{ct3}
\end{equation}

A definição do vértice original agora inclui o parâmetro de massa $\mu$ dado por

\begin{equation}
\parbox{8mm} {\begin{picture}(19,20) (324,-152)
    \SetWidth{1.0}
    \SetColor{Black}
    \Line(325,-134)(342,-151)
    \Line(325,-151)(342,-134)
    \Vertex(333.8,-142.6){1}
  \end{picture}} = (-g)\mu^{\varepsilon} .
\end{equation}

Os contra-termos $c_{\phi}$, $c_{m^2}$ e $c_g$ são escolhidos de tal modo que todos os termos divergentes são subtraídos e as funções de Green são finitas para $\varepsilon \longrightarrow 0$, ordem por ordem na teoria de perturbação.

Através de uma análise dimensional da equação (\ref{eneginteadd}), pode-se mostrar que os contra-termos são adimensionais. Na regularização dimensional, eles podem portanto, depender somente da constante de acoplamento adimensional $g$, ou de combinações adimensionais como $\frac{m^2}{\mu^2}$ ou $\frac{\vec{k}^2}{\mu^2}$. Acontece que a combinação $\frac{\vec{k}^2}{\mu^2}$ aparece somente em passos intermediários no processo de re\-nor\-ma\-li\-za\-ção como $log \left(\frac{\vec{k}^2}{\mu^2} \right)$. Isto é fundamental para o programa de renormalização que todos os termos não locais se cancelam na expressão final para os contra-termos. Isso implica que os contra-termos $c_{\phi}$, $c_{m^2}$ e $c_g$ dependem somente de $g$, $\varepsilon$ e $\frac{m^2}{\mu^2}$. No esquema de subtração mínima, a dependência de $\frac{m^2}{\mu^2}$ desaparece. Então, a única dependência dimensional dos diagramas dos contra-termos consiste nos fatores $\vec{k}^2$, $m^2$ e $\mu^2$ nas equações (\ref{ct1}), (\ref{ct2}) e (\ref{ct3}), e o cancelamento dos termos logarítmos são observados explicitamente na renormalização em {\it 2-loops}.

Assim, as constantes de renormalização são definidas como

\begin{eqnarray}
Z_{\phi} &\equiv & 1 + c_{\phi} , \\
Z_{m^2} &\equiv & 1 + c_{m^2} , \\
Z_{g} &\equiv & 1 + c_g .
\end{eqnarray}

Com essa definição, a energia funcional pode ser escrita como

\begin{equation}
E[\phi]=\int d^Dx \left[ \frac{1}{2}Z_{\phi}(\partial \phi)^2 + \frac{1}{2} Z_{m^2}m^2\phi^2 + \frac{\mu^{\varepsilon}g}{4!}Z_g\phi^4 \right] , \label{energiafuncional3}
\end{equation}
cujos coeficientes dependentes de $\varepsilon$.

As quantidades $\phi$, $m$ e $g$ na equação (\ref{eneginteadd}) são respectivamente, campo, massa e constante de acoplamento renormalizados. A energia funcional difere da original por um fator $Z_{\phi}$ no termo gradiente, e forma original da teoria é recuperada pela renormalização multiplicativa considerando as relações (\ref{massarenor}), (\ref{acoplarenor}) e (\ref{camporenor}), o que nos leva a obter

\begin{equation}
E[\phi]=E[\phi_B]=\int d^Dx \left[\frac{1}{2}(\partial \phi_B)^2 + \frac{1}{2}m_B^2\phi_B^2 + \frac{g_B}{4!}\phi_B^4 \right] . 
\end{equation}

As condições de normalização (\ref{condinor1}), (\ref{condinor2}) e (\ref{condinor3}) podem ser usadas em conexão com qualquer regularização de integrais divergentes. Esse procedimento tem uma importante vantagem, principalmente ao se tratar do estudo da região crítica em $D<4$ dimensões. Ela pode ser usada, não somente para remover as divergências ultravioletas da teoria quadridimensional, mas também pequenos valores finitos de $\varepsilon$ no ponto crítico quando nos referimos no caso de teorias de massa nula, isto é, em dimensões espaciais menores do que 4. Assim, as condições de normalização definem os contra-termos que, para $m^2 \neq 0$, dependem da massa.

Uma enorme simplificação surge pela existência do procedimento de regularização em que os contra-termos tornam-se independentes da massa $m$, exceto por um fator trivial geral $m^2$ em $c_{m^2}$. Isto é conhecido como esquema de {\bf Subtração Mínima}. Nesse esquema, os contra-termos adquirem a forma genérica (\ref{constanteexpa}), em que os coeficientes de $g^n$ consistem de termos de pólo puro $\frac{1}{\varepsilon^i}$, sem partes finitas para $\varepsilon \longrightarrow 0$. Os coeficientes são c-números e não contém a massa $m$ ou o parâmetro de massa $\mu$ introduzida no processo de regularização dimensional.

A ausência da dependência da massa é a origem para outra importante propriedade das constantes de renormalização no esquema de subtração mínima. Eles têm sempre a mesma expansão em potências da constante de acoplamento $g$, mesmo se estes fossem inicialmente definidos para transportar uma função analítica arbitrária $f(\varepsilon)$ com $f(0)=1$ como um fator.

Em outras palavras, se redefiníssemos

\begin{equation}
g\mu^{\varepsilon} \longrightarrow g\mu^{\varepsilon}f(\varepsilon)=g\mu^{\varepsilon}(1+f_1\varepsilon + f_2\varepsilon^2 +\cdots) ,
\end{equation}
acharíamos a mesma expressão (\ref{constanteexpa}) para as constantes de renormalização $Z_{\phi}$, $Z_{m^2}$ e $Z_g$ em potências do novo $g$. A razão para isso é que sempre podemos escrever $f(\varepsilon)=c^{\varepsilon}$ e absorver o fator $c$ no parâmetro de massa $\mu$. Desde que esse parâmetro de massa não apareça no final da expansão (\ref{constanteexpa}) e o mesmo para o parâmetro de massa redefinido $\mu c$.

Devido à invariância do final das expansões sobre uma reescala de $g \longrightarrow gf(\varepsilon)$, existe uma infinita variedade de esquemas de subtração que podem todos serem chamadas de mínimas, dependendo da escolha da função $f(\varepsilon)$. Todos os esquemas levam aos mesmos contra-termos e constantes de renormalização. Na estrita versão de Subtração Mínima, se expande todas as funções de $\varepsilon$ em cada integral de Feynman regularizada em potências de $\varepsilon$.

O esquema de Subtração Mínima é implementado com a ajuda de um operador $\mathcal{K}$, para a escolha dos termos de pólo puro da integral regularizada dimensionalmente, definido por

\begin{equation}
\mathcal{K} \sum_{i=-k}^{\infty}A_i\varepsilon^i = \sum_{i=-k}^{-1} A_i\varepsilon^i = \sum_{i=1}^{k}\frac{A_{-i}}{\varepsilon^i} ,
\end{equation}
onde, pela definição, $\mathcal{K}$ é um operador projeção dado por

\begin{equation}
\mathcal{K}^2=\mathcal{K} .
\end{equation}

Como exemplo, podemos observar a aplicação do operador $\mathcal{K}$ nos diagramas de Feynman $\scalebox{0.5}{\begin{picture}(30,17) (212,-153)
    \SetWidth{0.9}
    \SetColor{Black}
    \Arc(227,-144)(7.071,262,622)
    \Line(213,-152)(241,-152)
    \Vertex(227,-151.5){1.5}
  \end{picture}}$ e $\scalebox{0.5}{\begin{picture}(30,24) (218,-165)
    \SetWidth{1.0}
    \SetColor{Black}
    \Arc(233,-157)(8,252,612)
    \Line(241.5,-157)(246,-151)
    \Line(241.5,-157)(246,-163)
    \Line(224.5,-157)(220,-151)
    \Line(224.5,-157)(220,-163)
    \Vertex(224.5,-157){1.5}
    \Vertex(241.5,-157){1.5}
  \end{picture}}$ que resulta

\begin{equation}
\mathcal{K} \left(
 \begin{picture}(30,17) (212,-153)
    \SetWidth{0.9}
    \SetColor{Black}
    \Arc(227,-144)(7.071,262,622)
    \Line(213,-152)(241,-152)
    \Vertex(227,-151.5){1.5}
  \end{picture} \right) = m^2\left[\frac{g}{(4\pi)^2} \frac{2}{\varepsilon} \right] ,
\end{equation}

\begin{equation}
\mathcal{K} \left(
\parbox{8mm} {\begin{picture}(30,24) (218,-169)
    \SetWidth{1.0}
    \SetColor{Black}
    \Arc(233,-157)(8,252,612)
    \Line(241.5,-157)(246,-151)
    \Line(241.5,-157)(246,-163)
    \Line(224.5,-157)(220,-151)
    \Line(224.5,-157)(220,-163)
    \Vertex(224.5,-157){1.5}
    \Vertex(241.5,-157){1.5}
  \end{picture}} \hspace*{0.2cm} \right) = \mu^{\varepsilon}g \left[ \frac{g}{(4\pi)^2}\frac{2}{\varepsilon} \right].
\end{equation}

Ambos os termos de pólo em 1-{\it loop} são locais. Eles são proporcionais a $m^2$ para diagramas quadraticamente divergentes, e a $\mu^{\varepsilon}g$ para diagramas logaritmicamente divergentes.

Agora podemos executar a renormalização no esquema de Subtração Mínima para calcular as funções de vértices finitas de 2 e 4-pontos, $\overline{\Gamma}^{(2)}(\vec{k}_i)$ e $\overline{\Gamma}^{(4)}(\vec{k}_i)$ respectivamente, iniciando da energia funcional renormalizada (\ref{energiarenormalizada}), ou seja

\begin{equation}
E[\phi]=\int d^Dx \left[ \frac{1}{2}(\partial \phi)^2 + c_{\phi}\frac{1}{2}(\partial \phi)^2 + \frac{m^2}{2}\phi^2 + c_{m^2}\frac{m^2}{2}\phi^2 +\frac{\mu^{\varepsilon}g}{4!}\phi^4 + c_g\frac{\mu^{\varepsilon}g}{4!}\phi^4 \right] . \label{efunnor}
\end{equation}

Executando o cálculo na ordem de 1-{\it loop} (para a primeira ordem em $g$), os contra-termos que são necessários para tornar a função de vértice de 2-pontos finita também são da ordem em $g$. Assim, podemos escrever o resultado em termos diagramáticos como

\begin{eqnarray}
\overline{\Gamma}^{(2)}(\vec{k}_i)=\vec{k}^2 + m^2 -\left(\frac{1}{2}
\begin{picture}(30,17) (212,-153) 
    \SetWidth{0.9}
    \SetColor{Black}
    \Arc(227,-144)(7.071,262,622)
    \Line(213,-152)(241,-152)
    \Vertex(227,-151.5){1.5}
  \end{picture} +
  \begin{picture}(51,13) (111,-204)
    \SetWidth{1.0}
    \SetColor{Black}
    \Line(118,-198)(157,-198)
    \SetWidth{1.0}
    \Line(133,-192)(142,-203)
    \Line(142,-192)(133,-203)
  \end{picture} +
  \begin{picture}(68,14) (65,-167)
    \SetWidth{1.0}
    \SetColor{Black}
    \Line(84,-160)(115,-160)
    \Arc(100,-160)(3,270,630)
  \end{picture} + O(g^2) \right) .
\end{eqnarray}

Diagramaticamente, podemos observar apenas contribuições dos contra-termos da massa e do campo, $c_{m^2}$ e $c_{\phi}$. Logo, usando os resultados do apêndice C.2 temos

\begin{eqnarray}
\begin{picture}(51,13) (111,-204)
    \SetWidth{1.0}
    \SetColor{Black}
    \Line(118,-198)(157,-198)
    \SetWidth{1.0}
    \Line(133,-192)(142,-203)
    \Line(142,-192)(133,-203)
  \end{picture} = -m^2c_{m^2}^{1} = -\frac{1}{2}\mathcal{K}
  \left( 
  \begin{picture}(30,17) (212,-153) 
    \SetWidth{0.9}
    \SetColor{Black}
    \Arc(227,-144)(7.071,262,622)
    \Line(213,-152)(241,-152)
    \Vertex(227,-151.5){1.5}
  \end{picture} \right) = -m^2\frac{g}{(4\pi)^2}\frac{1}{\varepsilon} ,
\end{eqnarray}

\begin{eqnarray}
\begin{picture}(68,14) (65,-167)
    \SetWidth{1.0}
    \SetColor{Black}
    \Line(84,-160)(115,-160)
    \Arc(100,-160)(3,270,630)
  \end{picture} = -\vec{k}^2c_{\phi}^1=0 ,
\end{eqnarray}
onde o índice superior no contra-termo denota a ordem de aproximação em $g$. Assim observamos que não encontramos contribuições em primeira ordem para o contra-termo $c_{\phi}$. Portanto, escolhendo os contra-termos desta forma, o pólo $\frac{1}{\varepsilon}$ no diagrama em 1-{\it loop} é cancelado, e a função de vértice finita renormalizada é dada por

\begin{equation}
\overline{\Gamma}^{(2)}(\vec{k}_i)=\vec{k}^2 + m^2 -\left[\frac{1}{2}
\begin{picture}(30,17) (212,-153) 
    \SetWidth{0.9}
    \SetColor{Black}
    \Arc(227,-144)(7.071,262,622)
    \Line(213,-152)(241,-152)
    \Vertex(227,-151.5){1.5}
  \end{picture} - \frac{1}{2} \mathcal{K}
  \left( 
  \begin{picture}(30,17) (212,-153) 
    \SetWidth{0.9}
    \SetColor{Black}
    \Arc(227,-144)(7.071,262,622)
    \Line(213,-152)(241,-152)
    \Vertex(227,-151.5){1.5}
  \end{picture} \right) \right] + O(g^2) .
\end{equation}

A primeira correção perturbativa para a função de vértice $\overline{\Gamma}^{(4)}(\vec{k}_i)$ é da ordem de $g^2$. O pólo $\frac{1}{\varepsilon}$ na integral de Feynman é removida pelo contra-termo da constante de acoplamento $c_g$. Logo,

\begin{eqnarray}
\overline{\Gamma}^{(4)} = - \left(
\parbox{8mm}{\begin{picture}(19,20) (324,-152)
    \SetWidth{1.0}
    \SetColor{Black}
    \Line(325,-134)(342,-151)
    \Line(325,-151)(342,-134)
    \Vertex(333.8,-142.6){1}
  \end{picture}} + \frac{3}{2}
  \parbox{8mm}{\begin{picture}(30,24) (218,-169)
    \SetWidth{1.0}
    \SetColor{Black}
    \Arc(233,-157)(8,252,612)
    \Line(241.5,-157)(246,-151)
    \Line(241.5,-157)(246,-163)
    \Line(224.5,-157)(220,-151)
    \Line(224.5,-157)(220,-163)
    \Vertex(224.5,-157){1.5}
    \Vertex(241.5,-157){1.5}
  \end{picture}} \hspace*{0.2cm} +
 \parbox{8mm}{ \begin{picture}(19,20) (324,-152)
    \SetWidth{1.0}
    \SetColor{Black}
    \Line(325,-134)(342,-151)
    \Line(325,-151)(342,-134)
    \Vertex(333.8,-142.6){3}
  \end{picture} } \right) + O(g^3) .
\end{eqnarray}

Assim, temos apenas contribuições do contra-termo da constante de acoplamento $c_g$. Portanto,

\begin{eqnarray}
\parbox{8mm}{\begin{picture}(19,20) (324,-152)
    \SetWidth{1.0}
    \SetColor{Black}
    \Line(325,-134)(342,-151)
    \Line(325,-151)(342,-134)
    \Vertex(333.8,-142.6){3}
  \end{picture}} = -\mu^{\varepsilon}gc_g^1 = -\frac{3}{2} \mathcal{K} \left(
 \parbox{8mm}{ \begin{picture}(30,24) (218,-169)
    \SetWidth{1.0}
    \SetColor{Black}
    \Arc(233,-157)(8,252,612)
    \Line(241.5,-157)(246,-151)
    \Line(241.5,-157)(246,-163)
    \Line(224.5,-157)(220,-151)
    \Line(224.5,-157)(220,-163)
    \Vertex(224.5,-157){1.5}
    \Vertex(241.5,-157){1.5}
  \end{picture}} \hspace*{0.2cm} \right) = -\mu^{\varepsilon}g\frac{3g}{(4\pi)^2}\frac{1}{\varepsilon} ,
\end{eqnarray}
e assim obtemos a função de vértice finita dado por

\begin{equation}
\overline{\Gamma}^{(4)} = - \left[
\parbox{8mm}{\begin{picture}(19,20) (324,-152)
    \SetWidth{1.0}
    \SetColor{Black}
    \Line(325,-134)(342,-151)
    \Line(325,-151)(342,-134)
    \Vertex(333.8,-142.6){1}
  \end{picture}} + \frac{3}{2}
  \parbox{8mm}{\begin{picture}(30,24) (218,-169)
    \SetWidth{1.0}
    \SetColor{Black}
    \Arc(233,-157)(8,252,612)
    \Line(241.5,-157)(246,-151)
    \Line(241.5,-157)(246,-163)
    \Line(224.5,-157)(220,-151)
    \Line(224.5,-157)(220,-163)
    \Vertex(224.5,-157){1.5}
    \Vertex(241.5,-157){1.5}
  \end{picture}} \hspace*{0.2cm} -
    \frac{3}{2} \mathcal{K} \left(
  \parbox{8mm}{\begin{picture}(30,24) (218,-169)
    \SetWidth{1.0}
    \SetColor{Black}
    \Arc(233,-157)(8,252,612)
    \Line(241.5,-157)(246,-151)
    \Line(241.5,-157)(246,-163)
    \Line(224.5,-157)(220,-151)
    \Line(224.5,-157)(220,-163)
    \Vertex(224.5,-157){1.5}
    \Vertex(241.5,-157){1.5}
  \end{picture}} \hspace*{0.2cm} \right) \right] + O(g^3) .
\end{equation}

Extendendo o cálculo para a ordem de 2-{\it loops}, podemos encontrar as contribuições dos contra-termos na ordem em $g^2$ e obter as funções de vértices finitas na ordem de 2-{\it loops}. De modo análogo ao cálculo em 1-{\it loop}, temos que as funções de vértice de 2-pontos e 4-pontos, em termos diagramáticos, são representadas como

\begin{eqnarray}
\overline{\Gamma}^{(2)}(\vec{k}_i)=\vec{k}^2 + m^2 - \Biggl[ \frac{1}{2}
\begin{picture}(30,17) (212,-153)
    \SetWidth{0.9}
    \SetColor{Black}
    \Arc(227,-144)(7.071,262,622)
    \Line(213,-152)(241,-152)
    \Vertex(227,-151.5){1.5}
  \end{picture} +
  \begin{picture}(51,13) (111,-204)
    \SetWidth{1.0}
    \SetColor{Black}
    \Line(118,-198)(157,-198)
    \SetWidth{1.0}
    \Line(133,-192)(142,-203)
    \Line(142,-192)(133,-203)
  \end{picture} +
  \begin{picture}(68,14) (65,-167)
    \SetWidth{1.0}
    \SetColor{Black}
    \Line(84,-160)(115,-160)
    \Arc(100,-160)(3,270,630)
  \end{picture} + \frac{1}{4}
  \begin{picture}(28,30) (204,-138)
    \SetWidth{0.9}
    \SetColor{Black}
    \Arc(218,-129)(6.083,261,621)
    \Line(205,-136)(231,-136)
    \Vertex(218,-135.5){1.5}
    \Arc(218,-116)(6.083,261,621)
    \Vertex(218,-122){1.5}
  \end{picture} +  \frac{1}{2}
  \begin{picture}(16,21) (216,-193)
    \SetWidth{1.0}
    \SetColor{Black}
    \Arc(224,-182)(7,270,630)
    \Line(215,-190)(233,-190)
    \Vertex(224,-189.5){1.5}
    \Line(221,-172)(227,-179)
    \Line(227,-172)(221,-179)
  \end{picture} \nonumber \\
  + \frac{1}{6}
  \begin{picture}(29,18) (152,-178)
    \SetWidth{1.0}
    \SetColor{Black}
    \Arc(166,-173)(8.246,256,616)
    \Line(152,-173)(180,-173)
    \Vertex(158,-173){1.5}
    \Vertex(174,-173){1.5}
  \end{picture} + \frac{1}{2}
  \begin{picture}(30,17) (212,-153)
    \SetWidth{1.0}
    \SetColor{Black}
    \Arc(227,-144)(7.071,262,622)
    \Line(213,-152)(241,-152)
    \Vertex(227,-151){3}
  \end{picture} \Biggr] + O(g^3) ,
\end{eqnarray}

\begin{eqnarray}
\overline{\Gamma}^{(4)}(\vec{k}_i) = - \Biggl[
\parbox{8mm}{\begin{picture}(19,20) (324,-152)
    \SetWidth{1.0}
    \SetColor{Black}
    \Line(325,-134)(342,-151)
    \Line(325,-151)(342,-134)
    \Vertex(333.8,-142.6){1}
  \end{picture}} + \frac{3}{2}
  \parbox{8mm}{\begin{picture}(30,24) (218,-169)
    \SetWidth{1.0}
    \SetColor{Black}
    \Arc(233,-157)(8,252,612)
    \Line(241.5,-157)(246,-151)
    \Line(241.5,-157)(246,-163)
    \Line(224.5,-157)(220,-151)
    \Line(224.5,-157)(220,-163)
    \Vertex(224.5,-157){1.5}
    \Vertex(241.5,-157){1.5}
  \end{picture}} \hspace*{0.2cm} +
 \parbox{8mm}{ \begin{picture}(19,20) (324,-152)
    \SetWidth{1.0}
    \SetColor{Black}
    \Line(325,-134)(342,-151)
    \Line(325,-151)(342,-134)
    \Vertex(333.8,-142.6){3}
  \end{picture} }  + 3
  \parbox{8mm}{\begin{picture}(29,25) (216,-193)
    \SetWidth{0.9}
    \SetColor{Black}
    \Vertex(231,-172){1.5}
    \Arc(231,-181)(8.544,291,651)
    \Line(231,-172)(236,-167)
    \Line(217,-181)(244,-181)
    \Vertex(222.5,-181){1.5}
    \Vertex(239.5,-181){1.5}
    \Line(231,-172)(226,-167)
  \end{picture}}\hspace*{0.2cm} + \frac{3}{4}
  \parbox{8mm}{\begin{picture}(44,16) (142,-142)
    \SetWidth{1.0}
    \SetColor{Black}
    \Arc(156,-134)(7.28,254,614)
    \Arc(171,-134)(7.28,254,614)
    \Vertex(163.5,-134){1.5}
    \Line(178,-134)(185,-128)
    \Line(178,-134)(185,-141)
    \Line(149,-134)(143,-128)
    \Line(149,-134)(143,-141)
    \Vertex(149,-134){1.5}
    \Vertex(178,-134){1.5}
  \end{picture}}\hspace*{0.6cm} + \frac{3}{2}
  \parbox{8mm}{\begin{picture}(30,24) (218,-168)
    \SetWidth{1.0}
    \SetColor{Black}
    \Arc(233,-157)(8,252,612)
    \Arc(233,-145)(4.472,117,477)
    \Line(241,-157)(246,-151)
    \Line(241,-157)(246,-163)
    \Line(224,-157)(220,-151)
    \Line(224,-157)(220,-163)
    \Vertex(233,-149){1.5}
    \Vertex(224.7,-157){1.5}
    \Vertex(241.5,-157){1.5}
  \end{picture}} \hspace*{0.2cm}   + 3
  \parbox{8mm}{\begin{picture}(30,24) (218,-169)
    \SetWidth{1.0}
    \SetColor{Black}
    \Arc(233,-157)(8,252,612)
    \Line(241.5,-157)(246,-151)
    \Line(241.5,-157)(246,-163)
    \Line(224.5,-157)(220,-151)
    \Line(224.5,-157)(220,-163)
    \Vertex(224.5,-157){1.5}
    \Vertex(241.5,-157){3}
  \end{picture}}\hspace*{0.2cm} + 3
  \parbox{8mm}{\begin{picture}(30,24) (218,-169)
    \SetWidth{1.0}
    \SetColor{Black}
    \Arc(233,-157)(8,252,612)
    \Line(241.5,-157)(246,-151)
    \Line(241.5,-157)(246,-163)
    \Line(224.5,-157)(220,-151)
    \Line(224.5,-157)(220,-163)
    \Vertex(224.5,-157){1.5}
    \Vertex(241.5,-157){1.5}
    \Line(230,-145.5)(236,-152.5)
    \Line(236,-145.5)(230,-152.5)
  \end{picture}}  \hspace*{0.2cm} \Biggr] .
\end{eqnarray}

As contribuições dos contra-termos na ordem em $g$ são dados por

\begin{eqnarray}
\begin{picture}(51,13) (111,-204)
    \SetWidth{1.0}
    \SetColor{Black}
    \Line(118,-198)(157,-198)
    \SetWidth{1.0}
    \Line(133,-192)(142,-203)
    \Line(142,-192)(133,-203)
  \end{picture} +
  \begin{picture}(68,14) (65,-167)
    \SetWidth{1.0}
    \SetColor{Black}
    \Line(84,-160)(115,-160)
    \Arc(100,-160)(3,270,630)
  \end{picture} = -\mathcal{K} \left[
\frac{1}{2}
\begin{picture}(30,17) (212,-153)
    \SetWidth{0.9}
    \SetColor{Black}
    \Arc(227,-144)(7.071,262,622)
    \Line(213,-152)(241,-152)
    \Vertex(227,-151.5){1.5}
  \end{picture} 
   + \frac{1}{4}
  \begin{picture}(28,30) (204,-134)
    \SetWidth{0.9}
    \SetColor{Black}
    \Arc(218,-129)(6.083,261,621)
    \Line(205,-136)(231,-136)
    \Vertex(218,-135.5){1.5}
    \Arc(218,-116)(6.083,261,621)
    \Vertex(218,-122){1.5}
  \end{picture} +  \frac{1}{2}
  \begin{picture}(16,21) (216,-193)
    \SetWidth{1.0}
    \SetColor{Black}
    \Arc(224,-182)(7,270,630)
    \Line(215,-190)(233,-190)
    \Vertex(224,-189.5){1.5}
    \Line(221,-172)(227,-179)
    \Line(227,-172)(221,-179)
  \end{picture}  + \frac{1}{6}
  \begin{picture}(29,18) (152,-177)
    \SetWidth{1.0}
    \SetColor{Black}
    \Arc(166,-173)(8.246,256,616)
    \Line(152,-173)(180,-173)
    \Vertex(158,-173){1.5}
    \Vertex(174,-173){1.5}
  \end{picture} + \frac{1}{2}
  \begin{picture}(30,17) (212,-153)
    \SetWidth{1.0}
    \SetColor{Black}
    \Arc(227,-144)(7.071,262,622)
    \Line(213,-152)(241,-152)
    \Vertex(227,-151){3}
  \end{picture} \right] ,
\end{eqnarray}

\begin{eqnarray}
\parbox{8mm}{ \begin{picture}(19,20) (324,-152)
    \SetWidth{1.0}
    \SetColor{Black}
    \Line(325,-134)(342,-151)
    \Line(325,-151)(342,-134)
    \Vertex(333.8,-142.6){3}
  \end{picture} } = -\mathcal{K} \left[
\frac{3}{2}
  \parbox{8mm}{\begin{picture}(30,24) (218,-169)
    \SetWidth{1.0}
    \SetColor{Black}
    \Arc(233,-157)(8,252,612)
    \Line(241.5,-157)(246,-151)
    \Line(241.5,-157)(246,-163)
    \Line(224.5,-157)(220,-151)
    \Line(224.5,-157)(220,-163)
    \Vertex(224.5,-157){1.5}
    \Vertex(241.5,-157){1.5}
  \end{picture}} \hspace*{0.2cm}  + 3
  \parbox{8mm}{\begin{picture}(29,25) (216,-193)
    \SetWidth{0.9}
    \SetColor{Black}
    \Vertex(231,-172){1.5}
    \Arc(231,-181)(8.544,291,651)
    \Line(231,-172)(236,-167)
    \Line(217,-181)(244,-181)
    \Vertex(222.5,-181){1.5}
    \Vertex(239.5,-181){1.5}
    \Line(231,-172)(226,-167)
  \end{picture}}\hspace*{0.2cm} + \frac{3}{4}
  \parbox{8mm}{\begin{picture}(44,16) (142,-142)
    \SetWidth{1.0}
    \SetColor{Black}
    \Arc(156,-134)(7.28,254,614)
    \Arc(171,-134)(7.28,254,614)
    \Vertex(163.5,-134){1.5}
    \Line(178,-134)(185,-128)
    \Line(178,-134)(185,-141)
    \Line(149,-134)(143,-128)
    \Line(149,-134)(143,-141)
    \Vertex(149,-134){1.5}
    \Vertex(178,-134){1.5}
  \end{picture}}\hspace*{0.6cm} + \frac{3}{2}
  \parbox{8mm}{\begin{picture}(30,24) (218,-169)
    \SetWidth{1.0}
    \SetColor{Black}
    \Arc(233,-157)(8,252,612)
    \Arc(233,-145)(4.472,117,477)
    \Line(241,-157)(246,-151)
    \Line(241,-157)(246,-163)
    \Line(224,-157)(220,-151)
    \Line(224,-157)(220,-163)
    \Vertex(233,-149){1.5}
    \Vertex(224.7,-157){1.5}
    \Vertex(241.5,-157){1.5}
  \end{picture}} \hspace*{0.2cm}   + 3
  \parbox{8mm}{\begin{picture}(30,24) (218,-169)
    \SetWidth{1.0}
    \SetColor{Black}
    \Arc(233,-157)(8,252,612)
    \Line(241.5,-157)(246,-151)
    \Line(241.5,-157)(246,-163)
    \Line(224.5,-157)(220,-151)
    \Line(224.5,-157)(220,-163)
    \Vertex(224.5,-157){1.5}
    \Vertex(241.5,-157){3}
  \end{picture}}\hspace*{0.2cm} + 3
  \parbox{8mm}{\begin{picture}(30,24) (218,-169)
    \SetWidth{1.0}
    \SetColor{Black}
    \Arc(233,-157)(8,252,612)
    \Line(241.5,-157)(246,-151)
    \Line(241.5,-157)(246,-163)
    \Line(224.5,-157)(220,-151)
    \Line(224.5,-157)(220,-163)
    \Vertex(224.5,-157){1.5}
    \Vertex(241.5,-157){1.5}
    \Line(230,-145.5)(236,-152.5)
    \Line(236,-145.5)(230,-152.5)
  \end{picture}}\hspace*{0.2cm} \right] .
\end{eqnarray}

Portanto, usando o resultado dos diagramas do apêndice C.2, obtemos as contribuições dos contra-termos até a ordem em $g^2$, dados por

\begin{equation}
c_{\phi}=c_{\phi}^{1} + c_{\phi}^{2} = 0 -\frac{g^2}{(4\pi)^4 12 \varepsilon} ,
\end{equation}

\begin{equation}
c_{m^2}=c_{m^2}^{1} + c_{m^2}^{2}=\frac{g}{(4\pi)^2 \varepsilon} + \frac{g^2}{(4\pi)^4}\left( \frac{2}{\varepsilon^2} - \frac{1}{2\varepsilon} \right) ,
\end{equation}

\begin{equation}
c_g=c_g^1 + c_g^2=\frac{3g}{(4\pi)^2 \varepsilon} + \frac{g^2}{(4\pi)^4}\left( \frac{9}{\varepsilon^2} - \frac{3}{\varepsilon}\right) .
\end{equation}

\subsection{Cálculo das Constantes de Renormalização via Contra-Termos}

Através do método apresentado na subseção anterior, temos que até a ordem de $g^2$, obtemos funções de correlação finitos a partir da energia funcional (\ref{efunnor}), que pode ser escrito na forma da expressão (\ref{energiafuncional3}), onde as constantes de renormalização são dadas por

\begin{eqnarray}
Z_{\phi}(g,\varepsilon^{-1})=1+c_{\phi} = 1 + \frac{1}{\vec{k}^2}\frac{1}{6} \mathcal{K} 
\left( \parbox{8mm}{ \begin{picture}(29,18) (152,-182)
    \SetWidth{1.0}
    \SetColor{Black}
    \Arc(166,-173)(8.246,256,616)
    \Line(152,-173)(180,-173)
    \Vertex(158,-173){1.5}
    \Vertex(174,-173){1.5}
  \end{picture}} \hspace*{0.2cm}
\right)\Biggr|_{m^2=0} ,
\end{eqnarray}

\begin{eqnarray}
Z_{m^2}(g,\varepsilon^{-1})=1+c_{m^2} = 1 + \frac{1}{m^2} \Biggl[ \frac{1}{2} \mathcal{K}
\left( \begin{picture}(30,17) (212,-153) 
    \SetWidth{0.9}
    \SetColor{Black}
    \Arc(227,-144)(7.071,262,622)
    \Line(213,-152)(241,-152)
    \Vertex(227,-151.5){1.5}
  \end{picture} \right) +
  \frac{1}{4} \mathcal{K}
  \left( \parbox{8mm} {\begin{picture}(28,30) (204,-140)
    \SetWidth{0.9}
    \SetColor{Black}
    \Arc(218,-129)(6.083,261,621)
    \Line(205,-136)(231,-136)
    \Vertex(218,-135.5){1.5}
    \Arc(218,-116)(6.083,261,621)
    \Vertex(218,-122){1.5}
  \end{picture}} \hspace*{0.2cm}  \right) +
  \frac{1}{2} \mathcal{K}
  \left( \parbox{8mm} {\begin{picture}(16,21) (216,-193)
    \SetWidth{1.0}
    \SetColor{Black}
    \Arc(224,-182)(7,270,630)
    \Line(215,-190)(233,-190)
    \Vertex(224,-189.5){1.5}
    \Line(221,-172)(227,-179)
    \Line(227,-172)(221,-179)
  \end{picture}} \right) \nonumber \\
  + \frac{1}{2} \mathcal{K}
  \left( \begin{picture}(30,17) (212,-153)
    \SetWidth{1.0}
    \SetColor{Black}
    \Arc(227,-144)(7.071,262,622)
    \Line(213,-152)(241,-152)
    \Vertex(227,-151){3}
  \end{picture} \right) +
  \frac{1}{6} \mathcal{K}
  \left( \parbox{8mm}{ \begin{picture}(29,18) (152,-182)
    \SetWidth{1.0}
    \SetColor{Black}
    \Arc(166,-173)(8.246,256,616)
    \Line(152,-173)(180,-173)
    \Vertex(158,-173){1.5}
    \Vertex(174,-173){1.5}
  \end{picture}}\hspace*{0.2cm} \right) \Biggr|_{\vec{k}^2=0} \Biggr] ,
\end{eqnarray}

\begin{eqnarray}
Z_{g}(g,\varepsilon^{-1})=1+c_{g}=1 + \frac{1}{\mu^{\varepsilon}g} \Biggl[ \frac{3}{2} \mathcal{K}
\left( \parbox{8mm} { \begin{picture}(30,24) (218,-169)
    \SetWidth{1.0}
    \SetColor{Black}
    \Arc(233,-157)(8,252,612)
    \Line(241.5,-157)(246,-151)
    \Line(241.5,-157)(246,-163)
    \Line(224.5,-157)(220,-151)
    \Line(224.5,-157)(220,-163)
    \Vertex(224.5,-157){1.5}
    \Vertex(241.5,-157){1.5}
  \end{picture}}\hspace*{0.2cm}  \right) + 
  3 \mathcal{K} 
  \left(\parbox{8mm} { \begin{picture}(29,25) (216,-193)
    \SetWidth{0.9}
    \SetColor{Black}
    \Vertex(231,-172){1.5}
    \Arc(231,-181)(8.544,291,651)
    \Line(231,-172)(236,-167)
    \Line(217,-181)(244,-181)
    \Vertex(222.5,-181){1.5}
    \Vertex(239.5,-181){1.5}
    \Line(231,-172)(226,-167)
  \end{picture}}\hspace*{0.2cm} \right) +
  \frac{3}{4} \mathcal{K}
  \left(\parbox{8mm} { \begin{picture}(44,16) (142,-142)
    \SetWidth{1.0}
    \SetColor{Black}
    \Arc(156,-134)(7.28,254,614)
    \Arc(171,-134)(7.28,254,614)
    \Vertex(163.5,-134){1.5}
    \Line(178,-134)(185,-128)
    \Line(178,-134)(185,-141)
    \Line(149,-134)(143,-128)
    \Line(149,-134)(143,-141)
    \Vertex(149,-134){1.5}
    \Vertex(178,-134){1.5}
  \end{picture}}\hspace*{0.7cm} \right) \nonumber \\
  + \frac{3}{2} \mathcal{K}
  \left(\parbox{8mm} { \begin{picture}(30,24) (218,-165)
    \SetWidth{1.0}
    \SetColor{Black}
    \Arc(233,-157)(8,252,612)
    \Arc(233,-145)(4.472,117,477)
    \Line(241,-157)(246,-151)
    \Line(241,-157)(246,-163)
    \Line(224,-157)(220,-151)
    \Line(224,-157)(220,-163)
    \Vertex(233,-149){1.5}
    \Vertex(224.7,-157){1.5}
    \Vertex(241.5,-157){1.5}
  \end{picture}}\hspace*{0.2cm} \right) +
  3 \mathcal{K}
  \left(\parbox{8mm}{ \begin{picture}(30,24) (218,-169)
    \SetWidth{1.0}
    \SetColor{Black}
    \Arc(233,-157)(8,252,612)
    \Line(241.5,-157)(246,-151)
    \Line(241.5,-157)(246,-163)
    \Line(224.5,-157)(220,-151)
    \Line(224.5,-157)(220,-163)
    \Vertex(224.5,-157){1.5}
    \Vertex(241.5,-157){3}
  \end{picture}}\hspace*{0.2cm} \right) +
  3 \mathcal{K}
  \left( \parbox{8mm} { \begin{picture}(30,24) (218,-169)
    \SetWidth{1.0}
    \SetColor{Black}
    \Arc(233,-157)(8,252,612)
    \Line(241.5,-157)(246,-151)
    \Line(241.5,-157)(246,-163)
    \Line(224.5,-157)(220,-151)
    \Line(224.5,-157)(220,-163)
    \Vertex(224.5,-157){1.5}
    \Vertex(241.5,-157){1.5}
    \Line(230,-145.5)(236,-152.5)
    \Line(236,-145.5)(230,-152.5)
  \end{picture}}\hspace*{0.2cm} \right) \Biggr] .
  \end{eqnarray}

As constantes de renormalização são expansões na constante de acoplamento adimensional $g$, com coeficientes da expansão contendo somente termos de pólo na forma $\frac{1}{\varepsilon^i}$, onde $i$ varia de 1 a $n$ considerando o termo de ordem $g^n$. Como visto em seções anteriores, devemos lembrar que as divergências na teoria $\phi^4$ provêm exclusivamente dos diagramas de 2 e 4-pontos 1PI, onde todos as funções de vértice de $n$-pontos são finitos para $\varepsilon \longrightarrow 0$ até a ordem $g^2$, se a expansão perturbativa é produzida a partir da energia funcional (\ref{energiafuncional3}).

Como o estudo está sendo realizado em um campo de $N$ componentes, os resultados em 2-{\it loops} são extendidos pelos fatores de simetria. Daí temos que as constantes de renormalização são reescritas na forma

\begin{eqnarray}
Z_{\phi}(g,\varepsilon^{-1})=1+c_{\phi} = 1 + \frac{1}{\vec{k}^2}\frac{1}{6} \mathcal{K} 
\left( \parbox{8mm}{ \begin{picture}(29,18) (152,-182)
    \SetWidth{1.0}
    \SetColor{Black}
    \Arc(166,-173)(8.246,256,616)
    \Line(152,-173)(180,-173)
    \Vertex(158,-173){1.5}
    \Vertex(174,-173){1.5}
  \end{picture}} \hspace*{0.2cm}
\right) \Biggr|_{m^2=0} S_{\scalebox{0.3}{\begin{picture}(29,18) (152,-182)
    \SetWidth{1.0}
    \SetColor{Black}
    \Arc(166,-173)(8.246,256,616)
    \Line(152,-173)(180,-173)
    \Vertex(158,-173){1.5}
    \Vertex(174,-173){1.5}
  \end{picture}}} ,
\end{eqnarray}

\begin{eqnarray}
Z_{m^2}(g,\varepsilon^{-1})=1+c_{m^2} = 1 + \frac{1}{m^2} \Biggl[ \frac{1}{2} \mathcal{K}
\left( \begin{picture}(30,17) (212,-153) 
    \SetWidth{0.9}
    \SetColor{Black}
    \Arc(227,-144)(7.071,262,622)
    \Line(213,-152)(241,-152)
    \Vertex(227,-151.5){1.5}
  \end{picture} \right)S_{\scalebox{0.3}{\begin{picture}(30,17) (212,-153) 
    \SetWidth{0.9}
    \SetColor{Black}
    \Arc(227,-144)(7.071,262,622)
    \Line(213,-152)(241,-152)
    \Vertex(227,-151.5){1.5}
  \end{picture}}} +
  \frac{1}{4} \mathcal{K}
  \left( \parbox{8mm} {\begin{picture}(28,30) (204,-140)
    \SetWidth{0.9}
    \SetColor{Black}
    \Arc(218,-129)(6.083,261,621)
    \Line(205,-136)(231,-136)
    \Vertex(218,-135.5){1.5}
    \Arc(218,-116)(6.083,261,621)
    \Vertex(218,-122){1.5}
  \end{picture}} \hspace*{0.2cm}  \right)S_{\scalebox{0.3}{\begin{picture}(28,30) (204,-140)
    \SetWidth{0.9}
    \SetColor{Black}
    \Arc(218,-129)(6.083,261,621)
    \Line(205,-136)(231,-136)
    \Vertex(218,-135.5){1.5}
    \Arc(218,-116)(6.083,261,621)
    \Vertex(218,-122){1.5}
  \end{picture}}} +
  \frac{1}{2} \mathcal{K}
  \left( \parbox{8mm} {\begin{picture}(16,21) (216,-193)
    \SetWidth{1.0}
    \SetColor{Black}
    \Arc(224,-182)(7,270,630)
    \Line(215,-190)(233,-190)
    \Vertex(224,-189.5){1.5}
    \Line(221,-172)(227,-179)
    \Line(227,-172)(221,-179)
  \end{picture}} \right)S_{\scalebox{0.3}{\begin{picture}(16,21) (216,-193)
    \SetWidth{1.0}
    \SetColor{Black}
    \Arc(224,-182)(7,270,630)
    \Line(215,-190)(233,-190)
    \Vertex(224,-189.5){1.5}
    \Line(221,-172)(227,-179)
    \Line(227,-172)(221,-179)
  \end{picture}}} \nonumber \\
  + \frac{1}{2} \mathcal{K}
  \left( \begin{picture}(30,17) (212,-153)
    \SetWidth{1.0}
    \SetColor{Black}
    \Arc(227,-144)(7.071,262,622)
    \Line(213,-152)(241,-152)
    \Vertex(227,-151){3}
  \end{picture} \right)S_{\scalebox{0.3}{\begin{picture}(30,17) (212,-153)
    \SetWidth{1.0}
    \SetColor{Black}
    \Arc(227,-144)(7.071,262,622)
    \Line(213,-152)(241,-152)
    \Vertex(227,-151){3}
  \end{picture}}} +
  \frac{1}{6} \mathcal{K}
  \left( \parbox{8mm}{ \begin{picture}(29,18) (152,-182)
    \SetWidth{1.0}
    \SetColor{Black}
    \Arc(166,-173)(8.246,256,616)
    \Line(152,-173)(180,-173)
    \Vertex(158,-173){1.5}
    \Vertex(174,-173){1.5}
  \end{picture}}\hspace*{0.2cm} \right) \Biggr|_{\vec{k}^2=0}S_{\scalebox{0.3}{\begin{picture}(29,18) (152,-182)
    \SetWidth{1.0}
    \SetColor{Black}
    \Arc(166,-173)(8.246,256,616)
    \Line(152,-173)(180,-173)
    \Vertex(158,-173){1.5}
    \Vertex(174,-173){1.5}
  \end{picture}}} \Biggr] ,
\end{eqnarray}

\begin{eqnarray}
Z_{g}(g,\varepsilon^{-1})=1+c_{g}&=&1 + \frac{1}{\mu^{\varepsilon}g} \Biggl[ \frac{3}{2} \mathcal{K}
\left( \parbox{8mm} { \begin{picture}(30,24) (218,-169)
    \SetWidth{1.0}
    \SetColor{Black}
    \Arc(233,-157)(8,252,612)
    \Line(241.5,-157)(246,-151)
    \Line(241.5,-157)(246,-163)
    \Line(224.5,-157)(220,-151)
    \Line(224.5,-157)(220,-163)
    \Vertex(224.5,-157){1.5}
    \Vertex(241.5,-157){1.5}
  \end{picture}}\hspace*{0.2cm}  \right)S_{\scalebox{0.3}{\begin{picture}(30,24) (218,-169)
    \SetWidth{1.0}
    \SetColor{Black}
    \Arc(233,-157)(8,252,612)
    \Line(241.5,-157)(246,-151)
    \Line(241.5,-157)(246,-163)
    \Line(224.5,-157)(220,-151)
    \Line(224.5,-157)(220,-163)
    \Vertex(224.5,-157){1.5}
    \Vertex(241.5,-157){1.5}
  \end{picture}}} + 
  3 \mathcal{K} 
  \left(\parbox{8mm} { \begin{picture}(29,25) (216,-193)
    \SetWidth{0.9}
    \SetColor{Black}
    \Vertex(231,-172){1.5}
    \Arc(231,-181)(8.544,291,651)
    \Line(231,-172)(236,-167)
    \Line(217,-181)(244,-181)
    \Vertex(222.5,-181){1.5}
    \Vertex(239.5,-181){1.5}
    \Line(231,-172)(226,-167)
  \end{picture}}\hspace*{0.2cm} \right)S_{\scalebox{0.3}{\begin{picture}(29,25) (216,-193)
    \SetWidth{0.9}
    \SetColor{Black}
    \Vertex(231,-172){1.5}
    \Arc(231,-181)(8.544,291,651)
    \Line(231,-172)(236,-167)
    \Line(217,-181)(244,-181)
    \Vertex(222.5,-181){1.5}
    \Vertex(239.5,-181){1.5}
    \Line(231,-172)(226,-167)
  \end{picture}}} +
  \frac{3}{4} \mathcal{K}
  \left(\parbox{8mm} { \begin{picture}(44,16) (142,-142)
    \SetWidth{1.0}
    \SetColor{Black}
    \Arc(156,-134)(7.28,254,614)
    \Arc(171,-134)(7.28,254,614)
    \Vertex(163.5,-134){1.5}
    \Line(178,-134)(185,-128)
    \Line(178,-134)(185,-141)
    \Line(149,-134)(143,-128)
    \Line(149,-134)(143,-141)
    \Vertex(149,-134){1.5}
    \Vertex(178,-134){1.5}
  \end{picture}}\hspace*{0.7cm} \right)S_{\scalebox{0.3}{\begin{picture}(44,16) (142,-142)
    \SetWidth{1.0}
    \SetColor{Black}
    \Arc(156,-134)(7.28,254,614)
    \Arc(171,-134)(7.28,254,614)
    \Vertex(163.5,-134){1.5}
    \Line(178,-134)(185,-128)
    \Line(178,-134)(185,-141)
    \Line(149,-134)(143,-128)
    \Line(149,-134)(143,-141)
    \Vertex(149,-134){1.5}
    \Vertex(178,-134){1.5}
  \end{picture}}} \nonumber \\
  &&+ \frac{3}{2} \mathcal{K}
  \left(\parbox{8mm} { \begin{picture}(30,24) (218,-167)
    \SetWidth{1.0}
    \SetColor{Black}
    \Arc(233,-157)(8,252,612)
    \Arc(233,-145)(4.472,117,477)
    \Line(241,-157)(246,-151)
    \Line(241,-157)(246,-163)
    \Line(224,-157)(220,-151)
    \Line(224,-157)(220,-163)
    \Vertex(233,-149){1.5}
    \Vertex(224.7,-157){1.5}
    \Vertex(241.5,-157){1.5}
  \end{picture}}\hspace*{0.2cm} \right)S_{\scalebox{0.3}{\begin{picture}(30,24) (218,-167)
    \SetWidth{1.0}
    \SetColor{Black}
    \Arc(233,-157)(8,252,612)
    \Arc(233,-145)(4.472,117,477)
    \Line(241,-157)(246,-151)
    \Line(241,-157)(246,-163)
    \Line(224,-157)(220,-151)
    \Line(224,-157)(220,-163)
    \Vertex(233,-149){1.5}
    \Vertex(224.7,-157){1.5}
    \Vertex(241.5,-157){1.5}
  \end{picture}}} +
  3 \mathcal{K}
  \left(\parbox{8mm}{ \begin{picture}(30,24) (218,-169)
    \SetWidth{1.0}
    \SetColor{Black}
    \Arc(233,-157)(8,252,612)
    \Line(241.5,-157)(246,-151)
    \Line(241.5,-157)(246,-163)
    \Line(224.5,-157)(220,-151)
    \Line(224.5,-157)(220,-163)
    \Vertex(224.5,-157){1.5}
    \Vertex(241.5,-157){3}
  \end{picture}}\hspace*{0.2cm} \right)S_{\scalebox{0.3}{\begin{picture}(30,24) (218,-169)
    \SetWidth{1.0}
    \SetColor{Black}
    \Arc(233,-157)(8,252,612)
    \Line(241.5,-157)(246,-151)
    \Line(241.5,-157)(246,-163)
    \Line(224.5,-157)(220,-151)
    \Line(224.5,-157)(220,-163)
    \Vertex(224.5,-157){1.5}
    \Vertex(241.5,-157){3}
  \end{picture}}} +
  3 \mathcal{K}
  \left( \parbox{8mm} { \begin{picture}(30,24) (218,-169)
    \SetWidth{1.0}
    \SetColor{Black}
    \Arc(233,-157)(8,252,612)
    \Line(241.5,-157)(246,-151)
    \Line(241.5,-157)(246,-163)
    \Line(224.5,-157)(220,-151)
    \Line(224.5,-157)(220,-163)
    \Vertex(224.5,-157){1.5}
    \Vertex(241.5,-157){1.5}
    \Line(230,-145.5)(236,-152.5)
    \Line(236,-145.5)(230,-152.5)
  \end{picture}}\hspace*{0.2cm} \right)S_{\scalebox{0.3}{\begin{picture}(30,24) (218,-169)
    \SetWidth{1.0}
    \SetColor{Black}
    \Arc(233,-157)(8,252,612)
    \Line(241.5,-157)(246,-151)
    \Line(241.5,-157)(246,-163)
    \Line(224.5,-157)(220,-151)
    \Line(224.5,-157)(220,-163)
    \Vertex(224.5,-157){1.5}
    \Vertex(241.5,-157){1.5}
    \Line(230,-145.5)(236,-152.5)
    \Line(236,-145.5)(230,-152.5)
  \end{picture}}} \Biggr] .
  \end{eqnarray}

Usando os resultados dos diagramas de Feynman em expansão $\varepsilon$ apresentado no apêndice C.2 e considerando os fatores de simetria apresentados no apêndice C.3, podemos expressar as constantes de renormalização, até a ordem $g^2$, da seguinte maneira

\begin{equation}
Z_{\phi}(g,\varepsilon^{-1})=1-\frac{g^2}{(4\pi)^4}\frac{1}{\varepsilon}\frac{N+2}{36} , \label{cr1}
\end{equation}

\begin{equation}
Z_{m^2}(g,\varepsilon^{-1})=1+\frac{g}{(4\pi)^2}\frac{1}{\varepsilon}\frac{N+2}{3} + \frac{g^2}{(4\pi)^4}\left[ -\frac{1}{\varepsilon}\frac{N+2}{6} + \frac{1}{\varepsilon^2}\frac{(N+2)(N+5)}{9} \right] ,
\end{equation}

\begin{equation}
Z_g(g,\varepsilon^{-1})=1+\frac{g}{(4\pi)^2}\frac{1}{\varepsilon}\frac{N+8}{3} + \frac{g^2}{(4\pi)^4}\left[ \frac{1}{\varepsilon^2}\frac{(N+8)^2}{9} - \frac{1}{\varepsilon} \frac{5N+22}{9}\right] . \label{cr3}
\end{equation}

\section{Grupo de Renormalização}

O procedimento de renormalização mostrado anteriormente, elimina todas as divergências UV a partir das integrais de Feynman decorrentes de grandes momentos em $D=4-\varepsilon$ dimensões.  Isto é necessário para obter as funções de correlação finitas no limite $\varepsilon \longrightarrow 0$. Apresentaremos nessa seção as equações de Callan-Symanzik para as funções de vértice renormalizada, onde essas equações são úteis no estudo de teorias de campo massivos pois dão o comportamento das funções de vértice quando a massa renormalizada é variada, e tomam a mesma forma de equação do tipo grupo de renormalização em um determinado regime, que ocorre quando os momentos são muito grandes em comparação com a massa \cite{carvalho:2008,kleinert:2000}. E assim obtermos as propriedades críticas da teoria $\phi^4$.

\subsection{Equações de Callan-Symanzik e Funções do Grupo de Renormalização}

Diferenciando a função de vértice não-renormalizada $\overline{\Gamma}_B^{(n)}(\vec{k}_i,m_B,\lambda_B,\Lambda)$ com respeito a massa renormalizada $m$, sendo fixos $\lambda_B$ e $\Lambda$, temos que

\begin{equation}
m\frac{\partial}{\partial m}\overline{\Gamma}_B^{(n)}(\vec{k}_i,m_B,\lambda_B,\Lambda)\biggl|_{\lambda_B,\Lambda}=m\frac{\partial}{\partial m}m_B^2\biggl|_{\lambda_B,\Lambda}\overline{\Gamma}_B^{(1,n)}(0,\vec{k}_i,m_B,\lambda_B,\Lambda) , \label{eqcallsym}
\end{equation}
onde

\begin{equation}
\overline{\Gamma}_B^{(1,n)}(0,\vec{k}_i,m_B,\lambda_B,\Lambda) = \frac{\partial}{\partial m_B^2}\overline{\Gamma}_B^{(n)}(\vec{k}_i,m_B,\lambda_B,\Lambda) ,
\end{equation}
é a função de vértice associada com a função de correlação contendo um termo extra $-\phi^2(\vec{x})/2$ dentro do valor esperado

\begin{equation}
G^{(1,n)}(\vec{x},\vec{x}_1,\ldots ,\vec{x}_n) = -\frac{1}{2}\langle \phi^2(\vec{x})\phi(\vec{x}_1)\cdots \phi(\vec{x}_n)\rangle .
\end{equation}

Com a ajuda da constante de renormalização $Z_{\phi}$, temos que a função de correlação renormalizada nos dá

\begin{equation}
\overline{\Gamma}_B^{(n)}(\vec{k}_i,m_B,\lambda_B,\Lambda) = Z_{\phi}^{-n/2} \overline{\Gamma}^{(n)}(\vec{k}_i,m,g) , \hspace*{0.7cm} n\geq1 .
\end{equation}

Introduzindo a constante de renormalização $Z_2$, que torna a função de vértice composta finita no limite $\lambda_B \longrightarrow 0$, tem-se

\begin{equation}
\overline{\Gamma}_B^{(1,n)}(0,\vec{k}_i,m_B,\lambda_B,\Lambda)=Z_{\phi}^{-n/2}Z_2\overline{\Gamma}^{(1,n)}(0,\vec{k}_i,m,g) ,
\end{equation}
onde a constante $Z_2$ é fixada pela condição de normalização

\begin{equation}
\overline{\Gamma}^{(1,n)}(0,0,m,g)=1 . \label{condinorz2}
\end{equation}

Através da definição das funções auxiliares

\begin{equation}
\beta = m\frac{\partial g}{\partial m}\biggr|_{\lambda_B,\Lambda} ,
\end{equation}

\begin{equation}
\gamma = \frac{1}{2}Z_{\phi}^{-1}m\frac{\partial Z_{\phi}}{\partial m}\biggr|_{\lambda_B,\Lambda} ,
\end{equation}
pode-se reescrever a equação diferencial (\ref{eqcallsym}) na forma

\begin{equation}
\left( m\frac{\partial}{\partial m}+\beta \frac{\partial}{\partial g}-n\gamma \right) \overline{\Gamma}^{(n)}(\vec{k}_i,m,g)=Z_2m\frac{\partial m_B^2}{\partial m}\biggr|_{\lambda_B,\Lambda} \overline{\Gamma}^{(1,n)}(0,\vec{k}_i,m,g), \hspace*{0.4cm} n\geq 1 . \label{eqsyma}
\end{equation}

Usando as condições de normalização (\ref{condinor1}) e (\ref{condinorz2}) e fazendo $n=2$ na equação (\ref{eqsyma}), obtemos

\begin{equation}
(2-2\gamma)m^2=Z_2m\frac{\partial m_B^2}{\partial m}\biggr|_{\lambda_B,\Lambda} .
\end{equation}

Isso permite chegar na {\bf equação de Callan-Symanzik}, dada por

\begin{equation}
\left( m\frac{\partial}{\partial m}+\beta \frac{\partial}{\partial g}-n\gamma \right) \overline{\Gamma}^{(n)}(\vec{k}_i,m,g)=(2-2\gamma)m^2 \overline{\Gamma}^{(1,n)}(0,\vec{k}_i,m,g), \hspace*{0.4cm} n\geq 1 . \label{callansymansik}
\end{equation}

Em geral as funções adimensionais $\beta$ e $\gamma$ dependem de $g$ e $m$. A equação de Callan-Symanzik relaciona diversas teorias renormalizadas entre si, cujo parâmetro que varia é a escala de massa, indo para pequenas massas ou, equivalentemente, para grandes momentos. A partir dos resultados aproximados da equação homogênea da expressão (\ref{callansymansik}), pode-se deduzir o comportamento crítico da teoria, desde que a função $\beta$ seja nula em alguma força de acoplamento $g=g^{*}$.

Adicionando o parâmetro dimensional $\varepsilon$ na lista de argumentos, temos que as funções de correlação não-renormalizadas são dadas por

\begin{equation}
G_B^{(n)}(\vec{x}_1,\ldots, \vec{x}_n;m_B,\lambda_B,\varepsilon)=\langle \phi_B(\vec{x}_1)\cdots \phi_B(\vec{x}_n) \rangle ,
\end{equation}
e as funções de correlação renormalizadas são dadas por

\begin{equation}
G^{(n)}(\vec{x}_1,\ldots, \vec{x}_n;m,g,\mu,\varepsilon)=\langle \phi(\vec{x}_1)\cdots \phi(\vec{x}_n) \rangle .
\end{equation}

Através da renormalização multiplicativa, pode-se obter

\begin{equation}
G_B^{(n)}(\vec{x}_1,\ldots, \vec{x}_n;m_B,\lambda_B,\varepsilon)=Z_{\phi}^{n/2}(g(\mu),\varepsilon)G^{(n)}(\vec{x}_1,\ldots, \vec{x}_n;m,g,\mu,\varepsilon), \hspace*{0.4cm} n\geq 1 ,
\end{equation}
onde $Z_{\phi}(g(\mu),\varepsilon)$ é a constante de renormalização do campo definido por $\phi_B=Z_{\phi}^{1/2}\phi$. Assim, podemos obter uma relação para as funções de vértice renormalizada e não-renormalizada a partir de partes 1PI das funções de correlação de $n$-pontos conectadas, onde teremos

\begin{equation}
\overline{\Gamma}^{(n)}(\vec{k}_i;m,g,\mu , \varepsilon)=Z_{\phi}^{n/2}(g(\mu),\varepsilon)\overline{\Gamma}_B^{(n)}(\vec{k}_i;m_B,\lambda_B , \varepsilon), \hspace*{0.4cm} n\geq 1 . \label{funvertice2}
\end{equation}

Temos que os parâmetros renormalizados $g$, $m$ e $\phi$, definidos nas expressões (\ref{massarenor})-(\ref{camporenor}), dependem das quantidades não-renormalizadas e do parâmetro de massa $\mu$. Logo temos

\begin{equation}
\phi^2=Z_{\phi}^{-1}(g(\mu),\varepsilon)\phi_B^2 ,
\end{equation}

\begin{equation}
m^2=m^2(\mu)\equiv \frac{Z_{\phi}(g(\mu),\varepsilon)}{Z_{m^2}(g(\mu),\varepsilon)}m_B^2 ,
\end{equation}

\begin{equation}
g=g(\mu)\equiv \mu^{-\varepsilon}\frac{Z_{\phi}^2(g(\mu),\varepsilon)}{Z_g(g(\mu),\varepsilon)}\lambda_B .
\end{equation}

Assim, temos que as funções de vértice renormalizadas $\overline{\Gamma}^{(n)}(\vec{k}_i;m,g,\mu , \varepsilon)$ dependem de $\mu$ de duas maneiras: uma explicitamente, e uma via $g(\mu)$ e $m(\mu)$. A dependência explícita vem a partir dos fatores $\mu^{\varepsilon}$ que são gerados quando substituimos $\lambda$ por $\mu^{\varepsilon}g$ nas integrais de Feynman. Ao contrário, as funções de vértice não-renormalizadas $\overline{\Gamma}_B^{(n)}(\vec{k}_i;m_B,\lambda_B,\varepsilon)$ não dependem de $\mu$. As mudanças associadas às funções de vértice renormalizadas e de outros parâmetros renormalizados, devem ser relacionados uns aos outros de uma maneira específica. E essa relação é que assegura que a informação física nas funções renormalizadas permaneçam invariantes sobre as mudanças de $\mu$.

Agora aplicando o operador adimensional $\mu \frac{\partial}{\partial \mu}$ na equação (\ref{funvertice2}) e fixando os parâmetros não-renormalizados $m_B$ e $\lambda_B$, podemos obter

\begin{equation}
\left[ \mu \frac{\partial}{\partial \mu} + \mu \frac{\partial g}{\partial \mu}\biggr|_B \frac{\partial}{\partial g} - \frac{1}{2}n\mu \frac{\partial ln Z_{\phi}}{\partial \mu}\biggr|_B + \frac{\mu}{m^2}\frac{\partial m^2}{\partial \mu}\biggr|_B m^2\frac{\partial}{\partial m^2} \right] \overline{\Gamma}^{(n)}(\vec{k}_i;m,g,\mu, \varepsilon)=0 .
\end{equation}

Essa equação expressa a invariância de $\overline{\Gamma}^{(n)}(\vec{k}_i,m,g,\mu)$ sob uma transformação $(\mu, m(\mu),g(\mu)) \longrightarrow (\mu',m(\mu'),g(\mu'))$. Os observáveis do sistema do campo são invariantes sob uma mudança de escala de massa $\mu \longrightarrow \mu'$ se a constante de acoplamento $g(\mu)$ e massa $m(\mu)$ são adequadamente mudadas.

A dependência apropriada de $g$, $m$ e $Z_{\phi}$ em $\mu$ é descrito pelas {\bf funções do grupo de re\-nor\-ma\-li\-za\-ção}, dadas por

\begin{equation}
\gamma(m,g,\mu)=\mu \frac{\partial ln Z_{\phi}}{\partial \mu} \biggr|_B ,
\end{equation}

\begin{equation}
\gamma_m(m,g,\mu)=\frac{\mu}{m^2} \frac{\partial m^2}{\partial \mu} \biggr|_B ,
\end{equation}

\begin{equation}
\beta(m,g,\mu)=\mu \frac{\partial g}{\partial \mu} \biggr|_B .
\end{equation}

Assim a {\bf equação do grupo de renormalização} para as funções de vértice com $n \geq 1$ pode ser reescrita como

\begin{equation}
\left[\mu \frac{\partial}{\partial \mu} + \beta(m,g,\mu)\frac{\partial}{\partial g} - \frac{1}{2}n\gamma(m,g,\mu) + \gamma_m(m,g,\mu)m^2\frac{\partial}{\partial m^2} \right]\overline{\Gamma}^{(n)}(\vec{k}_i;m,g,\mu)=0 . \label{RGE}
\end{equation}

A solução da equação diferencial parcial (\ref{RGE}) é geralmente complicada, quando $\beta$, $\gamma$ e $\gamma_m$ dependem de $m$, $g$ e $\mu$. Uma importante propriedade do esquema de subtração mínima de 't Hooft e Veltman, é que os contra-termos independem da massa $m$, tendo dependência somente da constante de acoplamento $g$ e de $\varepsilon$ \cite{kleinert:2000}. Com isso, as funções do grupo de renormalização são independentes de $m$ e $\mu$, e dependem unicamente de $g$, logo as funções, sob o esquema de subtração mínima, podem ser reescritas como

\begin{equation}
\gamma(g)=\mu \frac{\partial ln Z_{\phi}}{\partial \mu} \biggr|_B ,  \label{fgr1}
\end{equation}

\begin{equation}
\gamma_m(g)=\frac{\mu}{m^2} \frac{\partial m^2}{\partial \mu} \biggr|_B , \label{fgr2}
\end{equation}

\begin{equation}
\beta(g)=\mu \frac{\partial g}{\partial \mu} \biggr|_B . \label{fgr3}
\end{equation}

Com isso, a equação do grupo de renormalização pode ser reescrita como

\begin{equation}
\left[\mu \frac{\partial}{\partial \mu} + \beta(g)\frac{\partial}{\partial g} - \frac{1}{2}n\gamma(g) + \gamma_m(g)m^2\frac{\partial}{\partial m^2} \right]\overline{\Gamma}^{(n)}(\vec{k}_i;m,g,\mu)=0  .
\end{equation}

\subsection{Cálculo das Funções do Grupo de Renormalização }

Serão calculadas as funções do grupo de renormalização aproveitando o fato que as constantes de renormalização dependem, através da subtração mínima, somente de $\mu$ via constante de acoplamento renormalizado $g(\mu)$.

Temos que podemos expressar as (\ref{fgr1})-(\ref{fgr3}) em termos de grandezas adimensionais, e assim podemos escrever de uma outra maneira as funções do grupo de renormalização  na forma

\begin{equation}
\beta(g)=-\varepsilon \left[\frac{\partial}{\partial g}ln(gZ_gZ_{\phi}^{-2}) \right]^{-1} ,
\end{equation}

\begin{equation}
\gamma(g)=\beta(g)\frac{\partial ln Z_{\phi}}{\partial g} , 
\end{equation}

\begin{equation}
\gamma_m(g)=\gamma(g) - \beta(g)\frac{\partial ln Z_{m^2}}{\partial g} .
\end{equation}

Fazendo então as seguintes expansões em termos da constante de acoplamento renormalizada

\begin{equation}
Z_{\phi}=1 + b_2g^2 + b_3g^3 , \label{crexpa1}
\end{equation}

\begin{equation}
Z_{m^2}=1 + c_1g + c_2g^2 ,
\end{equation}

\begin{equation}
Z_g=1 + a_1g + a_2g^2 , \label{crexpa3}
\end{equation}
no qual a expansão da constante de renormalização do campo, $Z_{\phi}$, é extendida até a ordem de $g^3$ para aplicarmos o mesmo conjunto de equações para o caso Lifshitz, em que o termo $b_1$ da expansão é omitido, pois não há contribuições em primeira ordem de $g$ para o campo. Logo, podemos escrever as funções do grupo de renormalização como

\begin{equation}
\beta(g)=-\varepsilon g[1-a_1g+2(a_1^2-a_2+2b_2)g^2] , \label{fgrexpa1}
\end{equation}

\begin{equation}
\gamma(g)=-\varepsilon g[2b_2g+(3b_3-2a_1b_2)g^2] ,
\end{equation}

\begin{equation}
\gamma_m(g)=\gamma(g)+\varepsilon g[c_1+(2c_2-c_1^2-a_1c_1)g] . \label{fgrexp3}
\end{equation}

Comparando as expansões das constantes de renormalização (\ref{crexpa1})-(\ref{crexpa3}) com os resultados (\ref{cr1})-(\ref{cr3}), podemos identificar os coeficientes das expansões dados por

\begin{equation}
a_1=\frac{1}{(4\pi)^2}\frac{(N+8)}{3\varepsilon} ,
\end{equation}

\begin{equation}
a_2=\frac{1}{(4\pi)^4}\left[\frac{(N+8)^2}{9\varepsilon^2} - \frac{5N+22}{9\varepsilon} \right] ,
\end{equation}

\begin{equation}
b_2=-\frac{1}{(4\pi)^4}\frac{N+2}{36\varepsilon} ,
\end{equation}

\begin{equation}
b_3=0 ,
\end{equation}

\begin{equation}
c_1=\frac{1}{(4\pi)^2}\frac{N+2}{3\varepsilon} ,
\end{equation}

\begin{equation}
c_2=\frac{1}{(4\pi)^4}\left[-\frac{N+2}{6\varepsilon} + \frac{(N+2)(N+5)}{9\varepsilon^2} \right] ,
\end{equation}
e assim, substituindo os coeficientes nas funções (\ref{fgrexpa1})-(\ref{fgrexp3}) e introduzindo a constante de acoplamento modificada $\overline{g} \equiv \frac{g}{(4\pi)^2}$ teremos

\begin{equation}
\beta(\overline{g})=-\varepsilon \overline{g} + \frac{N+8}{3}\overline{g}^2 - \frac{3N+14}{3}\overline{g}^3 , \label{fgr4}
\end{equation}

\begin{equation}
\gamma(\overline{g})=\frac{N+2}{18} \overline{g}^2 , \label{fgr5}
\end{equation}

\begin{equation}
\gamma_m(\overline{g})=\frac{N+2}{3}\overline{g}-\frac{5(N+2)}{18}\overline{g}^2 . \label{fgr6}
\end{equation}

\subsection{Expoentes Críticos}

Agora será calculado as propriedades críticas da teoria $\phi^4$ com simetria $O(N)$ na aproximação em 2-{\it loops}. Com o uso da equação (\ref{fgr4}) calculamos o ponto fixo não trivial $\overline{g}^{*}$ de $\beta(\overline{g}^{*})=0$ resultando em

\begin{equation}
\overline{g}^{*}=\frac{3}{(N+8)}\varepsilon \left[ 1 + \frac{3(3N+14)}{(N+8)^2}\varepsilon \right] . \label{pfixo}
\end{equation}

Esse ponto fixo é não-atrativo e com isso tem-se a necessidade de fixar o valor da constante de acoplamento renormalizada no seu valor do regime ultravioleta $\overline{g}^{*}$ para obtermos funções de vértice invariantes por transformações de escala. 

Os expoentes críticos $\eta$ e $\nu$ são calculados através das equações (\ref{fgr5}) e (\ref{fgr6}), no ponto fixo ultravioleta $g=\overline{g}^{*}$ como

\begin{eqnarray}
\eta &=&\gamma(\overline{g}^{*}) , \label{etaa}\\
\nu &=&\frac{1}{2-\gamma_m(\overline{g}^{*})} . \label{nuu}
\end{eqnarray}

Com isso, substituindo o ponto fixo (\ref{pfixo}) nas expressões (\ref{fgr5}) e (\ref{fgr6}), e com o auxílio das expressões (\ref{etaa}) e (\ref{nuu}), obtemos os expoentes $\eta$ e $\nu$ até a ordem de 2-{\it loops}, respectivamente, dado por

\begin{equation}
\eta=\frac{N+2}{2(N+8)^2}\varepsilon^2 ,
\end{equation}

\begin{equation}
\nu=\frac{1}{2} + \frac{N+2}{4(N+8)}\varepsilon + \frac{(N+2)(N^2 + 23N + 60)}{8(N+8)^3}\varepsilon^2 .
\end{equation}

Os outros expoentes críticos são calculados através das relações de escala (\ref{Rushbrooke})-(\ref{Josephson}).

\chapter{Método de BPHZ para Pontos de Lifshitz $m$-Axiais Anisotrópicos}

Neste capítulo abordaremos a renormalização e o estudo das propriedades críticas de sistemas competitivos que apresentam o comportamento crítico de Lifshitz. No capítulo anterior, expusemos com um pouco de detalhes a renormalização do modelo $N$-vetorial da teoria $\lambda\phi^4$, que descreve os sistemas livres de competição. A renormalização em sistemas competitivos é realizada de uma maneira similar à descrita no capítulo anterior, e destacaremos os principais resultados do Grupo de Renormalização aplicados no cenário com interações competitivas do tipo Lifshitz anisotrópico.

O diagrama de fases dos modelos com interações competitivas entre primeiros e segundos sítios vizinhos de spins apresenta o ponto de intersecção das fases denominado de ponto de Lifshitz. A representação deste modelo em variáveis contínuas é expressa através de uma modificação da teoria $\lambda\phi^4$ quando incluímos termos de derivadas de ordem superior ao longo de $m$ direções competitivas. A densidade de energia livre de Ginzburg-Landau para um sistema com competições anisotrópicas $(d\neq m)$ com um campo de parâmetro de ordem $N$-vetorial $\phi$ é escrita como

\begin{equation}
\mathcal{L}(\phi)=\frac{1}{2}|\nabla_{(d-m)}\phi |^2 + \frac{1}{2}|\nabla_{(m)}^2\phi|^2 + \frac{\delta_0}{2}|\nabla_{(m)}\phi|^2 + \frac{\mu_0^2}{2}\phi^2 + \frac{\lambda}{4!}\phi^4 ,
\end{equation}
onde  as constantes $\mu_0$ e $\lambda$ são a massa e a constante de acoplamento não-renormalizada, respectivamente. A região crítica do ponto de Lifshitz é definida em torno da temperatura crítica $T_L$ e a um determinado valor na razão $J_2/J_1$, o qual anula $\delta_0$. Portanto, podemos expressar a ação funcional para o comportamento crítico do tipo Lifshitz $m$-axial por meio da expressão

\begin{eqnarray}
E[\phi]=\int d^{d-m}x_{\bot}d^mx_{||}\mathcal{L}(\phi)=\int d^{d-m}qd^mk\frac{1}{2}\phi(-q,-k)[q^2 + (k^2)^2 + \mu_0^2]\phi(q,k)\nonumber \\
+ \int \mathcal{L}_{int}(\phi) ,
\end{eqnarray}
em que $\int \mathcal{L}_{int}(\phi)$ representa o setor de interação da ação.

O símbolo $x_{\bot}$ representa o espaço das coordenadas $\mathbb{R}^{d-m}$ ausente de competições, enquanto que $x_{||}$ é definido no subespaço competitivo $\mathbb{R}^m$. São definidas as dimensões das coordenadas por meio de $[x_{||}]=[x_{\bot}]^{1/2}=\Lambda^{-1/2}$, isto é equivalente a realizar uma redefinição dimensional do momento ao longo dos eixos competitivos, enquanto a condição $\delta_0=0$ é satisfeita. Neste caso, o elemento de integração desenvolve a dimensão $[d^{d-m}x_{\bot}d^mx_{||}]=\Lambda^{-(d-m/2)}$. Como a ação é mantida adimensional no sistema natural de medidas, a dimensão canônica da variável de campo é $[\phi]=\Lambda^{\frac{1}{2}\left(d-\frac{m}{2}\right)-1}$. E por seguinte, a dimensão canônica da constante de acoplamento é então definida através da relação

\begin{equation}
[\lambda]=\Lambda^{4+\frac{m}{2}-d} \equiv \Lambda^{d_c - d} .
\end{equation}

A dimensão que torna a constante de acoplamento adimensional é chamada de dimensão crítica $d_c$ da teoria. Daí introduzimos o parâmetro de regularização dimensional como

\begin{equation}
\varepsilon_L = d_c - d=4+\frac{m}{2}-d .
\end{equation}

Lembrando que $k$ é o momento definido ao longo das $m$ direções competitivas e $q$ é o momento ao longo das $d-m$ direções não competitivas.

\section{Equações de Callan-Symanzik e Funções de Renormalização}

Para sistemas competitivos, o tratamento do grupo de renormalização é semelhante àquele usado para sistemas sem competição. As integrais de Feynman para o caso das competições anisotrópicas envolvem duas escalas para os momentos externos. A definição do conjunto de condições de normalização são aplicados separadamente nos subespaços não-competitivo $\mathbb{R}^{d-m}$ e competitivo $\mathbb{R}^m$ semelhante as desenvolvido no capítulo 1 através

\begin{equation}
\overline{\Gamma}_{\tau}^{(2)}(0,m_{\tau},g_{\tau})=m_{\tau}^{2\tau}  , \label{condinor1lif}
\end{equation}

\begin{equation}
\frac{\partial}{\partial p_{\tau}^{2\tau}} \overline{\Gamma}_{\tau}^{(2)}(0,m_{\tau},g_{\tau})=1 , \label{condinor2lif}
\end{equation}

\begin{equation}
\overline{\Gamma}_{\tau}^{(4)}(0,m_{\tau},g_{\tau})=g_{\tau} , \label{condinor3lif}
\end{equation}
onde lembramos que fazemos as identificações $p_1=p$ e $p_2=K'$ para os momentos externos dos dois subespaços independentes.

Podemos obter as equações de Callan-Symanzik para o caso anisotrópico da mesma maneira que as obtemos para a teoria $\lambda\phi^4$ convencional do capítulo 2. Essas equações são então dadas por

\begin{equation}
\left( m_{\tau}\frac{\partial}{\partial m_{\tau}}+\beta_{\tau} \frac{\partial}{\partial g_{\tau}}-n\gamma_{\tau} \right) \overline{\Gamma}_{\tau}^{(n)}=(2-2\gamma_{\tau})m_{\tau}^{2\tau} \overline{\Gamma}^{(1,n)}_{\tau}, \hspace*{0.4cm} n\geq 1 .
\end{equation}

Similarmente ao que obtemos no capítulo 2, podemos escrever a equação do grupo de re\-nor\-ma\-li\-za\-ção para o caso Lifshitz na forma

\begin{equation}
\left[\mu_{\tau} \frac{\partial}{\partial \mu_{\tau}} + \beta_{\tau}(g_{\tau})\frac{\partial}{\partial g_{\tau}} - \frac{1}{2}n\gamma_{\tau}(g_{\tau}) + \gamma_{m_{\tau}}(g_{\tau})m_{\tau}^{2\tau}\frac{\partial}{\partial m_{\tau}^{2\tau}} \right]\overline{\Gamma}^{(n)}_{\tau}=0  ,
\end{equation}
onde as funções do grupo de renormalização são dadas por

\begin{equation}
\gamma_{\tau}(g_{\tau})=\mu_{\tau} \frac{\partial ln Z_{\phi(\tau)}}{\partial \mu_{\tau}} \biggr|_B , \label{fgr1lif}
\end{equation}

\begin{equation}
\gamma_{m_{\tau}}(g_{\tau})=\frac{\mu_{\tau}}{m_{\tau}^{2\tau}} \frac{\partial m_{\tau}^{2\tau}}{\partial \mu_{\tau}} \biggr|_B , \label{fgr2lif}
\end{equation}

\begin{equation}
\beta_{\tau}(g_{\tau})=\mu_{\tau} \frac{\partial g_{\tau}}{\partial \mu_{\tau}} \biggr|_B . \label{fgr3lif}
\end{equation}

O rótulo $\tau =1$ ou $\tau = 2$ nas variáveis diz respeito aos subespaços $\mathbb{R}^{d-m}$ e $\mathbb{R}^m$, respectivamente, em que é tratado as quantidades. A partir das funções do grupo de renormalização é possível obter os expoentes críticos no ponto fixo $\nu_{\tau}$ e $\eta_{\tau}$, onde veremos mais adiante. E também temos que, a partir das relações de escala pode-se obter o restante dos expoentes críticos.

\section{Cálculo das Constantes de Renormalização via Contra-Termos}

Métodos similares àqueles apresentados no capítulo 2, em referência ao cálculo das constantes de renormalização, serão aplicados agora para o caso Lifshitz. Onde calculamos a constante de renormalização do campo até a ordem $g_{\tau}^3$ e as constante da massa e de acoplamento até a ordem $g_{\tau}^2$. 

As constantes de renormalização para o caso Lifshitz, já considerando os fatores de simetria, são dados na forma

\begin{eqnarray}
Z_{\phi(\tau)}(g_{\tau},\varepsilon_L^{-1})= 1 + \frac{1}{p^2 + (K'^2)^2}\Biggl[ \frac{1}{6} \mathcal{K} 
\left( \parbox{8mm}{ \begin{picture}(29,18) (152,-182)
    \SetWidth{1.0}
    \SetColor{Black}
    \Arc(166,-173)(8.246,256,616)
    \Line(152,-173)(180,-173)
    \Vertex(158,-173){1.5}
    \Vertex(174,-173){1.5}
  \end{picture}} \hspace*{0.2cm}
\right) \Biggr|_{m_{\tau}^{2\tau}=0} S_{\scalebox{0.3}{\begin{picture}(29,18) (152,-182)
    \SetWidth{1.0}
    \SetColor{Black}
    \Arc(166,-173)(8.246,256,616)
    \Line(152,-173)(180,-173)
    \Vertex(158,-173){1.5}
    \Vertex(174,-173){1.5}
  \end{picture}}} +
\frac{1}{4} \mathcal{K} 
\left(\parbox{8mm} { \begin{picture}(30,22) (427,-102)
    \SetWidth{1.0}
    \SetColor{Black}
    \Arc(442,-91)(10.05,276,636)
    \Arc(429.125,-77.75)(13.558,-77.758,-18.268)
    \Arc[clock](454,-79)(12.748,-101.31,-168.69)
    \Vertex(442,-81){1.5}
    \Vertex(432,-91){1.5}
    \Vertex(452,-91){1.5}
    \Line(452,-91)(457,-91)
    \Line(432,-91)(427,-91)
  \end{picture}} \hspace*{0.2cm} \right) \Biggr|_{m_{\tau}^{2\tau}=0}S_{\scalebox{0.3}{\begin{picture}(30,22) (427,-102)
    \SetWidth{1.0}
    \SetColor{Black}
    \Arc(442,-91)(10.05,276,636)
    \Arc(429.125,-77.75)(13.558,-77.758,-18.268)
    \Arc[clock](454,-79)(12.748,-101.31,-168.69)
    \Vertex(442,-81){1.5}
    \Vertex(432,-91){1.5}
    \Vertex(452,-91){1.5}
    \Line(452,-91)(457,-91)
    \Line(432,-91)(427,-91)
  \end{picture}}} \nonumber \\
   +  \frac{1}{3} \mathcal{K}
  \left(\parbox{8mm} { \begin{picture}(29,18) (152,-182)
    \SetWidth{1.0}
    \SetColor{Black}
    \Arc(166,-173)(8.246,256,616)
    \Line(152,-173)(180,-173)
    \Vertex(158,-173){1.5}
    \Vertex(174,-173){2.8}
  \end{picture}} \hspace*{0.2cm} \right) S_{\scalebox{0.3}{\begin{picture}(29,18) (152,-182)
    \SetWidth{1.0}
    \SetColor{Black}
    \Arc(166,-173)(8.246,256,616)
    \Line(152,-173)(180,-173)
    \Vertex(158,-173){1.5}
    \Vertex(174,-173){2.8}
  \end{picture}}} \Biggr] ,
\end{eqnarray}

\begin{eqnarray}
Z_{m_{\tau}^{2\tau}}(g_{\tau},\varepsilon_L^{-1}) = 1 + \frac{1}{m_{\tau}^{2\tau}} \Biggl[ \frac{1}{2} \mathcal{K}
\left( \begin{picture}(30,17) (212,-153) 
    \SetWidth{0.9}
    \SetColor{Black}
    \Arc(227,-144)(7.071,262,622)
    \Line(213,-152)(241,-152)
    \Vertex(227,-151.5){1.5}
  \end{picture} \right)S_{\scalebox{0.3}{\begin{picture}(30,17) (212,-153) 
    \SetWidth{0.9}
    \SetColor{Black}
    \Arc(227,-144)(7.071,262,622)
    \Line(213,-152)(241,-152)
    \Vertex(227,-151.5){1.5}
  \end{picture}}} +
  \frac{1}{4} \mathcal{K}
  \left( \parbox{8mm} {\begin{picture}(28,30) (204,-140)
    \SetWidth{0.9}
    \SetColor{Black}
    \Arc(218,-129)(6.083,261,621)
    \Line(205,-136)(231,-136)
    \Vertex(218,-135.5){1.5}
    \Arc(218,-116)(6.083,261,621)
    \Vertex(218,-122){1.5}
  \end{picture}} \hspace*{0.2cm}  \right)S_{\scalebox{0.3}{\begin{picture}(28,30) (204,-140)
    \SetWidth{0.9}
    \SetColor{Black}
    \Arc(218,-129)(6.083,261,621)
    \Line(205,-136)(231,-136)
    \Vertex(218,-135.5){1.5}
    \Arc(218,-116)(6.083,261,621)
    \Vertex(218,-122){1.5}
  \end{picture}}} +
  \frac{1}{2} \mathcal{K}
  \left( \parbox{8mm} {\begin{picture}(16,21) (216,-193)
    \SetWidth{1.0}
    \SetColor{Black}
    \Arc(224,-182)(7,270,630)
    \Line(215,-190)(233,-190)
    \Vertex(224,-189.5){1.5}
    \Line(221,-172)(227,-179)
    \Line(227,-172)(221,-179)
  \end{picture}} \right)S_{\scalebox{0.3}{\begin{picture}(16,21) (216,-193)
    \SetWidth{1.0}
    \SetColor{Black}
    \Arc(224,-182)(7,270,630)
    \Line(215,-190)(233,-190)
    \Vertex(224,-189.5){1.5}
    \Line(221,-172)(227,-179)
    \Line(227,-172)(221,-179)
  \end{picture}}} \nonumber \\
  + \frac{1}{2} \mathcal{K}
  \left( \begin{picture}(30,17) (212,-153)
    \SetWidth{1.0}
    \SetColor{Black}
    \Arc(227,-144)(7.071,262,622)
    \Line(213,-152)(241,-152)
    \Vertex(227,-151){3}
  \end{picture} \right)S_{\scalebox{0.3}{\begin{picture}(30,17) (212,-153)
    \SetWidth{1.0}
    \SetColor{Black}
    \Arc(227,-144)(7.071,262,622)
    \Line(213,-152)(241,-152)
    \Vertex(227,-151){3}
  \end{picture}}} +
  \frac{1}{6} \mathcal{K}
  \left( \parbox{8mm}{ \begin{picture}(29,18) (152,-182)
    \SetWidth{1.0}
    \SetColor{Black}
    \Arc(166,-173)(8.246,256,616)
    \Line(152,-173)(180,-173)
    \Vertex(158,-173){1.5}
    \Vertex(174,-173){1.5}
  \end{picture}}\hspace*{0.2cm} \right) \Biggr|_{p^2 + (K'^2)^2=0}S_{\scalebox{0.3}{\begin{picture}(29,18) (152,-182)
    \SetWidth{1.0}
    \SetColor{Black}
    \Arc(166,-173)(8.246,256,616)
    \Line(152,-173)(180,-173)
    \Vertex(158,-173){1.5}
    \Vertex(174,-173){1.5}
  \end{picture}}} \Biggr] ,
\end{eqnarray}

\begin{eqnarray}
Z_{g_{\tau}}(g_{\tau},\varepsilon_L^{-1})&=&1 + \frac{1}{g_{\tau} \mu_{\tau}^{\tau \varepsilon_L}} \Biggl[ \frac{3}{2} \mathcal{K}
\left( \parbox{8mm} { \begin{picture}(30,24) (218,-169)
    \SetWidth{1.0}
    \SetColor{Black}
    \Arc(233,-157)(8,252,612)
    \Line(241.5,-157)(246,-151)
    \Line(241.5,-157)(246,-163)
    \Line(224.5,-157)(220,-151)
    \Line(224.5,-157)(220,-163)
    \Vertex(224.5,-157){1.5}
    \Vertex(241.5,-157){1.5}
  \end{picture}}\hspace*{0.2cm}  \right)S_{\scalebox{0.3}{\begin{picture}(30,24) (218,-169)
    \SetWidth{1.0}
    \SetColor{Black}
    \Arc(233,-157)(8,252,612)
    \Line(241.5,-157)(246,-151)
    \Line(241.5,-157)(246,-163)
    \Line(224.5,-157)(220,-151)
    \Line(224.5,-157)(220,-163)
    \Vertex(224.5,-157){1.5}
    \Vertex(241.5,-157){1.5}
  \end{picture}}} + 
  3 \mathcal{K} 
  \left(\parbox{8mm} {\begin{picture}(39,29) (156,-201)
    \SetWidth{0.8}
    \SetColor{Black}
    \Arc(179,-186)(11,270,630)
    \Line(168,-186)(157,-175)
    \Line(168,-186)(157,-197)
    \Arc(193.5,-186.5)(13.509,87.879,272.121)
    \Vertex(168,-186){1.5}
    \Vertex(184,-177){1.5}
    \Vertex(184,-196){1.5}
  \end{picture} }\hspace*{0.6cm} \right)S_{\scalebox{0.3}{\begin{picture}(39,29) (156,-201)
    \SetWidth{0.8}
    \SetColor{Black}
    \Arc(179,-186)(11,270,630)
    \Line(168,-186)(157,-175)
    \Line(168,-186)(157,-197)
    \Arc(193.5,-186.5)(13.509,87.879,272.121)
    \Vertex(168,-186){1.5}
    \Vertex(184,-177){1.5}
    \Vertex(184,-196){1.5}
  \end{picture}}} +
  \frac{3}{4} \mathcal{K}
  \left(\parbox{8mm} { \begin{picture}(44,16) (142,-142)
    \SetWidth{1.0}
    \SetColor{Black}
    \Arc(156,-134)(7.28,254,614)
    \Arc(171,-134)(7.28,254,614)
    \Vertex(163.5,-134){1.5}
    \Line(178,-134)(185,-128)
    \Line(178,-134)(185,-141)
    \Line(149,-134)(143,-128)
    \Line(149,-134)(143,-141)
    \Vertex(149,-134){1.5}
    \Vertex(178,-134){1.5}
  \end{picture}}\hspace*{0.7cm} \right)S_{\scalebox{0.3}{\begin{picture}(44,16) (142,-142)
    \SetWidth{1.0}
    \SetColor{Black}
    \Arc(156,-134)(7.28,254,614)
    \Arc(171,-134)(7.28,254,614)
    \Vertex(163.5,-134){1.5}
    \Line(178,-134)(185,-128)
    \Line(178,-134)(185,-141)
    \Line(149,-134)(143,-128)
    \Line(149,-134)(143,-141)
    \Vertex(149,-134){1.5}
    \Vertex(178,-134){1.5}
  \end{picture}}} \nonumber \\
  &&+ \frac{3}{2} \mathcal{K}
  \left(\parbox{8mm} { \begin{picture}(30,24) (218,-167)
    \SetWidth{1.0}
    \SetColor{Black}
    \Arc(233,-157)(8,252,612)
    \Arc(233,-145)(4.472,117,477)
    \Line(241,-157)(246,-151)
    \Line(241,-157)(246,-163)
    \Line(224,-157)(220,-151)
    \Line(224,-157)(220,-163)
    \Vertex(233,-149){1.5}
    \Vertex(224.7,-157){1.5}
    \Vertex(241.5,-157){1.5}
  \end{picture}}\hspace*{0.2cm} \right)S_{\scalebox{0.3}{\begin{picture}(30,24) (218,-167)
    \SetWidth{1.0}
    \SetColor{Black}
    \Arc(233,-157)(8,252,612)
    \Arc(233,-145)(4.472,117,477)
    \Line(241,-157)(246,-151)
    \Line(241,-157)(246,-163)
    \Line(224,-157)(220,-151)
    \Line(224,-157)(220,-163)
    \Vertex(233,-149){1.5}
    \Vertex(224.7,-157){1.5}
    \Vertex(241.5,-157){1.5}
  \end{picture}}} +
  3 \mathcal{K}
  \left(\parbox{8mm}{ \begin{picture}(30,24) (218,-169)
    \SetWidth{1.0}
    \SetColor{Black}
    \Arc(233,-157)(8,252,612)
    \Line(241.5,-157)(246,-151)
    \Line(241.5,-157)(246,-163)
    \Line(224.5,-157)(220,-151)
    \Line(224.5,-157)(220,-163)
    \Vertex(224.5,-157){1.5}
    \Vertex(241.5,-157){3}
  \end{picture}}\hspace*{0.2cm} \right)S_{\scalebox{0.3}{\begin{picture}(30,24) (218,-169)
    \SetWidth{1.0}
    \SetColor{Black}
    \Arc(233,-157)(8,252,612)
    \Line(241.5,-157)(246,-151)
    \Line(241.5,-157)(246,-163)
    \Line(224.5,-157)(220,-151)
    \Line(224.5,-157)(220,-163)
    \Vertex(224.5,-157){1.5}
    \Vertex(241.5,-157){3}
  \end{picture}}} +
  3 \mathcal{K}
  \left( \parbox{8mm} { \begin{picture}(30,24) (218,-169)
    \SetWidth{1.0}
    \SetColor{Black}
    \Arc(233,-157)(8,252,612)
    \Line(241.5,-157)(246,-151)
    \Line(241.5,-157)(246,-163)
    \Line(224.5,-157)(220,-151)
    \Line(224.5,-157)(220,-163)
    \Vertex(224.5,-157){1.5}
    \Vertex(241.5,-157){1.5}
    \Line(230,-145.5)(236,-152.5)
    \Line(236,-145.5)(230,-152.5)
  \end{picture}}\hspace*{0.2cm} \right)S_{\scalebox{0.3}{\begin{picture}(30,24) (218,-169)
    \SetWidth{1.0}
    \SetColor{Black}
    \Arc(233,-157)(8,252,612)
    \Line(241.5,-157)(246,-151)
    \Line(241.5,-157)(246,-163)
    \Line(224.5,-157)(220,-151)
    \Line(224.5,-157)(220,-163)
    \Vertex(224.5,-157){1.5}
    \Vertex(241.5,-157){1.5}
    \Line(230,-145.5)(236,-152.5)
    \Line(236,-145.5)(230,-152.5)
  \end{picture}}} \Biggr] .
  \end{eqnarray}

Usando os resultados dos diagramas de Feynman em expansão $\varepsilon_L$ apresentado no apêndice C.1 e considerando os fatores de simetria apresentados no apêndice C.3, podemos expressar as constantes de renormalização para o caso Lifshitz na forma

\begin{equation}
Z_{\phi(\tau)}(g_{\tau},\varepsilon_L^{-1})=1-\frac{(N+2)}{144 \varepsilon_L}g_{\tau}^2 - \frac{(N+2)(N+8)}{1296 \varepsilon_L^2}\left( 1 - \frac{1}{4}\varepsilon_L \right)g_{\tau}^3 , \label{crlif1}
\end{equation}

\begin{equation}
Z_{m_{\tau}^{2\tau}}(g_{\tau},\varepsilon_L^{-1})=1+\frac{(N+2)}{6 \varepsilon_L}g_{\tau} + \left[ \frac{(N+2)(N+5)}{36 \varepsilon_L^2} - \frac{(N+2)}{24 \varepsilon_L} \right]g_{\tau}^2 ,
\end{equation}

\begin{equation}
Z_{g_{\tau}}(g_{\tau},\varepsilon_L^{-1})=1+\frac{(N+8)}{6 \varepsilon_L}g_{\tau} + \left[ \frac{(N+8)^2}{36 \varepsilon_L^2} - \frac{5N+22}{36\varepsilon_L}\right]g_{\tau}^2 . \label{crlif3}
\end{equation}

\section{Cálculo das Funções do Grupo de Renormalização }

Agora iremos calcular as funções do grupo de renormalização como realizado no capítulo 2, para o caso Lifshitz.

Podemos expressar as equações (\ref{fgr1lif})-(\ref{fgr3lif}) em termos de grandezas adimensionais, e assim podemos escrever de uma outra maneira as funções do grupo de renormalização  na forma

\begin{equation}
\beta_{\tau}(g_{\tau})=-\tau \varepsilon_L \left[\frac{\partial}{\partial g_{\tau}}ln(g_{\tau}Z_{g_{\tau}}Z_{\phi(\tau)}^{-2}) \right]^{-1} ,
\end{equation}

\begin{equation}
\gamma_{\tau}(g_{\tau})=\beta_{\tau}(g_{\tau})\frac{\partial ln Z_{\phi(\tau)}}{\partial g_{\tau}} ,
\end{equation}

\begin{equation}
\gamma_{m_{\tau}}(g_{\tau})=\gamma_{\tau}(g_{\tau}) - \beta_{\tau}(g_{\tau})\frac{\partial ln Z_{m_{\tau}^{2\tau}}}{\partial g_{\tau}} .
\end{equation}

Fazendo então as seguintes expansões em termos da constante de acoplamento renormalizada

\begin{equation}
Z_{\phi(\tau)}=1 + b_{2\tau}g_{\tau}^2 + b_{3\tau}g_{\tau}^3 , \label{crexpa1lif}
\end{equation}

\begin{equation}
Z_{m_{\tau}^{2\tau}}=1 + c_{1\tau}g_{\tau} + c_{2\tau}g_{\tau}^2 ,
\end{equation}

\begin{equation}
Z_{g_{\tau}}=1 + a_{1\tau}g_{\tau} + a_{2\tau}g_{\tau}^2 , \label{crexpa3lif}
\end{equation}
podemos escrever as funções do grupo de renormalização como

\begin{equation}
\beta_{\tau}(g_{\tau})=-\tau \varepsilon_L g_{\tau}[1-a_{1\tau}g_{\tau}+2(a_{1\tau}^2-a_{2\tau}+2b_{2\tau})g_{\tau}^2] , \label{fgrexpa1lif}
\end{equation}

\begin{equation}
\gamma_{\tau}(g_{\tau})=-\tau \varepsilon_L g_{\tau}[2b_{2\tau}g_{\tau}+(3b_{3\tau}-2a_{1\tau}b_{2\tau})g_{\tau}^2] ,
\end{equation}

\begin{equation}
\gamma_{m_{\tau}}(g_{\tau})=\gamma_{\tau}(g_{\tau})+ \tau \varepsilon_L g_{\tau}[c_{1\tau}+(2c_{2\tau}-c_{1\tau}^2-a_{1\tau}c_{1\tau})g_{\tau}] . \label{fgrexp3lif}
\end{equation}

Comparando as expansões das constantes de renormalização (\ref{crexpa1lif})-(\ref{crexpa3lif}) com os resultados (\ref{crlif1})-(\ref{crlif3}), podemos identificar os coeficientes das expansões dados por

\begin{equation}
a_{1\tau}=\frac{N+8}{6\varepsilon_L} ,
\end{equation}

\begin{equation}
a_{2\tau}=\frac{(N+8)^2}{36 \varepsilon_L^2} - \frac{5N+22}{36 \varepsilon_L} ,
\end{equation}

\begin{equation}
b_{2\tau}=-\frac{N+2}{144 \varepsilon_L} ,
\end{equation}

\begin{equation}
b_{3\tau}=-\frac{(N+2)(N+8)}{1296 \varepsilon_L^2}\left(1-\frac{1}{4}\varepsilon_L  \right) ,
\end{equation}

\begin{equation}
c_{1\tau}=\frac{N+2}{6 \varepsilon_L} ,
\end{equation}

\begin{equation}
c_{2\tau}=\frac{(N+2)(N+5)}{36\varepsilon_L^2} - \frac{N+2}{24 \varepsilon_L} ,
\end{equation}
e assim, substituindo os coeficientes nas funções (\ref{fgrexpa1lif})-(\ref{fgrexp3lif}) teremos

\begin{equation}
\beta_{\tau}(g_{\tau})=\tau \left[-\varepsilon_L g_{\tau} + \frac{N+8}{6}g_{\tau}^2 - \frac{3N+14}{12}g_{\tau}^3  \right] , \label{fgr4lif}
\end{equation}

\begin{equation}
\gamma_{\tau}(g_{\tau})= \tau \left[\frac{N+2}{72}g_{\tau}^2 - \frac{(N+2)(N+8)}{1728}g_{\tau}^3   \right] , \label{fgr5lif}
\end{equation}

\begin{equation}
\gamma_{m_{\tau}}(g_{\tau})= \tau \left[\frac{N+2}{6}g_{\tau} - \frac{5(N+2)}{72}g_{\tau}^2 -  \frac{(N+2)(N+8)}{1728}g_{\tau}^3   \right] . \label{fgr6lif}
\end{equation}

\section{Cálculo dos Expoentes Críticos Anisotrópicos}

Nesta seção determinaremos as propriedades críticas de sistemas do tipo Lifshitz pela adaptação do método da teoria $\lambda\phi^4$ com simetria $O(N)$ mostrada no capítulo 2. Incluída a discussão da seção anterior, calcularemos na aproximação até a ordem 3 em número de {\it loops} para o expoente $\eta_{\tau}$ e o expoente $\nu_{\tau}$ até a ordem 2 em número de {\it loops}. Com o uso da equação (\ref{fgr4lif}) calculamos o ponto fixo não trivial $g_{\tau}^{*}$ de $\beta_{\tau}(g_{\tau}^{*})=0$ resultando em

\begin{equation}
g_{\tau}^{*}=\frac{6}{(N+8)}\varepsilon_L \left[1 + \frac{3(3N+ 14)}{(N+8)^2}\varepsilon_L  \right]  . \label{pfixolif}
\end{equation}

Os expoentes críticos $\eta_{\tau}$ e $\nu_{\tau}$ são calculados através das equações (\ref{fgr5lif}) e (\ref{fgr6lif}), no ponto fixo ultravioleta $g_{\tau}=g_{\tau}^{*}$. Logo temos

\begin{eqnarray}
\eta_{\tau} &=&\gamma_{\tau}(g_{\tau}^{*}) , \label{etaalif}\\
\nu_{\tau} &=&\frac{1}{2\tau-\gamma_{m_{\tau}}(g_{\tau}^{*})} . \label{nuulif}
\end{eqnarray}

Com isso, substituindo o ponto fixo (\ref{pfixolif}) nas expressões (\ref{fgr5lif}) e (\ref{fgr6lif}), e com o auxílio das expressões (\ref{etaalif}) e (\ref{nuulif}), obtemos os expoentes $\eta_{\tau}$ até a ordem 3 em número de {\it loops} e $\nu_{\tau}$ até a ordem 2 em número de {\it loops}, respectivamente, dado por

\begin{equation}
\eta_{\tau}=\frac{\tau}{2} \frac{N+2}{(N+8)^2}\varepsilon_L^2 \left[ 1+ \left( \frac{6(3N+14)}{(N+8)^2}-\frac{1}{4} \right)\varepsilon_L \right] ,
\end{equation}

\begin{equation}
\nu_{\tau}=\frac{1}{\tau} \left[ \frac{1}{2} + \frac{N+2}{4(N+8)}\varepsilon_L + \frac{(N+2)(N^2 + 23N + 60)}{8(N+8)^3}\varepsilon_L^2 \right] .
\end{equation}

Os expoentes críticos $\eta_{\tau}$ e $\nu_{\tau}$ concordam com os expoentes encontrados utilizando outras técnicas, confirmando assim o caráter universal desses expoentes. Eles só dependem do número $N$ de componentes de parâmetro de ordem e da dimensão espacial $d$ do sistema. Assim, sistemas que possuem os mesmos parâmetros $(N,d,m)$ pertencem à mesma classe de universalidade.

\chapter{Conclusão e Perspectivas}

Neste trabalho investigamos o comportamento crítico de sistemas físicos com interações competitivas que apresentam pontos de Lifshitz $m$-axiais. Para esse estudo usamos as técnicas de Teoria Quântica de Campos Escalares Massivos renormalizada em momentos externos não nulos, com interações do tipo $\lambda \phi^4$ para obtermos uma expansão perturbativa para as funções de vértice de 2-pontos até a ordem de 3-{\it loops} e de 4-pontos até a ordem de 2-{\it loops}. 

Essas funções de vértice foram regularizadas usando o método de regularização dimensional de polos dimensionais e renormalizadas usando o método de subtração mínina, onde calculamos as integrais de Feynman, para o caso anisotrópico, usando a aproximação ortogonal. Uma das vantagens para utilizarmos o método de subtração mínima foi que não precisamos calcular as integrais logarítimicas que aparecem no processo de resolução, pois estas são canceladas no decorrer do cálculo, e uma outra vantagem está ligada a não especificação da escala de momento $k$, onde não precisamos inserir o conceito de ponto de simetria \cite{amit:1978} como é feito em condições de normalização.  No entanto, as  divergências encontradas nas integrais de Feynman devido ao esquema de regularização, foram removidas pela  adição dos vértices de contra-termos correspondentes às amplitudes superficialmente divergentes, caracterizando o método BPHZ \cite{kleinert:2000} usado neste trabalho.

Os expoentes críticos calculados neste trabalho são exemplo de grandezas que exibem um caráter universal, ou seja, são independentes das características microscópicas dos sistemas considerados. A propriedade de universalidade dos expoentes críticos é uma consequência natural da aplicação das técnicas do grupo de renormalização aos sistemas que exibem transições de fase, o que engloba também os sistemas com interações competitivas do tipo Lifshitz.

Através das ideias do Grupo de Renormalização, foram definidas as funções de Wilson que originam os pontos fixos, e a partir dessas funções e dos pontos fixos, calculamos os expoentes críticos anisotrópicos $\eta_{\tau}$, até a ordem três no número de {\it loops}, e $\nu_{\tau}$ até a ordem dois no número de {\it loops}, que caracterizam o comportamento crítico do tipo Lifshitz $m$-axial.  Os expoentes calculados estão em perfeita concordância com os correspondentes expoentes calculados anteriormente usando outros métodos \cite{carvalho:2008,messias:2010}, dando confiança sobre os resultados encontrados e em contrapartida, os nossos resultados confirmam a veracidade dos expoentes críticos anisotrópicos encontrados em \cite{carvalho:2008,messias:2010}. Este fato confirma a hipótese de universalidade dos expoentes críticos que são os mesmos independentemente da teoria utilizada para calculá-los.
 
Uma extensão desse trabalho é aplicar o mesmo método BPHZ para o estudo do comportamento crítico de Lifshitz isotrópico aproximado e exato, e estendê-lo para o estudo de caráter geral. Levando a confirmar os resultados anteriormente obtidos por outros métodos.

\apendice
\chapter{}

Nesse apêndice serão mostradas algumas expressões que são utilizadas para o cálculo de integrais D-dimensionais. 


\section{Coordenadas Polares e Superfície de uma Esfera em D Dimensões}

Vamos considerar o espaço de Minkowski d-dimensional, onde temos uma dimensão temporal e $(D-1)$ dimensões espaciais. Iremos considerar o sistema em coordenadas polares esféricas, onde podemos representar um ponto qualquer no espaço, na forma

\begin{equation}
P=(P_0,\vec{r})=(P_0,r,\phi ,\theta_1 ,\theta_2 ,\ldots ,\theta_{D-3}).
\end{equation}

\begin{figure}[htb]
\begin{center}
\includegraphics[scale=0.45]{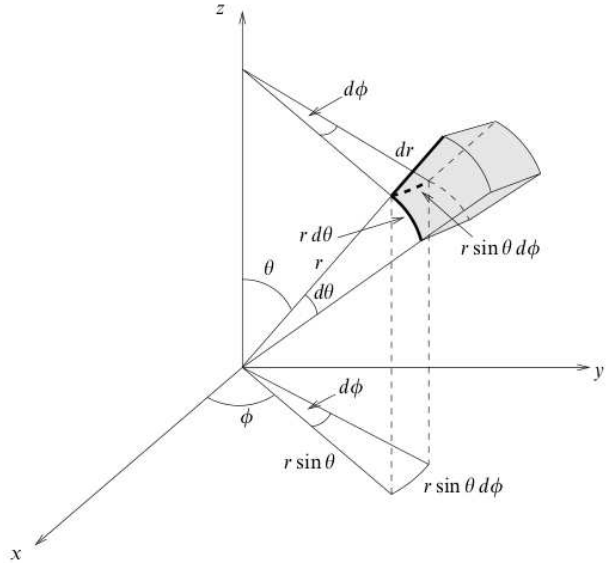}
\caption{Elemento de volume tridimensional em coordenadas polares esféricas.}
\label{fig:elemento3d}
\end{center}
\end{figure}

Geometricamente, como mostra a figura (\ref{fig:elemento3d}), o elemento de volume tridimensional é dado por

\begin{equation}
d^3x=r^2sin\theta drd\theta d\phi ,
\end{equation}

Agora vamos generalizar o conceito para coordenadas esféricas em 4D. Para isso iremos considerar um conjunto de coordenadas $x,y,z$ e $t$, onde a distância de um ponto qualquer no espaço à origem tenha norma $r$. Projetando esse vetor em $t$ e no $R^3$, formado por $x,y$ e $z$, teremos como coordenadas desse vetor 

\begin{eqnarray}
rcos\theta_2 ,\\
rsin\theta_2 .
\end{eqnarray}

Escolhendo $\theta_2$ de tal forma que $rcos\theta_2$ esteja projetado em $t$ e $rsin\theta_2$ esteja em $R^3$, temos que aplicando novamente as coordenadas esféricas para esse novo vetor em $R^3$, iremos obter o conjunto de coordenadas para o $R^4$ dado por

\begin{eqnarray}
x&=&rsin\theta_2 sin\theta_1 cos\phi , \label{xx} \\
y&=&rsin\theta_2 sin\theta_1 sin\phi , \label{y} \\
z&=&rsin\theta_2 cos\theta_1 , \label{z} \\
t&=&rcos\theta_2 . \label{t}
\end{eqnarray}

Assim, o elemento de volume em coordenas esféricas do $R^4$ será dado por

\begin{equation}
d^4x=\frac{\partial(x,y,z,t)}{\partial(r,\theta_1, \theta_2, \phi)}drd\theta_1 d\theta_2 d\phi , \label{elemento4d}
\end{equation}
onde o jacobiano dessa transformação é dado por

$$
\frac{\partial(x,y,z,t)}{\partial(r,\theta_1, \theta_2, \phi)} = \left|
\begin{array}{cccc}
\frac{\partial x }{\partial r} & \frac{\partial x }{\partial \theta_1} & \frac{\partial x }{\partial \theta_2} & \frac{\partial x }{\partial \phi} \\
\frac{\partial y }{\partial r} & \frac{\partial y }{\partial \theta_1} & \frac{\partial y }{\partial \theta_2} & \frac{\partial y }{\partial \phi} \\
\frac{\partial z }{\partial r} & \frac{\partial z }{\partial \theta_1} & \frac{\partial z }{\partial \theta_2} & \frac{\partial z }{\partial \phi} \\
\frac{\partial t }{\partial r} & \frac{\partial t }{\partial \theta_1} & \frac{\partial t }{\partial \theta_2} & \frac{\partial t }{\partial \phi} \\
\end{array}
\right| .
$$

Usando as coordenadas (\ref{xx})-(\ref{t}) e aplicando as derivadas correspondentes, iremos encontrar o determinante do jacobiano dado por

\begin{equation}
\frac{\partial(x,y,z,t)}{\partial(r,\theta_1, \theta_2, \phi)}=r^3sin\theta_1 sin^2\theta_2 . \label{jacob}
\end{equation}

Portanto, substituindo o resultado (\ref{jacob}) na expressão (\ref{elemento4d}), obtemos o elemento de volume em coordenadas esféricas para 4D, dado por

\begin{equation}
d^4x=r^3sin\theta_1 sin^2\theta_2 drd\theta_1 d\theta_2 d\phi .
\end{equation}

Generalizando para $D$ dimensões, temos que

\begin{eqnarray}
d^DP&=&dP_0dr(rd\theta_1 rd\theta_2 \cdots rd\theta_{D-3})(rsin\theta_1 sin^2\theta_2 \cdots sin^{D-3}\theta_{D-3})d\phi ,\\
d^DP&=&dP_0r^{D-2}drd\phi sin\theta_1 d\theta_1 sin^2\theta_2 d\theta_2 \cdots sin^{D-3}\theta_{D-3} , \nonumber \\
d^DP&=&dP_0r^{D-2}drd\phi \displaystyle \prod_{k=1}^{D-3} sin^k\theta_k d\theta_k . \label{elevol}
\end{eqnarray}

Através do resultado (\ref{elevol}), podemos obter o elemento de volume D-dimensional para o espaço euclidiano fazendo a troca da dimensão temporal pela dimensão espacial, assim obtemos

\begin{eqnarray}
dP_0\equiv rsin\theta d\theta , \\
d^DP=r^{D-1}drd\phi \displaystyle \prod_{k=1}^{D-2} sin^k\theta_k d\theta_k ,\label{vol}
\end{eqnarray}
onde os limites de integração são 

$$
\left\{
\begin{array}{ccccc}
0 &<& r &<& +\infty , \\
0 &<& \phi &<& 2\pi ,\\
0 &<& \theta_k &<& \pi .\\
\end{array}
\right.
$$

Ao calcular integrais D-dimensionais, podemos denotar parte dela como sendo uma integral direcional, cujo resultado é uma superfície de uma esfera de raio unitário em $D$ dimensões:

\begin{equation}
S_D= \int d\phi \displaystyle \prod_{k=1}^{D-2} sin^k\theta_k d\theta_k .
\end{equation}

Usando os limites de integração obtemos

\begin{equation}
S_D= 2\pi \displaystyle \prod_{k=1}^{D-2} \int_0^\pi sin^k\theta_k d\theta_k .\label{int1}
\end{equation}

Usando a fórmula integral

\begin{equation}
\int_0^{\frac{\pi}{2}}(sin\theta)^{2n-1}(cos\theta)^{2m-1}d\theta = \frac{1}{2}\frac{\Gamma(n) \Gamma (m)}{\Gamma(n+m)}, \hspace*{0.3cm} Re(n)>0,\hspace*{0.1cm} Re(m)>0,
\end{equation}
e fazendo $m=\frac{1}{2}$ e $n=\frac{k+1}{2}$, iremos obter

\begin{equation}
\int_0^{\pi}sin^k\theta_k d\theta_k =2 \int_0^{\frac{\pi}{2}}sin^k\theta_k d\theta_k = \sqrt{\pi}\frac{\Gamma(\frac{k+1}{2})}{\Gamma(\frac{k+2}{2})} . \label{int2}
\end{equation}

Usando o resultado (\ref{int2}) na expressão (\ref{int1}), temos

\begin{equation}
S_D= 2\pi^{\frac{D}{2}} \displaystyle \prod_{k=1}^{D-2} \frac{\Gamma(\frac{k+1}{2})}{\Gamma(\frac{k+2}{2})} = \frac{2\pi^{\frac{D}{2}}}{\Gamma(\frac{D}{2})} .
\end{equation}

Portanto, a superfície de uma esfera unitária  em $D$ dimensões é

\begin{equation}
S_D = \frac{2\pi^{\frac{D}{2}}}{\Gamma(\frac{D}{2})} . \label{sd}
\end{equation}


\section{Expansão da Função Gama}

A função Gama pode ser definida pela integral

\begin{equation}
\Gamma(z)=\int_0^{\infty}dt t^{z-1}e^{-t} . \label{gam}
\end{equation}

Essa integral tem pólo para $z=0$ e inteiros negativos. Aplicando uma integração por partes, iremos obter a identidade

\begin{equation}
\Gamma(z+1)=z \int_0^{\infty}dt t^{z-1}e^{-t} = z\Gamma(z) , \label{gama}
\end{equation}
o que demonstra que a função Gama é uma generalização de um fatorial de uma variável arbitrária $z$
complexa.

A função Digama de Euler é definida pela relação

\begin{equation}
\psi(z)=\frac{d}{dz} ln\Gamma(z) = \frac{\Gamma'(z)}{\Gamma(z)} . \label{digama}
\end{equation}

Usando a expressão (\ref{gama}), e usando a definição

\begin{equation}
\Gamma(z)= \displaystyle \lim_{n \to \infty}\frac{n!n^z}{z(z+1)(z+2)\cdots(z+n)} ,
\end{equation}
iremos obter

\begin{equation}
\Gamma(z+1)= \displaystyle \lim_{n \to \infty}\frac{n!n^z}{(z+1)(z+2)\cdots(z+n)} . \label{gama2}
\end{equation}

Aplicando o logaritmo na expressao (\ref{gama2}), temos

\begin{eqnarray}
ln\Gamma (z+1) &=& \displaystyle \lim_{n \to \infty}ln \Biggl[\frac{n! n^z}{(z+1)(z+2)\cdots(z+n)}\Biggr] \\
&=& \displaystyle \lim_{n \to \infty}\Bigl[ln[n!n^z]-ln[(z+1)(z+2)\cdots(z+n)]\Bigr] \nonumber \\
&=& \displaystyle\lim_{n \to \infty} \Bigl[ ln(n!) + zln(n)-ln(z+1)-ln(z+2)-\cdots -ln(z+n) \Bigr],
\end{eqnarray} 
derivando em relação a $z$, temos

\begin{eqnarray}
\frac{d}{dz}ln\Gamma(z+1)&=&\displaystyle \lim_{n \to \infty}\Biggl[ ln(n) - \frac{1}{z+1} - \frac{1}{z+2} - \cdots - \frac{1}{z+n} \Biggr] = \psi(z+1) \\
&=& \displaystyle \lim_{n \to \infty}\Biggl[ ln(n) + \displaystyle\sum_{r=1}^{n} \frac{1}{r} -\displaystyle\sum_{r=1}^{n} \frac{1}{r} -  \frac{1}{z+1} - \frac{1}{z+2} - \cdots - \frac{1}{z+n} \Biggr] \nonumber \\
&=& \displaystyle \lim_{n \to \infty} \Biggl[ ln(n) - \displaystyle\sum_{r=1}^{n} \frac{1}{r} \Biggr] + \displaystyle\sum_{n=1}^{\infty} \Biggl( \frac{1}{n} - \frac{1}{z+n} \Biggr) .
\end{eqnarray}

Definindo a constante de Euler-Mascheroni como

\begin{equation}
\gamma= \displaystyle \lim_{n \to \infty} \Biggl[ \displaystyle\sum_{r=1}^{n} \frac{1}{r} - ln(n)\Biggr] ,
\end{equation}
temos 

\begin{equation}
\psi(z+1)=-\gamma + \displaystyle\sum_{n=1}^{\infty} \frac{z}{n(n+z)} ,
\end{equation}
fazendo $z=0$, obtemos

\begin{equation}
\psi(1)=-\gamma . \label{constante}
\end{equation}

Agora vamos considerar a função $\Gamma(1+\varepsilon)$, onde expandindo em série de Taylor e usando as relações  (\ref{digama}) e (\ref{constante}), temos 

\begin{eqnarray}
\Gamma(1+\varepsilon)&=&\Gamma(1)+\Gamma'(1)+O(\varepsilon^2) \\
&=& 1+\psi(1)\varepsilon + O(\varepsilon^2) \nonumber \\ 
&=& 1-\gamma \varepsilon + O(\varepsilon^2) . \label{gama3}
\end{eqnarray}

Fazendo $z=\varepsilon$ na expressão (\ref{gama}) e usando o resultado da expressão (\ref{gama3}), obtemos

\begin{equation}
\Gamma(\varepsilon)=\frac{1}{\varepsilon}-\gamma + O(\varepsilon) . \label{eps}
\end{equation}

Novamente usando a expressão (\ref{gama}) e usando o resultado (\ref{eps}), temos

\begin{eqnarray}
\Gamma(-1+\varepsilon)&=&(-1+\varepsilon)^{-1}\Gamma(\varepsilon) \\
&=& -(1+ \varepsilon + \varepsilon^2 + \cdots)\biggl( \frac{1}{\varepsilon}-\gamma + O(\varepsilon) \biggr) \nonumber \\
&=& - \biggl( \frac{1}{\varepsilon} +1 -\gamma + O(\varepsilon) \biggr) .
\end{eqnarray}

De modo análogo, temos

\begin{eqnarray}
\Gamma(-2+\varepsilon)&=&(-2+\varepsilon)^{-1}\Gamma(-1+\varepsilon) \\
&=& \frac{(-1)^2}{2} (1-\frac{\varepsilon}{2})^{-1} \biggl( \frac{1}{\varepsilon} +1 -\gamma + O(\varepsilon) \biggr) \nonumber \\
&=& \frac{(-1)^2}{2} \biggl( 1 + \frac{\varepsilon}{2} + \frac{\varepsilon^2}{4} + \cdots   \biggr)     \biggl( \frac{1}{\varepsilon} +1 -\gamma + O(\varepsilon) \biggr) \nonumber \\
&=& \frac{(-1)^2}{2} \biggl[ \frac{1}{\varepsilon} + (1+ \frac{1}{2} -\gamma) +  O(\varepsilon)  \biggr] .
\end{eqnarray}

Generalizando, obtemos

\begin{eqnarray}
\Gamma(-n+\varepsilon)=\frac{(-1)^n}{n!}\biggl[\frac{1}{\varepsilon} + \psi(n+1) +O(\varepsilon)      \biggr] ,\hspace*{0.3cm} \psi(n+1)= -\gamma + \displaystyle\sum_{r=1}^{n}\frac{1}{r} . \label{expgam}
\end{eqnarray}

\chapter{}

Nesse apêndice será mostrado algumas fórmulas que são necessárias para o cálculo dos diagramas de Feynman em qualquer dimensão.

\section{ Parametrização de Feynman}

Uma fórmula muito útil para o cálculo dos diagramas de Feynman, é a que pode ser encontrada usando o método da parametrização de Feynman. Onde cada intervalo de integração sobre esses parâmetros de Feynman, $x_i$, são dados por

\begin{equation}
0 \leq x_i \leq 1 ; \hspace*{0.3cm} x_1 + x_2 + \cdots x_{n-1} \leq 1 .
\end{equation}

Para demonstrar a expressão, vamos considerar a parametrização-$\alpha$ \cite{grozin:2007} dada por

\begin{equation}
\frac{1}{a_{1}^{\alpha_1}}=\frac{1}{\Gamma(\alpha_1)} \displaystyle\int_{0}^{+\infty}e^{-a_1\beta_1}\beta_1^{\alpha_1-1}d\beta_1 ,
\end{equation}
logo temos que

\begin{eqnarray}
\frac{1}{a_{1}^{\alpha_1}a_{2}^{\alpha_2}} &=& \Biggl[ \frac{1}{\Gamma(\alpha_1)} \displaystyle\int_{0}^{+\infty}e^{-a_1\beta_1}\beta_1^{\alpha_1-1}d\beta_1 \Biggr]  \Biggl[  \frac{1}{\Gamma(\alpha_2)} \displaystyle\int_{0}^{+\infty}e^{-a_2\beta_2}\beta_2^{\alpha_2-1}d\beta_2 \Biggr] \\
&=& \frac{1}{\Gamma(\alpha_1)\Gamma(\alpha_2)} \displaystyle\int_{0}^{+\infty} e^{-a_1\beta_1 -a_2\beta_2 } \beta_{1}^{\alpha_1 -1}\beta_{2}^{\alpha_2 -1}d\beta_1 d\beta_2 . \label{integral1}
\end{eqnarray}

Considerando os parâmetros $\beta_1 =\eta x_1$, $\beta_2 =\eta(1-x_1)$ e $\eta=\beta_1 + \beta_2$, temos que a transformação jacobiana é dada por

\begin{equation}
d\beta_1 d\beta_2 = J(x_1,\eta) d\eta dx_1 ,
\end{equation} 
onde 

$$
J(x_1,\eta) = \left|
\begin{array}{cc}
\frac{\partial \beta_1 }{\partial x_1} & \frac{\partial \beta_1 }{\partial \eta}  \\
\frac{\partial \beta_2 }{\partial x_1} & \frac{\partial \beta_2 }{\partial \eta}  \\
\end{array}
\right| = \eta ,
$$
obtendo, portanto

\begin{equation}
d\beta_1 d\beta_2 = \eta d\eta dx_1 . \label{jacobiano}
\end{equation}

Substituindo o resultado (\ref{jacobiano}) na expressão (\ref{integral1}), teremos

\begin{eqnarray}
\frac{1}{a_{1}^{\alpha_1}a_{2}^{\alpha_2}} &=& \frac{1}{\Gamma(\alpha_1)\Gamma(\alpha_2)} \displaystyle\int_{0}^{1} \displaystyle\int_{0}^{+ \infty} e^{-a_1\eta x_1 - a_2 \eta (1-x_1)}(\eta x_1)^{\alpha_1 -1} \bigl[\eta(1- x_1) \bigr]^{\alpha_2 - 1}\eta d\eta dx_1 \\
&=& \frac{1}{\Gamma(\alpha_1)\Gamma(\alpha_2)} \displaystyle\int_{0}^{1} \displaystyle\int_{0}^{+ \infty} e^{-\eta[a_1 x_1 + a_2(1- x_1)]}\eta^{\alpha_1 + \alpha_2 -1}x_1^{\alpha_1 -1}(1-x_1)^{\alpha_2 - 1}d\eta dx_1 . \label{integral2}
\end{eqnarray}

Usando a definição da função Gama (\ref{gam})

\begin{equation}
\Gamma(\alpha_1)=\int_0^{+\infty}dt t^{\alpha_1 -1}e^{-t} ,
\end{equation}
fazendo a mudança de variável $\eta=\frac{t}{[a_1 x_1 + a_2(1-x_1)]}$, obtemos

\begin{eqnarray}
\eta^{\alpha_1 + \alpha_2 -1} &=& \frac{t^{\alpha_1 + \alpha_2 -1}}{[a_1x_1 + a_2(1-x_1)]^{\alpha_1 + \alpha_2 -1}} \label{eta1} , \\ 
d\eta &=& \frac{dt}{[a_1x_1 + a_2(1-x_1)]} , \label{deta}
\end{eqnarray}
e substituindo os resultados (\ref{eta1}) e (\ref{deta}) na expressão (\ref{integral2}), temos

\begin{eqnarray}
\frac{1}{a_1^{\alpha_1} a_2^{\alpha_2}} &=& \frac{1}{\Gamma(\alpha_1)\Gamma(\alpha_2)} \int_{0}^{1} \int_{0}^{+\infty} e^{-t} \frac{t^{\alpha_1 + \alpha_2 -1}}{[a_1x_1 + a_2(1-x_1)]^{\alpha_1 + \alpha_2 -1}} \frac{1}{[a_1x_1 + a_2(1-x_1)]} \times \nonumber\\
&& dt x_1^{\alpha_1 -1}(1-x_1)^{\alpha_2 - 1}dx_1 \nonumber \\
&=& \frac{1}{\Gamma(\alpha_1)\Gamma(\alpha_2)}\int_{0}^{1} \int_{0}^{+\infty}e^{-t} t^{\alpha_1 + \alpha_2 -1}dt \frac{x_1^{\alpha_1 -1}(1-x_1)^{\alpha_2 -1}}{[a_1x_1 + a_2(1-x_1)]^{\alpha_1 + \alpha_2}}dx_1 .
\end{eqnarray}

Portanto, para dois termos teremos

\begin{eqnarray}
\frac{1}{a_1^{\alpha_1} a_2^{\alpha_2}} = 
\frac{\Gamma(\alpha_1 + \alpha_2)}{\Gamma(\alpha_1)\Gamma(\alpha_2)}\int_{0}^{1}
\frac{x_1^{\alpha_1 -1}(1-x_1)^{\alpha_2 -1}}{[a_1x_1 + a_2(1-x_1)]^{\alpha_1 + \alpha_2}}dx_1 . \label{paramefey}
\end{eqnarray}

De modo análogo, podemos encontrar a expressão para três termos, dada por

\begin{eqnarray}
\frac{1}{a_1^{\alpha_1} a_2^{\alpha_2} a_3^{\alpha_3}} = 
\frac{\Gamma(\alpha_1 + \alpha_2 + \alpha_3)}{\Gamma(\alpha_1)\Gamma(\alpha_2)\Gamma(\alpha_3)}\int_{0}^{1}
\frac{x_1^{\alpha_1 -1} x_2^{\alpha_2 -1} (1-x_1-x_2)^{\alpha_3 -1}}{[a_1x_1 + a_2x_2 + a_3(1-x_1 - x_2)]^{\alpha_1 + \alpha_2 + \alpha_3}}dx_1dx_2 .
\end{eqnarray}

Generalizando, podemos encontrar uma expressão geral para $n$ termos, ou seja

\begin{eqnarray}
\frac{1}{a_1^{\alpha_1} a_2^{\alpha_2} \cdots a_n^{\alpha_n}} &=& 
\frac{\Gamma(\alpha_1 + \alpha_2 + \cdots + \alpha_n)}{\Gamma(\alpha_1)\Gamma(\alpha_2) \cdots \Gamma(\alpha_n)} \times \nonumber \\
&& \int_{0}^{1}
\frac{x_1^{\alpha_1 -1} x_2^{\alpha_2 -1} \cdots x_{n-1}^{\alpha_{n-1} -1} (1-x_1-x_2 - \cdots - x_{n-1})^{\alpha_n -1}}{[a_1x_1 + a_2x_2 + \cdots + a_{n-1}x_{n-1} + a_n(1-x_1 - x_2 - \cdots -x_{n-1})]^{\alpha_1 + \alpha_2 + \cdots +  \alpha_n}} \times \nonumber \\
&& dx_1dx_2\cdots dx_{n-1} . \label{parametrizafey}
\end{eqnarray}

\section{Fórmulas Integrais}

Considerando o espaço euclidiano $D$-dimensional, vamos encontrar uma expressão para um caso simples e geral da integral de Feynman, 

\begin{eqnarray}
I(D,\alpha;k)=\int \frac{d^Dq}{(2\pi)^D}\frac{1}{(q^2 + 2qp +m^2)^{\alpha}} ,
\end{eqnarray}
onde será usado o processo de regularização dimensional de 't Hooft e Veltman \cite{kleinert:2000}.

Através do resultado (\ref{vol}) temos

\begin{eqnarray}
I(D,\alpha;k) &=& \int \frac{1}{(2\pi)^D}\frac{1}{(q^2 + 2qp +m^2)^{\alpha}} q^{D-1}dqd\phi \prod_{v=1}^{D-2}sin^v\theta_v d\theta_v \nonumber \\
&=& \frac{2\pi}{(2\pi)^D} \prod_{v=1}^{D-2} \int_{0}^{\pi} sin^v\theta_v d\theta_v \int_{0}^{\infty}\frac{q^{D-1}}{(q^2 + 2qp +m^2)^{\alpha}}dq .
\end{eqnarray}

 Usando a expressão (\ref{int1}) temos que

\begin{eqnarray}
I(D,\alpha;k)=\frac{S_D}{(2\pi)^D}\int_{0}^{\infty}\frac{q^{D-1}}{(q^2 + 2qp +m^2)^{\alpha}}dq ,
\end{eqnarray}
onde $S_D$ é a superfície de uma esfera unitária.

Usando a função Beta de Euler dada por

\begin{eqnarray}
B(x,y)=\frac{\Gamma(x)\Gamma(y)}{\Gamma(x+y)}=2\int_{0}^{\infty}t^{2x-1}(1-t^2)^{-x-y}dt ,
\end{eqnarray}
e fazendo as mudanças de variáveis $x=\frac{1+\beta}{2}$, $y=\alpha-\frac{1+\beta}{2}$ e $t=\frac{s}{u}$, obtemos

\begin{eqnarray}
\int_{0}^{\infty}\frac{s^\beta}{(s^2+u^2)^\alpha}ds = \frac{\Gamma(\frac{1+\beta}{2})\Gamma(\alpha - \frac{1+\beta}{2})}{2(u^2)^{\alpha -(\frac{1+\beta}{2})}\Gamma(\alpha)} . \label{betainte}
\end{eqnarray}

Através da transformação

\begin{eqnarray}
q^2+2qp+m^2= q^2+2qp+m^2+p^2-p^2=(q+p)^2+m^2-p^2 ,
\end{eqnarray}
onde teremos uma equivalência $q'=q+p\equiv q$, podemos obter

\begin{eqnarray}
I(D,\alpha;k)=\frac{S_D}{(2\pi)^D}\int_{0}^{\infty}\frac{q^{D-1}}{[q^2 +(m^2-k^2)]^{\alpha}}dq . \label{ifey}
\end{eqnarray}

Através da comparação do resultado (\ref{ifey}) com a expressão (\ref{betainte}), observarmos que $s=q$, $u^2=m^2 - p^2$ e $\beta=D-1$, temos que

\begin{eqnarray}
I(D,\alpha;k)&=&\frac{S_D}{(2\pi)^D} \frac{\Gamma(\frac{1+D-1}{2})\Gamma(\alpha - \frac{1+D-1}{2})}{2(m^2-p^2)^{\alpha -(\frac{1+D-1}{2})}\Gamma(\alpha)} \nonumber \\
&=& \frac{S_D}{(2\pi)^D} \frac{\Gamma(\frac{D}{2})\Gamma(\alpha - \frac{D}{2})}{2(m^2-p^2)^{\alpha -\frac{D}{2}}\Gamma(\alpha)} .
\end{eqnarray}
usando o resultado (\ref{sd}) obtemos a expressão final da integral

\begin{eqnarray}
\int \frac{d^Dq}{(2\pi)^D}\frac{1}{(q^2 + 2qp +m^2)^{\alpha}}=\frac{1}{(4\pi)^{\frac{D}{2}}} \frac{\Gamma(\alpha - \frac{D}{2})(m^2-p^2)^{\frac{D}{2} -\alpha}}{\Gamma(\alpha)} . \label{inteklei}
\end{eqnarray}

O resultado encontrado na expressão (\ref{inteklei}) será utilizado para o cálculo dos diagramas de Feynman relacionadas ao estudo das constantes de renormalização do capítulo 2.

Para o cálculo das constantes de renormalização para o caso Lifshitz no capítulo 3, iremos redefinir o fator geométrico $S_D$ de modo que

\begin{eqnarray}
\int d^Dq \frac{1}{(q^2 + 2qp +m^2)^{\alpha}} =  S_D \frac{1}{2} \frac{\Gamma(\frac{D}{2})\Gamma(\alpha - \frac{D}{2})}{\Gamma(\alpha)}(m^2-p^2)^{\frac{D}{2} -\alpha} , \label{intamit}
\end{eqnarray}
onde 

\begin{eqnarray}
S_D=\frac{1}{2^{D-1}\pi^{\frac{D}{2}} \Gamma(\frac{D}{2})} .
\end{eqnarray}

\chapter{}

Nesse apêndice serão mostrados os resultados dos cálculos das integrais de Feynman em regularização dimensional com os resultados encontrados nos Apêndices A e B.


\section{Cálculo de Integrais de Feynman: Pontos de Lifshitz m-axiais}
 
 Calcularemos agora integrais de Feynman em regularização dimensional para o caso Lifshitz. No caso anisotrópico, as integrais são calculadas usando uma aproximação conhecida como a\-pro\-xi\-ma\-ção ortogonal \cite{leite2:2003}. Esta aproximação é utilizada no cálculo de integrais onde seus propagadores possuem momentos com potências quárticas e consiste em fazermos
 
\begin{equation}
[(k+K')^2]^2\approx[k^2+K'^2]^2=(k^2)^2+2k^2K'^2+(K'^2)^2 .
\end{equation}
 
Nessa aproximação, ao integrarmos nos momentos quárticos, usamos a fórmula \cite{messias:2010}

\begin{equation}
\int d^mk\frac{1}{[(k^2)^2 + 2ak^2 + m^2]^\beta}\cong S_m \frac{1}{4} \frac{\Gamma(m/4)\Gamma(\beta - m/4)}{\Gamma(\beta)}(m^2 - a^2)^{m/4-\beta} , \label{intlif1}
\end{equation} 
onde $S_m$ é a área da hiperesfera $m$-dimensional.

 
\subsection{Caso Anisotrópico Aproximado}

Lembrando que no caso anisotrópico a dimensão crítica é $d_c=4+\frac{m}{2}$ e o parâmetro de expansão $\varepsilon_L$ nesse caso é definido por $\varepsilon_L=4+\frac{m}{2}-d$ e iremos, nos resultados finais, desprezar os termos de $O(\varepsilon_L)$.

O primeiro diagrama associado a integral de Feynman na ordem de um {\it loop} que contribui para a função de dois pontos é dada por

\begin{eqnarray}
\begin{picture}(30,17) (212,-153)
    \SetWidth{0.9}
    \SetColor{Black}
    \Arc(227,-144)(7.071,262,622)
    \Line(213,-152)(241,-152)
    \Vertex(227,-151.5){1.5}
  \end{picture}=
  -\lambda_{\tau}\int d^{d-m}qd^mk\frac{1}{q^2+(k^2)^2+m_{\tau}^{2\tau}} .
\end{eqnarray}

Para resolvê-la, iremos primeiramente integrar em $q$ fazendo $p=0$, $\alpha=1$ e $m^2=(k^2)^2+m_{\tau}^{2\tau}$ na expressão (\ref{intamit}), que implica na expressão

\begin{eqnarray}
\begin{picture}(30,17) (212,-153)
    \SetWidth{0.9}
    \SetColor{Black}
    \Arc(227,-144)(7.071,262,622)
    \Line(213,-152)(241,-152)
    \Vertex(227,-151.5){1.5}
  \end{picture}=
  -\lambda_{\tau}S_{d-m}\frac{1}{2}\Gamma\left(\frac{d-m}{2}\right)\Gamma\left(1-\frac{d-m}{2}\right)\int d^mk\frac{1}{\left[(k^2)^2+m_{\tau}^{2\tau}\right]^{1-\frac{d-m}{2}}} .
\end{eqnarray}

Fazendo $a=0$ no resultado (\ref{intlif1}) para resolver a integral em $k$, e introduzindo a constante de acoplamento adimensional $g_{\tau}$ na forma

\begin{equation}
g_{\tau}\equiv \lambda_{\tau}\mu^{d-\frac{m}{2}-4}=\lambda_{\tau}\mu^{-\tau \varepsilon_L} , \label{relalif}
\end{equation}
podemos obter o resultado da integral em termos de $g_{\tau}$ e $\varepsilon_L$ dado por

\begin{eqnarray}
\begin{picture}(30,17) (212,-153)
    \SetWidth{0.9}
    \SetColor{Black}
    \Arc(227,-144)(7.071,262,622)
    \Line(213,-152)(241,-152)
    \Vertex(227,-151.5){1.5}
  \end{picture}=
-\frac{g_{\tau} \mu_{\tau}^{\tau \varepsilon_L}}{2}\left[\frac{1}{4}S_{d-m}S_m\Gamma\left(\frac{d-m}{2}\right)\Gamma(m/4)\right]\Gamma(1-d/2+m/4)(m_{\tau}^{2\tau})^{-1+d/2-m/4} ,
\end{eqnarray}
que através da relação $\frac{m}{4}-\frac{d}{2}=\frac{\varepsilon_L}{2}-2$, nos leva a

\begin{eqnarray}
\begin{picture}(30,17) (212,-153)
    \SetWidth{0.9}
    \SetColor{Black}
    \Arc(227,-144)(7.071,262,622)
    \Line(213,-152)(241,-152)
    \Vertex(227,-151.5){1.5}
  \end{picture}=
  -\frac{g_{\tau} \mu_{\tau}^{\tau \varepsilon_L} m_{\tau}^{2\tau}}{2} \left[\frac{1}{4} S_{d-m}S_m \Gamma\left(2-\frac{m}{4}-\frac{\varepsilon_L}{2}\right)\Gamma\left(\frac{m}{4}\right) \right] \Gamma\left(-1+\frac{\varepsilon_L}{2}\right) (m_{\tau}^{2\tau})^{-\varepsilon_L /2} .
  \end{eqnarray}

Expandindo a função Gama entre colchetes na forma

\begin{equation}
\Gamma(a+bx)=\Gamma(a)[1+bx\psi(a)+O(x^2)] , \label{expalif}
\end{equation}

Teremos

\begin{eqnarray}
\begin{picture}(30,17) (212,-153)
    \SetWidth{0.9}
    \SetColor{Black}
    \Arc(227,-144)(7.071,262,622)
    \Line(213,-152)(241,-152)
    \Vertex(227,-151.5){1.5}
  \end{picture}=
  -\frac{g_{\tau} m_{\tau}^{2\tau}}{2} \left[\frac{1}{4} S_{d-m}S_m\Gamma\left(2-\frac{m}{4}\right)\Gamma\left(\frac{m}{4}\right)\right] \left(1-\frac{\varepsilon_L}{2}\psi\left(2-\frac{m}{4}\right) \right) \Gamma \left(-1 + \frac{\varepsilon_L}{2} \right) \left(\frac{m_{\tau}^{2\tau}}{\mu_{\tau}^{2\tau}} \right)^{-\varepsilon_L /2} .
\end{eqnarray}

Agora absorvendo o fator angular

\begin{equation}
\frac{1}{4} S_{d-m}S_m\Gamma\left(2-\frac{m}{4}\right)\Gamma\left(\frac{m}{4}\right) , \label{fatorlif}
\end{equation}
em uma redefinição da constante de acoplamento, e usando as expansões em $\varepsilon_L$ de acordo com

\begin{equation}
\Gamma\left(-1+\frac{\varepsilon_L}{2}\right)=-\frac{2}{\varepsilon_L}\left[1+\frac{\varepsilon_L}{2}\psi(2) + O(\varepsilon_L^2)\right] ,
\end{equation}

\begin{equation}
\left(\frac{m_{\tau}^{2\tau}}{\mu_{\tau}^{2\tau}} \right)^{-\varepsilon_L /2}=1-\frac{\varepsilon_L}{2}ln\left(\frac{m_{\tau}^{2\tau}}{\mu_{\tau}^{2\tau}} \right) + O(\varepsilon_L^2) , \label{expalif2}
\end{equation}
obtemos 
 
\begin{eqnarray}
\begin{picture}(30,17) (212,-153)
    \SetWidth{0.9}
    \SetColor{Black}
    \Arc(227,-144)(7.071,262,622)
    \Line(213,-152)(241,-152)
    \Vertex(227,-151.5){1.5}
  \end{picture}=
  \frac{g_{\tau} m_{\tau}^{2\tau}}{\varepsilon_L} \left[ 1-\frac{\varepsilon_L}{2}\psi\left(2-\frac{m}{4}\right) \right] \left[1+\frac{\varepsilon_L}{2}\psi(2) + O(\varepsilon_L^2)\right] \left[1-\frac{\varepsilon_L}{2}ln\left(\frac{m_{\tau}^{2\tau}}{\mu_{\tau}^{2\tau}} \right) + O(\varepsilon_L^2) \right] .
\end{eqnarray} 

Fazendo $n=1$ na relação $\psi(n+1)-\psi(n)=1/n$ e definindo

\begin{equation}
[i_2]_m=1+\frac{1}{2}\left[\psi(1) - \psi\left(2-\frac{m}{4}\right)   \right] , \label{deflif}
\end{equation}

Obtemos o resultado final para o primeiro diagrama de Feynman na ordem de um {\it loop}

\begin{eqnarray}
\begin{picture}(30,17) (212,-153)
    \SetWidth{0.9}
    \SetColor{Black}
    \Arc(227,-144)(7.071,262,622)
    \Line(213,-152)(241,-152)
    \Vertex(227,-151.5){1.5}
  \end{picture}=
 \frac{g_{\tau} m_{\tau}^{2\tau}}{\varepsilon_L} \left[1+\left([i_2]_m - \frac{1}{2}\right)\varepsilon_L - \frac{\varepsilon_L}{2}ln\left(\frac{m_{\tau}^{2\tau}}{\mu_{\tau}^{2\tau}}\right)\right] . \label{intliftad}
\end{eqnarray}

Agora calcularemos o segundo diagrama associado a integral de Feynman na ordem de um {\it loop} que contribui para a função de quatro pontos, dado por

\begin{eqnarray}
\begin{picture}(30,24) (218,-161)
    \SetWidth{0.8}
    \SetColor{Black}
    \Arc(233,-157)(8,252,612)
    \Line(241.5,-157)(246,-151)
    \Line(241.5,-157)(246,-163)
    \Line(224.5,-157)(220,-151)
    \Line(224.5,-157)(220,-163)
    \Vertex(224.5,-157){1.5}
    \Vertex(241.5,-157){1.5}
  \end{picture}=
  \lambda_{\tau}^2 \int d^{d-m}qd^mk \frac{1}{[q^2 + (k^2)^2 + m_{\tau}^{2\tau}]\{(q+p)^2 + [(k+K')^2]^2+ m_{\tau}^{2\tau} \}} .
\end{eqnarray}

Usando parâmetros de Feynman, com $\alpha_1=\alpha_2=1$, $a_1=(q+p)^2+[(k+K')^2]^2 + m_{\tau}^{2\tau}$ e $a_2=q^2 + (k^2)^2 + m_{\tau}^{2\tau}$ na expressão (\ref{paramefey}), obtemos

\begin{eqnarray}
\begin{picture}(30,24) (218,-161)
    \SetWidth{0.8}
    \SetColor{Black}
    \Arc(233,-157)(8,252,612)
    \Line(241.5,-157)(246,-151)
    \Line(241.5,-157)(246,-163)
    \Line(224.5,-157)(220,-151)
    \Line(224.5,-157)(220,-163)
    \Vertex(224.5,-157){1.5}
    \Vertex(241.5,-157){1.5}
  \end{picture}=
  \lambda_{\tau}^2 \int_{0}^{1} dx \int d^mk \int d^{d-m}q \frac{1}{\{q^2 + 2qpx + p^2x +2k^2K'^2x + (K'^2)^2x + (k^2)^2 + m_{\tau}^{2\tau}  \}^2} .
\end{eqnarray}

Usando o resultado da expressão (\ref{intamit}), resolvemos a integral em $q$, e obtemos

\begin{eqnarray}
\begin{picture}(30,24) (218,-161)
    \SetWidth{0.8}
    \SetColor{Black}
    \Arc(233,-157)(8,252,612)
    \Line(241.5,-157)(246,-151)
    \Line(241.5,-157)(246,-163)
    \Line(224.5,-157)(220,-151)
    \Line(224.5,-157)(220,-163)
    \Vertex(224.5,-157){1.5}
    \Vertex(241.5,-157){1.5}
  \end{picture}&=&
  \lambda_{\tau}^2 \frac{1}{2} S_{d-m} \Gamma\left(\frac{d-m}{2}\right) \Gamma\left(2-\frac{d-m}{2}\right) \int_{0}^{1} dx \times \nonumber \\
 && \int d^mk \frac{1}{[(k^2)^2 + 2k^2K'^2x + p^2x(1-x) + (K'^2)^2x + m_{\tau}^{2\tau}]^{2-\frac{d-m}{2}}} . \label{intelifh}
\end{eqnarray}

Através do resultado (\ref{intlif1}) teremos

\begin{eqnarray}
\begin{picture}(30,24) (218,-161)
\SetWidth{0.8}
    \SetColor{Black}
    \Arc(233,-157)(8,252,612)
    \Line(241.5,-157)(246,-151)
    \Line(241.5,-157)(246,-163)
    \Line(224.5,-157)(220,-151)
    \Line(224.5,-157)(220,-163)
    \Vertex(224.5,-157){1.5}
    \Vertex(241.5,-157){1.5}
  \end{picture}&=&
\frac{\lambda_{\tau}^2}{2} \left[\frac{1}{4}S_{d-m}S_m \Gamma\left(\frac{d-m}{2}\right) \Gamma \left(\frac{m}{4}\right) \right]\Gamma \left(2-\frac{d}{2}+\frac{m}{4}\right)\times \nonumber \\
 &&\int_{0}^{1} dx \{[p^2+(K'^2)^2]x(1-x)+m_{\tau}^{2\tau}\}^{-2+d/2-m/4} . \label{intlif2}
\end{eqnarray} 

Considerando as relações 

\begin{eqnarray}
2-\frac{d}{2}+\frac{m}{4}&=&\frac{\varepsilon_L}{2} \label{relalif2} , \\
\Gamma\left(\frac{d-m}{2}\right)&=&\Gamma\left(2-\frac{m}{4}-\frac{\varepsilon_L}{2}\right) , \label{relalif3}
\end{eqnarray} 
podemos reescrever a expressão (\ref{intlif2}) como

\begin{eqnarray}
\begin{picture}(30,24) (218,-161)
\SetWidth{0.8}
    \SetColor{Black}
    \Arc(233,-157)(8,252,612)
    \Line(241.5,-157)(246,-151)
    \Line(241.5,-157)(246,-163)
    \Line(224.5,-157)(220,-151)
    \Line(224.5,-157)(220,-163)
    \Vertex(224.5,-157){1.5}
    \Vertex(241.5,-157){1.5}
  \end{picture}&=&
\frac{\lambda_{\tau}^2}{2} \left[\frac{1}{4}S_{d-m}S_m \Gamma\left(2-\frac{m}{4}-\frac{\varepsilon_L}{2}  \right) \Gamma \left(\frac{m}{4}\right) \right]\Gamma \left(\frac{\varepsilon_L}{2}\right)\times \nonumber \\
 &&\int_{0}^{1} dx \{[p^2+(K'^2)^2]x(1-x)+m_{\tau}^{2\tau}\}^{-\varepsilon_L /2} .
\end{eqnarray}

Considerando a expansão em $\varepsilon_L$

\begin{eqnarray}
\Gamma\left(\frac{\varepsilon_L}{2}\right)=\frac{2}{\varepsilon_L}+\psi(1)+O(\varepsilon_L) , \label{expalif3}
\end{eqnarray}
com a expansão (\ref{expalif}) dentro dos colchetes e a relação (\ref{relalif}), obtemos

\begin{eqnarray}
\begin{picture}(30,24) (218,-161)
\SetWidth{0.8}
    \SetColor{Black}
    \Arc(233,-157)(8,252,612)
    \Line(241.5,-157)(246,-151)
    \Line(241.5,-157)(246,-163)
    \Line(224.5,-157)(220,-151)
    \Line(224.5,-157)(220,-163)
    \Vertex(224.5,-157){1.5}
    \Vertex(241.5,-157){1.5}
  \end{picture}&=&
  \frac{g_{\tau}^2 \mu_{\tau}^{\tau \varepsilon_L}}{2} \left[ \frac{1}{4} S_{d-m}S_m\Gamma         \left(2-\frac{m}{4}\right)\Gamma \left(\frac{m}{4}\right) \right] \left[1-\frac{\varepsilon_L}{2}\psi\left(2-\frac{m}{4}\right) +O(\varepsilon_L^2) \right]\times \nonumber \\ 
 && \left[ \frac{2}{\varepsilon_L} +\psi(1) + O(\varepsilon_L)    \right] 
\int_{0}^{1}dx \left\{ \frac{[p^2 + (K'^2)^2]x(1-x)+m_{\tau}^{2\tau}}{\mu_{\tau}^{2\tau}} \right\}^{-\varepsilon_L /2} .
\end{eqnarray}

Absorvendo o fator geométrico (\ref{fatorlif}) em uma redefinição da constante de acoplamento e usando a definição (\ref{deflif}), obtemos uma expressão final dada por

\begin{eqnarray}
\begin{picture}(30,24) (218,-161)
\SetWidth{0.8}
    \SetColor{Black}
    \Arc(233,-157)(8,252,612)
    \Line(241.5,-157)(246,-151)
    \Line(241.5,-157)(246,-163)
    \Line(224.5,-157)(220,-151)
    \Line(224.5,-157)(220,-163)
    \Vertex(224.5,-157){1.5}
    \Vertex(241.5,-157){1.5}
  \end{picture}=
 g_{\tau} \mu_{\tau}^{\tau \varepsilon_L} \frac{g_{\tau} }{\varepsilon_L} \left\{1+ ([i_2]_m -1)\varepsilon_L - \frac{\varepsilon_L}{2} \int_{0}^{1}dx ln\left\{ \frac{[p^2+(K'^2)^2]x(1-x)+m_{\tau}^{2\tau})}{\mu_{\tau}^{2\tau}} \right\}    \right\} . \label{intlif5}
\end{eqnarray}

Podemos observar que o prefator $g_{\tau} \mu_{\tau}^{\tau \varepsilon_L}$ é a constante de acoplamento $\lambda_{\tau}$ que deve ser renormalizada, e portanto não é expandida em termos de $\varepsilon_L$. Somente a expressão atrás dele contribue para a constante de renormalização $Z_{g_{\tau}}$.

O terceiro diagrama associado a integral de Feynman na ordem de dois {\it loops} que contribui para função de dois pontos é dado por

\begin{eqnarray}
\begin{picture}(28,30) (204,-138)
    \SetWidth{0.8}
    \SetColor{Black}
    \Arc(218,-129)(6.083,261,621)
    \Line(205,-136)(231,-136)
    \Vertex(218,-135.5){1.5}
    \Arc(218,-116)(6.083,261,621)
    \Vertex(218,-122){1.5}
  \end{picture}=
  \lambda_{\tau}^2\int d^{d-m}q_1d^{d-m}q_2d^mk_1d^mk_2\frac{1}{[q_1^2+(k_1^2)^2+m_{\tau}^{2\tau}]^2 [q_2^2 + (k_2^2)^2 + m_{\tau}^{2\tau}]} , \label{diagfeyn}
\end{eqnarray}
que podemos reescrever na forma

\begin{eqnarray}
\begin{picture}(28,30) (204,-138)
    \SetWidth{0.8}
    \SetColor{Black}
    \Arc(218,-129)(6.083,261,621)
    \Line(205,-136)(231,-136)
    \Vertex(218,-135.5){1.5}
    \Arc(218,-116)(6.083,261,621)
    \Vertex(218,-122){1.5}
  \end{picture}=
  \left[\lambda_{\tau}\int d^{d-m}q_1d^mk_1\frac{1}{[q_1^2+(k_1^2)^2+m_{\tau}^{2\tau}]^2}   \right]   \left[\lambda_{\tau}\int d^{d-m}q_2d^mk_2\frac{1}{[q_2^2 + (k_2^2)^2 + m_{\tau}^{2\tau}]}  \right] .
\end{eqnarray}

Para resolver a integral do primeiro colchete, vamos usar o resultado (\ref{intamit}) para resolver a integral em $q_1$, onde teremos

\begin{eqnarray}
\lambda_{\tau}\int d^{d-m}q_1d^mk_1\frac{1}{[q_1^2+(k_1^2)^2+m_{\tau}^{2\tau}]^2} &=& \lambda_{\tau}S_{d-m}\frac{1}{2}\Gamma \left(\frac{d-m}{2}\right)\Gamma \left(2-\frac{d-m}{2}\right) \times \nonumber \\
&&\int d^mk_1 \frac{1}{[(k_1^2)^2 + m_{\tau}^{2\tau}]^{2-\frac{d-m}{2}}} .
\end{eqnarray}

Usando agora o resultado (\ref{intlif1}) para resolver a integral em $k_1$ e a relação (\ref{relalif}), obtemos 

\begin{eqnarray}
\lambda_{\tau}\int d^{d-m}q_1d^mk_1\frac{1}{[q_1^2+(k_1^2)^2+m_{\tau}^{2\tau}]^2} &=& \frac{g_{\tau} \mu_{\tau}^{\tau \varepsilon_L}}{2} \left[ \frac{1}{4}S_{d-m}S_m \Gamma  \left(\frac{d-m}{2}\right) \Gamma \left(\frac{m}{4}\right)    \right] \times \nonumber \\
&&\Gamma \left(2-\frac{d}{2} + \frac{m}{4}  \right) (m_{\tau}^{2\tau})^{-2+\frac{d}{2} -\frac{m}{4}} .
\end{eqnarray}

E através das relações (\ref{relalif2}), (\ref{relalif3}) e a expansão (\ref{expalif}) dentro do colchete, temos que

\begin{eqnarray}
\lambda_{\tau}\int d^{d-m}q_1d^mk_1\frac{1}{[q_1^2+(k_1^2)^2+m_{\tau}^{2\tau}]^2} &=& \frac{g_{\tau}}{2} \left[\frac{1}{4} S_{d-m}S_m \Gamma \left(2-\frac{m}{4}\right) \Gamma \left(\frac{m}{4}\right) \right] \left(1- \frac{\varepsilon_L}{2}\psi\left(2-\frac{m}{4}\right)      \right) \times \nonumber \\
&&\Gamma \left(\frac{\varepsilon_L}{2}\right) \left(\frac{m_{\tau}^{2\tau}}{\mu_{\tau}^{2\tau}} \right)^{-\varepsilon_L /2} .
\end{eqnarray}

E com os resultados das expansões em $\varepsilon_L$ (\ref{expalif2}), (\ref{expalif3}) e com o auxílio da definição (\ref{deflif}) juntamente com a absorção de (\ref{fatorlif}) na redefinição da constante de acoplamento, obtemos

\begin{eqnarray}
\lambda_{\tau}\int d^{d-m}q_1d^mk_1\frac{1}{[q_1^2+(k_1^2)^2+m_{\tau}^{2\tau}]^2}= \frac{g_{\tau}}{\varepsilon_L}\left[1-(1-[i_2]_m)\varepsilon_L - \frac{\varepsilon_L}{2}ln\left(\frac{m_{\tau}^{2\tau}}{\mu_{\tau}^{2\tau}}\right) \right] . \label{intlif3}
\end{eqnarray}

De acordo com a expressão (\ref{intliftad}), a integral do segundo colchete é dado por

\begin{eqnarray}
\lambda_{\tau}\int d^{d-m}q_2d^mk_2\frac{1}{[q_2^2 + (k_2^2)^2 + m_{\tau}^{2\tau}]}=-\frac{g_{\tau} m_{\tau}^{2\tau}}{\varepsilon_L} \left[1+\left([i_2]_m - \frac{1}{2}\right)\varepsilon_L - \frac{\varepsilon_L}{2}ln\left(\frac{m_{\tau}^{2\tau}}{\mu_{\tau}^{2\tau}}\right)\right] . \label{intlif4}
\end{eqnarray}

Com os resultados (\ref{intlif3}) e (\ref{intlif4}), podemos obter uma expressão final de (\ref{diagfeyn}) dado por

\begin{eqnarray}
\begin{picture}(28,30) (204,-138)
    \SetWidth{0.8}
    \SetColor{Black}
    \Arc(218,-129)(6.083,261,621)
    \Line(205,-136)(231,-136)
    \Vertex(218,-135.5){1.5}
    \Arc(218,-116)(6.083,261,621)
    \Vertex(218,-122){1.5}
  \end{picture}=
  -\frac{g_{\tau}^2 m_{\tau}^{2\tau}}{\varepsilon_L^2} \left[ 1+ \left(2[i_2]_m -\frac{3}{2}\right)\varepsilon_L - \varepsilon_L ln\left(\frac{m_{\tau}^{2\tau}}{\mu_{\tau}^{2\tau}}\right) \right] .
\end{eqnarray}

O quarto diagrama associado a integral de Feynman na ordem de dois {\it loops} que contribui para a função de quatro pontos é dado por

\begin{eqnarray}
\begin{picture}(44,16) (142,-138)
    \SetWidth{0.8}
    \SetColor{Black}
    \Arc(156,-134)(7.28,254,614)
    \Arc(171,-134)(7.28,254,614)
    \Vertex(163.5,-134){1.5}
    \Line(178,-134)(185,-128)
    \Line(178,-134)(185,-141)
    \Line(149,-134)(143,-128)
    \Line(149,-134)(143,-141)
    \Vertex(149,-134){1.5}
    \Vertex(178,-134){1.5}
  \end{picture}&=&
  -\lambda_{\tau}^3\int d^{d-m}q_1d^{d-m}q_2d^mk_1d^mk_2 \frac{1}{[q_1^2 +(k_1^2)^2 + m_{\tau}^{2\tau} ] [q_2^2 + (k_2^2)^2 + m_{\tau}^{2\tau} ]} \times \nonumber \\
 && \frac{1}{\{(q_1+p)^2 + [(k_1+K')^2]^2 + m_{\tau}^{2\tau}\} \{(q_2+p)^2 + [(k_2+K')^2]^2 +m_{\tau}^{2\tau} \}} .
\end{eqnarray}

Podemos reescrever a última integral como

\begin{eqnarray}
\begin{picture}(44,16) (142,-138)
    \SetWidth{0.8}
    \SetColor{Black}
    \Arc(156,-134)(7.28,254,614)
    \Arc(171,-134)(7.28,254,614)
    \Vertex(163.5,-134){1.5}
    \Line(178,-134)(185,-128)
    \Line(178,-134)(185,-141)
    \Line(149,-134)(143,-128)
    \Line(149,-134)(143,-141)
    \Vertex(149,-134){1.5}
    \Vertex(178,-134){1.5}
  \end{picture}&=&
-\frac{1}{\lambda_{\tau}} \left[\lambda_{\tau}^2 \int d^{d-m}q_1d^mk_1 \frac{1}{[q_1^2 + (k_1^2)^2 + m_{\tau}^{2\tau}]\{(q_1+p)^2 + [(k_1+K')^2]^2 + m_{\tau}^{2\tau}\} } \right] \times \nonumber \\
&&\left[\lambda_{\tau}^2 \int d^{d-m}q_2d^mk_2 \frac{1}{[q_2^2 + (k_2^2)^2 + m_{\tau}^{2\tau}]\{(q_2+p)^2 + [(k_2+K')^2]^2 + m_{\tau}^{2\tau}\} } \right] .
\end{eqnarray}

Usando a relação (\ref{relalif}) e o resultado (\ref{intlif5}) temos que

\begin{eqnarray}
\begin{picture}(44,16) (142,-138)
    \SetWidth{0.8}
    \SetColor{Black}
    \Arc(156,-134)(7.28,254,614)
    \Arc(171,-134)(7.28,254,614)
    \Vertex(163.5,-134){1.5}
    \Line(178,-134)(185,-128)
    \Line(178,-134)(185,-141)
    \Line(149,-134)(143,-128)
    \Line(149,-134)(143,-141)
    \Vertex(149,-134){1.5}
    \Vertex(178,-134){1.5}
  \end{picture}&=&
-\frac{1}{g_{\tau} \mu_{\tau}^{\tau \varepsilon_L}} \left\{ \frac{g_{\tau}^2 \mu_{\tau}^{\tau \varepsilon_L}}{\varepsilon_L} \left\{1+ ([i_2]_m -1)\varepsilon_L - \frac{\varepsilon_L}{2} \int_{0}^{1}dx ln\left\{ \frac{[p^2+(K'^2)^2]x(1-x)+m_{\tau}^{2\tau}}{\mu_{\tau}^{2\tau}} \right\}    \right\} \right\}\times \nonumber \\
&&\left\{ \frac{g_{\tau}^2 \mu_{\tau}^{\tau \varepsilon_L}}{\varepsilon_L} \left\{1+ ([i_2]_m -1)\varepsilon_L - \frac{\varepsilon_L}{2} \int_{0}^{1}dx ln\left\{ \frac{[p^2+(K'^2)^2]x(1-x)+m_{\tau}^{2\tau}}{\mu_{\tau}^{2\tau}} \right\}    \right\} \right\} .
\end{eqnarray}

Multiplicando os termos, obtemos a expressão final dada por

\begin{eqnarray}
\begin{picture}(44,16) (142,-138)
    \SetWidth{0.8}
    \SetColor{Black}
    \Arc(156,-134)(7.28,254,614)
    \Arc(171,-134)(7.28,254,614)
    \Vertex(163.5,-134){1.5}
    \Line(178,-134)(185,-128)
    \Line(178,-134)(185,-141)
    \Line(149,-134)(143,-128)
    \Line(149,-134)(143,-141)
    \Vertex(149,-134){1.5}
    \Vertex(178,-134){1.5}
  \end{picture}=
 -\frac{g_{\tau}^3 \mu_{\tau}^{\tau \varepsilon_L}}{\varepsilon_L^2} \left\{1+ 2([i_2]_m -1)\varepsilon_L -\varepsilon_L \int_{0}^{1}dx ln\left\{ \frac{[p^2+(K'^2)^2]x(1-x)+m_{\tau}^{2\tau}}{\mu_{\tau}^{2\tau}} \right\}    \right\} .
\end{eqnarray}

O quinto diagrama associado a integral de Feynman na ordem de dois {\it loops} que contribui para a função de quatro pontos é dado por

\begin{eqnarray}
\begin{picture}(30,24) (218,-161)
    \SetWidth{0.8}
    \SetColor{Black}
    \Arc(233,-157)(8,252,612)
    \Arc(233,-145)(4.472,117,477)
    \Line(241,-157)(246,-151)
    \Line(241,-157)(246,-163)
    \Line(224,-157)(220,-151)
    \Line(224,-157)(220,-163)
    \Vertex(233,-149){1.5}
    \Vertex(224.7,-157){1.5}
    \Vertex(241.5,-157){1.5}
  \end{picture}&=&
  -\lambda_{\tau}^3 \int d^{d-m}q_1d^{d-m}q_2d^mk_1d^mk_2 \frac{1}{[q_1^2 + (k_1^2)^2 + m_{\tau}^{2\tau}]^2 \left\{(q_1 + p)^2 + [(k_1 + K')^2]^2 + m_{\tau}^{2\tau} \right\} } \times \nonumber \\
&& \frac{1}{[q_2^2 + (k_2^2)^2 + m_{\tau}^{2\tau}]} ,
\end{eqnarray}
que pode ser reescrito na forma

\begin{eqnarray}
\begin{picture}(30,24) (218,-161)
    \SetWidth{0.8}
    \SetColor{Black}
    \Arc(233,-157)(8,252,612)
    \Arc(233,-145)(4.472,117,477)
    \Line(241,-157)(246,-151)
    \Line(241,-157)(246,-163)
    \Line(224,-157)(220,-151)
    \Line(224,-157)(220,-163)
    \Vertex(233,-149){1.5}
    \Vertex(224.7,-157){1.5}
    \Vertex(241.5,-157){1.5}
  \end{picture}&=&
  \left[\lambda_{\tau}^2 \int d^{d-m}q_1d^mk_1 \frac{1}{[q_1^2 + (k_1^2)^2 + m_{\tau}^{2\tau}]^2 \left\{(q_1 + p)^2 +[(k_1 + K')^2]^2 + m_{\tau}^{2\tau} \right\} } \right] \times \nonumber \\
 && \left[-\lambda_{\tau} \int d^{d-m}q_2d^mk_2 \frac{1}{q_2^2 + (k_2^2)^2 + m_{\tau}^{2\tau}}    \right] .
\end{eqnarray}

Vamos obter um resultado para a integral do primeiro colchete usando os parâmetros de Feynman, onde fazendo $\alpha_1=1$ e $\alpha_2=2$ com $a_1=\left\{(q_1+p)^2 + [(k_1+K')^2]^2 + m_{\tau}^{2\tau}\right\}$ e $a_2=[q_1^2 + (k_1^2)^2 + m_{\tau}^{2\tau}]$ na expressão (\ref{paramefey}), obtemos

\begin{eqnarray}
&&\int d^{d-m}q_1 \frac{1}{[q_1^2 + (k_1^2)^2 + m_{\tau}^{2\tau}]^2 \left\{(q_1 + p)^2 +[(k_1 + K')^2]^2 + m_{\tau}^{2\tau} \right\} } = \Gamma(3)\int_{0}^{1}dx(1-x)\times \nonumber \\
&&\int d^{d-m}q_1\frac{1}{\left\{q_1^2 + 2q_1px + p^2x + (k_1^2)^2 + 2k_1^2K'^2x + (K'^2)^2x + m_{\tau}^{2\tau} \right\}^3 } .
\end{eqnarray}

Usando o resultado (\ref{intamit}), resolvemos a integral em $q_1$, ou seja

\begin{eqnarray}
&&\int d^{d-m}q_1 \frac{1}{[q_1^2 + (k_1^2)^2 + m_{\tau}^{2\tau}]^2 \left\{(q_1 + p)^2 +[(k_1 + K')^2]^2 + m_{\tau}^{2\tau} \right\} } = \frac{1}{2}S_{d-m} \Gamma \left(\frac{d-m}{2}  \right)\times \nonumber \\
&& \Gamma \left(3-\frac{d-m}{2} \right)\int_{0}^{1}dx (1-x) \frac{1}{[(k_1^2)^2 + 2k_1^2K'^2x + p^2x(1-x)+(K'^2)^2x + m_{\tau}^{2\tau} ]^{3-\frac{d-m}{2}}} .
\end{eqnarray}

E através do resultado (\ref{intlif1}), resolvemos a integral em $k_1$. Usando a relação (\ref{relalif}) e absorvendo o fator geométrico (\ref{fatorlif}) na redefinição da constante de acoplamento, obtemos uma expressão final para a integral do primeiro colchete dado por

\begin{eqnarray}
&&\lambda_{\tau}^2 \int d^{d-m}q_1d^mk_1 \frac{1}{[q_1^2 + (k_1^2)^2 + m_{\tau}^{2\tau}]^2 \left\{(q_1 + p)^2 +[(k_1 + K')^2]^2 + m_{\tau}^{2\tau} \right\} } = \frac{g_{\tau}^2 \mu_{\tau}^{2\tau \varepsilon_L}}{2} \times \nonumber \\
&&\left( 1- \frac{\varepsilon_L}{2}\psi\left(2-\frac{m}{4}\right) \right)\Gamma \left(1+\frac{\varepsilon_L}{2}\right) \int_{0}^{1}dx\frac{(1-x)}{\left\{[p^2+(K'^2)^2]x(1-x)+m_{\tau}^{2\tau} \right\}}[1+O(\varepsilon_L)] .
\end{eqnarray}

Após calcularmos o resultado do primeiro colchete, podemos observar que o resultado da integral do segundo colchete é dado pela expressão (\ref{intliftad}), onde podemos chegar ao resultado final do diagrama dado por

\begin{eqnarray}
\begin{picture}(30,24) (218,-161)
    \SetWidth{0.8}
    \SetColor{Black}
    \Arc(233,-157)(8,252,612)
    \Arc(233,-145)(4.472,117,477)
    \Line(241,-157)(246,-151)
    \Line(241,-157)(246,-163)
    \Line(224,-157)(220,-151)
    \Line(224,-157)(220,-163)
    \Vertex(233,-149){1.5}
    \Vertex(224.7,-157){1.5}
    \Vertex(241.5,-157){1.5}
  \end{picture}=
  \frac{g_{\tau}^3 \mu_{\tau}^{\tau \varepsilon_L}}{2\varepsilon_L} \int_{0}^{1} dx \frac{m_{\tau}^{2\tau}(1-x)}{ \left\{ [p^2+(K'^2)^2]x(1-x) + m_{\tau}^{2\tau} \right\}} .
\end{eqnarray}

Calcularemos agora o sexto diagrama associado a integral de Feynman na ordem de dois {\it loops} que contribui para a função de quatro pontos que é dado por

\begin{eqnarray}
\begin{picture}(39,29) (156,-192)
    \SetWidth{0.8}
    \SetColor{Black}
    \Arc(179,-186)(11,270,630)
    \Line(168,-186)(157,-175)
    \Line(168,-186)(157,-197)
    \Arc(193.5,-186.5)(13.509,87.879,272.121)
    \Vertex(168,-186){1.5}
    \Vertex(184,-177){1.5}
    \Vertex(184,-196){1.5}
  \end{picture}&=&
  -\lambda_{\tau}^3 \int d^{d-m}q_1d^{d-m}q_2d^mk_1d^mk_2 \frac{1}{[q_1^2 + (k_1^2)^2 + m_{\tau}^{2\tau}][q_2^2 + (k_2^2)^2 + m_{\tau}^{2\tau}]} \times \nonumber \\
 &&\frac{1}{\left\{(q_1-q_2+p)^2 + [(k_1-k_2+K')^2]^2 + m_{\tau}^{2\tau}   \right\} }\times \nonumber \\
&& \frac{1}{\left\{(p-q_1)^2 + [(K'-k_1)^2]^2 + m_{\tau}^{2\tau} \right\}} .
\end{eqnarray}
que podemos reescrever na forma

\begin{eqnarray}
\begin{picture}(39,29) (156,-192)
    \SetWidth{0.8}
    \SetColor{Black}
    \Arc(179,-186)(11,270,630)
    \Line(168,-186)(157,-175)
    \Line(168,-186)(157,-197)
    \Arc(193.5,-186.5)(13.509,87.879,272.121)
    \Vertex(168,-186){1.5}
    \Vertex(184,-177){1.5}
    \Vertex(184,-196){1.5}
  \end{picture}&=&
  \left[-\lambda_{\tau} \int d^{d-m}q_1d^mk_1\frac{1}{[q_1^2 + (k_1^2)^2 + m_{\tau}^{2\tau}] \left\{(p-q_1)^2 + [(K'-k_1)^2]^2 + m_{\tau}^{2\tau} \right\} } \right]\times \nonumber \\
&& \Biggl[ \lambda_{\tau}^2 \int d^{d-m}q_2d^mk_2 \frac{1}{\left\{(q_1-q_2+p)^2 + [(k_1-k_2+K')^2]^2 + m_{\tau}^{2\tau} \right\} } \times \nonumber \\
&&  \frac{1}{[q_2^2 + (k_2^2)^2 + m_{\tau}^{2\tau}]} \Biggr] . \label{intlif6}
\end{eqnarray}
 
 Temos que através dos parâmetros de Feynman e usando os resultados (\ref{intamit}) e (\ref{intlif1}), com as relações (\ref{relalif}), (\ref{expalif}), (\ref{relalif2}) e (\ref{relalif3}), podemos colocar a integral do segundo colchete na forma
 
\begin{eqnarray}
&&\lambda_{\tau}^2 \int d^{d-m}q_2d^mk_2 \frac{1}{[q_2^2 + (k_2^2)^2 + m_{\tau}^{2\tau}] \left\{(q_1-q_2+p)^2 + [(k_1-k_2+K')^2]^2 + m_{\tau}^{2\tau} \right\} } = \frac{g_{\tau}^2 \mu_{\tau}^{\tau \varepsilon_L}}{\varepsilon_L} \times \nonumber \\
&&[1+([i_2]_m-1)\varepsilon_L] \int_{0}^{1}dx \left\{\frac{\mu_{\tau}^{2\tau}}{[(p+q_1)^2+(K'^2+k_1^2)^2]x(1-x) + m_{\tau}^{2\tau}}\right\}^{\varepsilon_L /2} . \label{intlif7}
\end{eqnarray}
 
Substituindo o resultado (\ref{intlif7}) na expressão (\ref{intlif6}), temos que

\begin{eqnarray}
\begin{picture}(39,29) (156,-192)
    \SetWidth{0.8}
    \SetColor{Black}
    \Arc(179,-186)(11,270,630)
    \Line(168,-186)(157,-175)
    \Line(168,-186)(157,-197)
    \Arc(193.5,-186.5)(13.509,87.879,272.121)
    \Vertex(168,-186){1.5}
    \Vertex(184,-177){1.5}
    \Vertex(184,-196){1.5}
  \end{picture}&=&
  -\frac{g_{\tau}^{3} \mu_{\tau}^{3\tau \varepsilon_L}}{\varepsilon_L} [1+ ([i_2]_m-1)\varepsilon_L] \int_{0}^{1}dx[x(1-x)]^{-\varepsilon_L /2} I(p,K') , \label{intlif8}
\end{eqnarray}
 
onde

\begin{eqnarray}
I(q_1,k_1)&=&\int d^{d-m}q_1d^mk_1\frac{1}{[q_1^2 + (k_1^2)^2 + m_{\tau}^{2\tau}] \left\{(p-q_1)^2 + [(K'-k_1)^2]^2 + m_{\tau}^{2\tau} \right\} } \times \nonumber \\
 &&\left\{ \frac{1}{[(p+q_1)^2 + (K'^2 + k_1^2)^2] + \frac{m_{\tau}^{2\tau}}{x(1-x)}} \right\}^{\varepsilon_L /2} .
\end{eqnarray} 
 
Para resolver a integral $I(p,K')$, iremos usar os parâmetros de Feynman para obter a relação

\begin{eqnarray}
I(p,K')&=&\int d^{d-m}q_1d^mk_1 \int_{0}^{1}dz \frac{1}{ \left\{ \left\{(p-q_1)^2 + [(K'-k_1)^2]^2 + m_{\tau}^{2\tau} \right\}z + [q_1^2 + (k_1^2)^2 + m_{\tau}^{2\tau}](1-z) \right\}^2 }\times \nonumber \\
&&\frac{1}{\left[(p+q_1)^2 + (K'^2+k_1^2)^2 + \frac{m_{\tau}^{2\tau}}{x(1-x)}\right]^{\varepsilon_L /2}} .
\end{eqnarray}
 
Usando novamente os parâmetros de Feynman e usando o resultado (\ref{intamit}) para resolver a integral em $q_1$, podemos obter

\begin{eqnarray}
I(p,K')&=&\frac{1}{2}S_{d-m} \frac{\Gamma \left(\frac{d-m}{2}\right) \Gamma \left(2+\frac{\varepsilon_L -d +m}{2}\right)}{\Gamma(\varepsilon_L /2)} \int_{0}^{1}dyy(1-y)^{\varepsilon_L /2 - 1} \int_{0}^{1}dz \int d^mk_1 \Biggl[p^2[zy+(1-y)] +  \nonumber \\ &&(k_1^2)^2y-2K'^2k_1^2zy + (K'^2)^2zy + (K'^2 + k_1^2)^2(1-y) + m_{\tau}^{2\tau}\left[y + \frac{1-y}{x(1-x)}  \right] - \nonumber \\
&&[pzy-(1-y)p]^2 \Biggr]^{\frac{d-m}{2} -2-\varepsilon_L /2} .
\end{eqnarray}

Podemos observar que o único termo de pólo vem do ponto de singularidade na integral em $y=1$, quando tem-se $\varepsilon_L \rightarrow 0$. Neste caso, podemos fazer $y=1$ no colchete e usar o resultado (\ref{intlif1}) para resolver a integral em $k_1$, e assim obter uma resultado final para $I(p,K')$ dado por

\begin{eqnarray}
I(p,K')=\frac{1}{2} \frac{\Gamma(\varepsilon_L)}{\Gamma(\varepsilon_L /2)} \left[\frac{1}{4} S_{d-m}S_m \Gamma \left(\frac{d-m}{2}\right) \Gamma \left(\frac{m}{4}\right) \right] \int_{0}^{1}dy y(1-y)^{\frac{\varepsilon_L}{2} -1}\times \nonumber \\
 \int_{0}^{1}dz \left\{[p^2 + (K'^2)^2]z(1-z)+m_{\tau}^{2\tau}  \right\}^{-\varepsilon_L} . \label{intlif9}
\end{eqnarray}
 
Substituindo o resultado (\ref{intlif9}) na expressão (\ref{intlif8}) e usando a expansão (\ref{expalif}), obtemos

\begin{eqnarray}
\begin{picture}(39,29) (156,-192)
    \SetWidth{0.8}
    \SetColor{Black}
    \Arc(179,-186)(11,270,630)
    \Line(168,-186)(157,-175)
    \Line(168,-186)(157,-197)
    \Arc(193.5,-186.5)(13.509,87.879,272.121)
    \Vertex(168,-186){1.5}
    \Vertex(184,-177){1.5}
    \Vertex(184,-196){1.5}
  \end{picture}&=&
  -\frac{g_{\tau}^3 \mu_{\tau}^{\tau \varepsilon_L}}{2\varepsilon_L} \frac{\Gamma(\varepsilon_L)}{\Gamma(\varepsilon_L /2)} \left[\frac{1}{4} S_{d-m}S_m \Gamma \left(2 - \frac{m}{4} \right)\Gamma \left(\frac{m}{4} \right) \right] \left(1-\frac{\varepsilon_L}{2} \psi \left(2-\frac{m}{4}\right)  \right)\times \nonumber \\
 && [1+([i_2]_m-1)\varepsilon_L ] \int_{0}^{1}dx[x(1-x)]^{-\varepsilon_L /2} \int_{0}^{1}dy y(1-y)^{\frac{\varepsilon_L}{2}-1} (\mu_{\tau}^{2\tau})^{\varepsilon_L} \times \nonumber \\
&& \int_{0}^{1}dz\left\{[p^2 + (K'^2)^2]z(1-z)+m_{\tau}^{2\tau}  \right\}^{-\varepsilon_L} .
\end{eqnarray}
 
Absorvendo o fator (\ref{fatorlif}) na redefinição da constante de acoplamento e usando as relações
 
\begin{equation}
\left\{[p^2 + (K'^2)^2]z(1-z)+m_{\tau}^{2 \tau} \right\}^{-\varepsilon_L}=1-\varepsilon_L ln\left\{[p^2 + (K'^2)^2]z(1-z)+m_{\tau}^{2 \tau}  \right\} ,
\end{equation}

\begin{equation}
\int_{0}^{1}dx[x(1-x)]^{-\varepsilon_L /2} = \frac{\Gamma \left(1-\frac{\varepsilon_L}{2}\right) \Gamma \left(1-\frac{\varepsilon_L}{2}\right)}{\Gamma(2-\varepsilon_L)} ,
\end{equation}

\begin{equation}
\int_{0}^{1}dy y(1-y)^{\frac{\varepsilon_L}{2}-1}=\frac{\Gamma(\varepsilon_L /2)}{\Gamma(2+\varepsilon_L /2)} ,
\end{equation}

\begin{equation}
(\mu_{\tau}^{2\tau})^{\varepsilon_L}=1 + \varepsilon_L ln(\mu_{\tau}^{2\tau}) ,
\end{equation}

\begin{equation}
\frac{\Gamma(\varepsilon_L)}{\Gamma(\varepsilon_L /2)}=\frac{1}{2}(1+\varepsilon_L \psi(1))\left(1+\frac{\varepsilon_L}{2}\psi(1)\right)^{-1} ,
\end{equation}
 
 podemos obter um resultado final para o sexto diagrama de Feynman dado por

\begin{eqnarray}
\begin{picture}(39,29) (156,-192)
    \SetWidth{0.8}
    \SetColor{Black}
    \Arc(179,-186)(11,270,630)
    \Line(168,-186)(157,-175)
    \Line(168,-186)(157,-197)
    \Arc(193.5,-186.5)(13.509,87.879,272.121)
    \Vertex(168,-186){1.5}
    \Vertex(184,-177){1.5}
    \Vertex(184,-196){1.5}
  \end{picture}=
  -\frac{g_{\tau}^3 \mu_{\tau}^{\tau \varepsilon_L}}{2\varepsilon_L^2} \left\{1+ \left(2[i_2]_m - \frac{3}{2} \right)\varepsilon_L - \varepsilon_L \int_{0}^{1}dz \left\{ \frac{[p^2 + (K'^2)^2]z(1-z)+ m_{\tau}^{2 \tau}}{\mu_{\tau}^{2 \tau}} \right\} \right\} .
\end{eqnarray}
 
Iremos calcular agora o sétimo diagrama de Feynman na ordem de dois {\it loops} que contribui para a função de dois pontos, conhecido como {\it sunset}, que está associado a integral

\begin{eqnarray}
\begin{picture}(29,18) (152,-178)
    \SetWidth{1.0}
    \SetColor{Black}
    \Arc(166,-173)(8.246,256,616)
    \Line(152,-173)(180,-173)
    \Vertex(158,-173){1.5}
    \Vertex(174,-173){1.5}
  \end{picture}&=&
  \lambda_{\tau}^{2} \int d^{d-m}q_1 d^{d-m}q_2 d^mk_1 d^mk_2 \frac{1}{\left[q_1^2 + (k_1^2)^2 +m_{\tau}^{2\tau}\right]\left[q_2^2 + (k_2^2)^2 + m_{\tau}^{2\tau}\right]}\times \nonumber \\
&&\frac{1}{\left\{(q_1 + q_2 + p)^2 + [(k_1 + k_2 + K')^2]^2 +m_{\tau}^{2\tau} \right\} } . \label{intelsunset}
\end{eqnarray}
 
O cálculo desse diagrama é bastante difícil devido a divergências na integral paramêtrica. Nesse caso, iremos usar integrais parciais, conhecido como método da "parcial p" desenvolvido por 't Hooft e Veltman \cite{kleinert:2000}, onde inserimos no integrando o operador diferencial unitário

\begin{eqnarray}
\frac{1}{2\left(d-\frac{m}{2}\right)} \left[\frac{\partial q_1^{\mu}}{\partial q_1^{\mu}} + \frac{\partial q_2^{\mu}}{\partial q_2^{\mu}} + \frac{1}{2} \frac{\partial k_1^{\mu}}{\partial k_1^{\mu}}  + \frac{1}{2} \frac{\partial k_2^{\mu}}{\partial k_2^{\mu}} \right]=1 .
\end{eqnarray} 

Inserindo essa identidade na expressão (\ref{intelsunset}) e usando a integração parcial em que o termo de superfície é desconsiderado, nós podemos colocar na forma
 
\begin{eqnarray}
\begin{picture}(29,18) (152,-178)
    \SetWidth{1.0}
    \SetColor{Black}
    \Arc(166,-173)(8.246,256,616)
    \Line(152,-173)(180,-173)
    \Vertex(158,-173){1.5}
    \Vertex(174,-173){1.5}
  \end{picture}=
  -\frac{\lambda_{\tau}^2}{\left[\left(d-\frac{m}{2}\right)-3\right]} \left[3m_{\tau}^{2\tau} A_{\tau}(p,K') + B_{\tau}(p,K')  \right] , \label{sunsetlif}
\end{eqnarray}

onde

\begin{eqnarray}
A_{\tau}(p,K')&=&\int d^{d-m}q_1 d^{d-m}q_2 d^mk_1 d^mk_2 \frac{1}{\left[q_1^2 + (k_1^2)^2 +m_{\tau}^{2\tau}\right]\left[q_2^2 + (k_2^2)^2 + m_{\tau}^{2\tau}\right]}\times \nonumber \\
&&\frac{1}{\left\{(q_1 + q_2 + p)^2 + [(k_1 + k_2 + K')^2]^2 +m_{\tau}^{2\tau} \right\}^2 } ,\\
B_{\tau}(p,K')&=&\int d^{d-m}q_1 d^{d-m}q_2 d^mk_1 d^mk_2 \frac{p(q_1 + q_2 + p) + K'(k_1 + k_2 + K')(k_1 + k_2 + K')^2}{\left[q_1^2 + (k_1^2)^2 +m_{\tau}^{2\tau}\right]\left[q_2^2 + (k_2^2)^2 + m_{\tau}^{2\tau}\right]}\times \nonumber \\
&&\frac{1}{\left\{(q_1 + q_2 + p)^2 + [(k_1 + k_2 + K')^2]^2 +m_{\tau}^{2\tau} \right\}^2 } .
\end{eqnarray}

Vamos primeiro calcular a integral de $A_{\tau}(p,K')$ fazendo as mudanças de variáveis $-q=q_1 + q_2 +p $ e $-k=k_1 + k_2 + K' $, ou seja,

\begin{eqnarray}
A_{\tau}(p,K')&=&\left[\int d^{d-m}qd^mk \frac{1}{[q^2 + (k^2)^2 + m_{\tau}^{2\tau}]^2} \right] \Biggl[ \int d^{d-m}q_1 d^mk_1 \frac{1}{[q_1^2 + (k_1^2)^2 + m_{\tau}^{2\tau}]} \times \nonumber \\
&& \frac{1}{ \left\{(q+q_1+p)^2 + [(k+k_1+K')^2]^2 + m_{\tau}^{2\tau} \right\} }  \Biggr] . \label{intelif10}
\end{eqnarray}

Usando o resultado da integral na expressão (\ref{intelifh}), podemos resolver a integral em $q_1$  e através do resultado (\ref{intlif1}) podemos resolver a integral em $k_1$ no segundo colchete. Absorvendo o fator geométrico (\ref{fatorlif}) na redefinição da constante de acoplamento, obtemos uma expressão final para a integral em $q_1$, dada por

\begin{eqnarray}
&& \int d^{d-m}q_1 d^mk_1 \frac{1}{[q_1^2 + (k_1^2)^2 + m_{\tau}^{2\tau}] \left\{(q+q_1+p)^2 + [(k+k_1+K')^2]^2 + m_{\tau}^{2\tau} \right\}} =  \frac{1}{2} \Gamma(\varepsilon_L /2) \times \nonumber \\
&& \left(1-\frac{\varepsilon_L}{2}\psi \left(2-\frac{m}{4}\right) \right) \int_{0}^{1} dx \left\{ \left\{(q+p)^2 + [(k+K')^2]^2 \right\} x(1-x) + m_{\tau}^{2\tau} \right\}^{-\varepsilon_L /2} . \label{intelif11}
\end{eqnarray}

Substituindo o resultado (\ref{intelif11}) na expressão (\ref{intelif10}), e usando os parâmetros de Feynman, obtemos

\begin{eqnarray}
&& A_{\tau}(p,K')=\frac{1}{2} \left(1-\frac{\varepsilon_L}{2}\psi \left(2-\frac{m}{4}\right) \right) \Gamma \left(2+\frac{\varepsilon_L}{2} \right) \int_{0}^{1}dx[x(1-x)]^{-\varepsilon_L /2} \int_{0}^{1}dy y(1-y)^{\frac{\varepsilon_L}{2}-1} \times \nonumber \\
&&\int d^{d-m}q d^mk \frac{1}{ \left\{ [q^2 + (k^2)^2 + m_{\tau}^{2\tau}]y + \left\{ (q+p)^2 + [(k+K')^2]^2 + \frac{m_{\tau}^{2\tau}}{x(1-x)} \right\}(1-y)\right\}^{2+\frac{\varepsilon_L}{2}} } .
\end{eqnarray}

Podemos observar que somente a integração em $y$ é singular para $\varepsilon_L \rightarrow 0$. A singularidade vem do ponto onde $y=1$, onde podemos fazer essa condição no denominador da integral em $q$ e $k$ e resolvê-la usando os resultados (\ref{intamit}) e (\ref{intlif1}) com o auxílio das relações

\begin{eqnarray}
\int_{0}^{1}dx[x(1-x)]^{-\varepsilon_L /2}=(1-\varepsilon_L)^{-1} ,\\
\int_{0}^{1} y(1-y)^{\frac{\varepsilon_L}{2}-1}=\frac{2}{\varepsilon_L}\left(1+\frac{\varepsilon_L}{2}\right)^{-1} ,
\end{eqnarray} 

E assim obter uma expressão final para $A_{\tau}(p,K')$ dado por

\begin{eqnarray}
A_{\tau}(p,K')=\frac{1}{2\varepsilon_L^2}\left[1+\left(2[i_2]_m-\frac{3}{2} \right) \varepsilon_L \right](m_{\tau}^{2\tau})^{-\varepsilon_L} . \label{apk}
\end{eqnarray}

Há muitas maneiras de se calcular $B_{\tau}(p,K')$, a maioria delas envolve expressões complicadas. A maneira mais viável é usando o fato que $B_{\tau}(p,K')$ pode ser reescrito na forma

\begin{eqnarray}
B_{\tau}(p,K')&=&-\left[\frac{p}{2} \frac{\partial}{\partial p} + \frac{K'}{4} \frac{\partial}{\partial K'}  \right] \int d^{d-m}q_1 d^{d-m}q_2 d^mk_1 d^mk_2 \frac{1}{[q_1^2 + (k_1^2)^2 + m_{\tau}^{2\tau}][q_2^2 + (k_2^2)^2 + m_{\tau}^{2\tau}]} \times \nonumber \\
&& \frac{1}{\left\{(q_1 + q_2 + p)^2 + [(k_1 + k_2 + K')^2]^2 + m_{\tau}^{2\tau}  \right\}} .
\end{eqnarray}

Através do resultado (\ref{intelif11}) obtemos uma expressão para a integral em $q_1$ e $k_1$, e usando os parâmetros de Feynman obtemos

\begin{eqnarray}
B_{\tau}(p,K')=-\frac{1}{2}\left(1-\frac{\varepsilon_L}{2}\psi \left(2-\frac{m}{4}\right)\right) \Gamma \left(1+\frac{\varepsilon_L}{2}\right) \left[\frac{p}{2} \frac{\partial}{\partial p} + \frac{K'}{4} \frac{\partial}{\partial K'}  \right] \times \nonumber \\
 \int_{0}^{1}dx[x(1-x)]^{- \varepsilon_L /2} \int_{0}^{1}dy y^{\frac{\varepsilon_L}{2}-1} \times \nonumber \\
 \int d^{d-m}q_2d^mk_2 \frac{1}{\left\{[q_2^2 + (k_2^2)^2 + m_{\tau}^{2\tau}](1-y) + (q_2+p)^2y + [(k_2 + K')^2]^2 + \frac{m_ {\tau}^{2\tau}}{x(1-x)}y\right\}^{1+\frac{\varepsilon_L}{2}}} . \label{intelif12}
\end{eqnarray}

Usando os resultados (\ref{intamit}) e (\ref{intlif1}) e absorvendo o fator geométrico (\ref{fatorlif}) na redefinição da constante de acoplamento, podemos obter uma expressão para a integral de $q_2$ e $k_2$ dada por

\begin{eqnarray}
&&\int d^{d-m}q_2d^mk_2 \frac{1}{\left\{[q_2^2 + (k_2^2)^2 + m_{\tau}^{2\tau}](1-y) + (q_2+p)^2y + [(k_2 + K')^2]^2 + \frac{m_ {\tau}^{2\tau}}{x(1-x)}y\right\}^{1+\frac{\varepsilon_L}{2}}}=\frac{1}{2}\times \nonumber \\
&& \frac{\Gamma(-1+\varepsilon_L)}{\Gamma \left(1+\frac{\varepsilon_L}{2}\right)} \left(1-\frac{\varepsilon_L}{2}\psi \left(2-\frac{m}{4}\right) \right) \left\{[p^2 + (K'^2)^2]y(1-y) + m_{\tau}^{2\tau}\left[(1-y) + \frac{y}{x(1-x)}  \right]   \right\}^{1-\varepsilon_L} \label{intelif13}
\end{eqnarray}

Substituindo o resultado (\ref{intelif13}) na expressão (\ref{intelif12}) e usando as relações 

\begin{equation}
[x(1-x)]^{-\frac{\varepsilon_L}{2}}=1-\frac{\varepsilon_L}{2}lnx(1-x) ,
\end{equation}

\begin{equation}
y^{\frac{\varepsilon_L}{2}}=1+\frac{\varepsilon_L}{2}lny ,
\end{equation}

\begin{equation}
\int_{0}^{1} dx[x(1-x)]^{-\frac{\varepsilon_L}{2}}=1+\varepsilon_L ,
\end{equation}

\begin{equation}
\int_{0}^{1}dy y^{\frac{\varepsilon_L}{2}}(1-y)=\frac{1}{2}\left(1-\frac{3}{4}\varepsilon_L \right) ,
\end{equation}

podemos obter uma expressão final para $B_{\tau}(p,K')$ dada por

\begin{eqnarray}
 B_{\tau}(p,K')&=&\frac{1}{\mu_{\tau}^{2\tau \varepsilon_L}} \frac{[p^2 + (K'^2)^2]}{8\varepsilon_L} \Biggl\{1+ \left(2[i_2]_m - \frac{7}{4}\right)\varepsilon_L -2\varepsilon_L \int_{0}^{1}dxdy(1-y) \times \nonumber \\
 && ln \Biggl\{ \frac{[p^2 + (K'^2)^2]y(1-y)}{\mu_{\tau}^{2\tau}} + \frac{m_{\tau}^{2\tau}}{\mu_{\tau}^{2\tau}}\left[(1-y) + \frac{y}{x(1-x)}  \right]  \Biggr\}   \Biggr\} . \label{bpk}
\end{eqnarray}

Substituindo os resultados (\ref{apk}) e (\ref{bpk}) na expressão (\ref{sunsetlif}), obtemos o resultado final para o sétimo diagrama de Feynman, cujo resultado é escrito como

\begin{eqnarray}
\begin{picture}(29,18) (152,-178)
    \SetWidth{1.0}
    \SetColor{Black}
    \Arc(166,-173)(8.246,256,616)
    \Line(152,-173)(180,-173)
    \Vertex(158,-173){1.5}
    \Vertex(174,-173){1.5}
  \end{picture}&=&
  -g_{\tau}^2 \Bigg\{ \frac{3m_{\tau}^{2\tau}}{2\varepsilon_L^2} \left[1+\left(2[i_2]_m - \frac{1}{2}\right)\varepsilon_L - \varepsilon_L ln \left(\frac{m_{\tau}^{2\tau}}{\mu_{\tau}^{2\tau}}\right)    \right] + \frac{p^2 + (K'^2)^2}{8\varepsilon_L} \Biggl\{1+ \left(2[i_2]_m-\frac{3}{4}\right)\varepsilon_L - \nonumber \\
&&  2\varepsilon_L \int_{0}^{1}dxdy(1-y) ln \Biggl\{ \frac{[p^2 + (K'^2)^2]y(1-y)}{\mu_{\tau}^{2\tau}} + \frac{m_{\tau}^{2\tau}}{\mu_{\tau}^{2\tau}} \Biggl[(1-y) + \frac{y}{x(1-x)} \Biggr] \Biggr\} \Biggr\}  \Biggr\} .
\end{eqnarray}

Agora iremos calcular o oitavo diagrama de Feynman na ordem de três {\it loops} que contribui para a função de dois pontos, que é associado à integral

\begin{eqnarray}
\begin{picture}(30,22) (427,-96)
    \SetWidth{1.0}
    \SetColor{Black}
    \Arc(442,-91)(10.05,276,636)
    \Arc(429.125,-77.75)(13.558,-77.758,-18.268)
    \Arc[clock](454,-79)(12.748,-101.31,-168.69)
    \Vertex(442,-81){1.5}
    \Vertex(432,-91){1.5}
    \Vertex(452,-91){1.5}
    \Line(452,-91)(457,-91)
    \Line(432,-91)(427,-91)
  \end{picture}&=&
  -\lambda_{\tau}^3 \int d^{d-m}q_1 d^{d-m}q_2 d^{d-m}q_3 d^mk_1 d^mk_2 d^mk_3 \frac{1}{[q_1^2+ (k_1^2)^2 + m_{\tau}^{2\tau}][q_2^2 +(k_2^2)^2 + m_{\tau}^{2\tau}]} \times \nonumber \\
&& \frac{1}{[q_3^2 +(k_3^2)^2 + m_{\tau}^{2\tau}]\left\{(q_1 + q_2 + p)^2 + [(k_1 + k_2 + K')^2]^2 + m_{\tau}^{2\tau} \right\}}  \times \nonumber \\
&& \frac{1}{\left\{(q_1 + q_3 + p)^2 + [(k_1 + k_3 + K')^2]^2 + m_{\tau}^{2\tau} \right\}} . \label{intelif14}
\end{eqnarray} 

O cálculo desse diagrama também é bastante difícil devido a divergências na integral paramêtrica. Nesse caso, iremos novamente usar o método da "parcial p" através da identidade

\begin{eqnarray}
\frac{1}{3(d-\frac{m}{2})} \left[\frac{\partial q_1^{\mu}}{\partial q_1^{\mu}} + \frac{\partial q_2^{\mu}}{\partial q_2^{\mu}} + \frac{\partial q_3^{\mu}}{\partial q_3^{\mu}} +\frac{1}{2}\frac{\partial k_1^{\mu}}{\partial k_1^{\mu}}+ \frac{1}{2}\frac{\partial k_2^{\mu}}{\partial k_2^{\mu}} +\frac{1}{2}\frac{\partial k_3^{\mu}}{\partial k_3^{\mu}}   \right] =1 . \label{ident2}
\end{eqnarray}

Usando a identidade (\ref{ident2}) na integral (\ref{intelif14}), podemos colocá-la na forma

\begin{eqnarray}
\begin{picture}(30,22) (427,-96)
    \SetWidth{1.0}
    \SetColor{Black}
    \Arc(442,-91)(10.05,276,636)
    \Arc(429.125,-77.75)(13.558,-77.758,-18.268)
    \Arc[clock](454,-79)(12.748,-101.31,-168.69)
    \Vertex(442,-81){1.5}
    \Vertex(432,-91){1.5}
    \Vertex(452,-91){1.5}
    \Line(452,-91)(457,-91)
    \Line(432,-91)(427,-91)
  \end{picture}=
  \frac{2\lambda_{\tau}^3}{[3(d-m/2)-10]} \left[5m_{\tau}^{2\tau} C_{\tau}(p,K') + 2D_{\tau}(p,K') \right] , \label{intelif15}
\end{eqnarray}

onde

\begin{eqnarray}
C_{\tau}(p,K')&=& \int d^{d-m}q_1 d^{d-m}q_2 d^{d-m}q_3 d^mk_1 d^mk_2 d^mk_3 \frac{1}{[q_1^2 + (k_1^2)^2 + m_{\tau}^{2\tau}][q_2^2 + (k_2^2)^2 + m_{\tau}^{2\tau}] }\times \nonumber \\
&&\frac{1}{[q_3^2 + (k_3^2)^2 + m_{\tau}^{2\tau}] \left\{(q_1 + q_2 + p)^2 + [(k_1 + k_2 + K')^2]^2 + m_{\tau}^{2\tau}  \right\}} \times \nonumber \\
&& \frac{1}{\left\{(q_1 + q_3 + p)^2 + [(k_1 + k_2 + K')^2]^2 + m_{\tau}^{2\tau}  \right\}^2} ,
\end{eqnarray}

\begin{eqnarray}
D_{\tau}(p,K')&=& \int d^{d-m}q_1 d^{d-m}q_2 d^{d-m}q_3 d^mk_1 d^mk_2 d^mk_3 \frac{1}{[q_1^2 + (k_1^2)^2 + m_{\tau}^{2\tau}][q_2^2 + (k_2^2)^2 + m_{\tau}^{2\tau}] }\times \nonumber \\
&&\frac{p(q_1 + q_3 + p) + K'(k_1 + k_3 + K')(k_1 + k_3 + K')^2}{[q_3^2 + (k_3^2)^2 + m_{\tau}^{2\tau}] \left\{(q_1 + q_2 + p)^2 + [(k_1 + k_2 + K')^2]^2 + m_{\tau}^{2\tau}  \right\}} \times \nonumber \\
&& \frac{1}{\left\{(q_1 + q_3 + p)^2 + [(k_1 + k_3 + K')^2]^2 + m_{\tau}^{2\tau}  \right\}^2} .
\end{eqnarray}

Fazendo a condição $m_{\tau}^{2\tau}=0$ na expressão (\ref{intelif15}), que contribui para o cálculo da constante de renormalização $Z_{\phi_{(\tau)}}$, iremos apenas encontrar uma expressão final para $D_{\tau}(p,K')$. Logo podemos reescrever $D_{\tau}(p,K')$ na forma

\begin{eqnarray}
&& D_{\tau}(p,K')=- \left[\frac{p}{2}\frac{\partial}{\partial p} + \frac{K'}{4} \frac{\partial}{\partial K'} \right] \left[d^{d-m}q_1 d^mk_1 \frac{1}{[q_1^2 + (k_1^2)^2 + m_{\tau}^{2\tau}]}  \right] \times \nonumber \\
&& \left[\int d^{d-m}q_2d^mk_2 \frac{1}{[q_2^2 + (k_2^2)^2 + m_{\tau}^{2\tau}] \left\{(q_1 + q_2 + p)^2 + [(k_1 + k_2 + K')^2]^2 + m_{\tau}^{2\tau}  \right\}}  \right] \times \nonumber \\
&& \left[\int d^{d-m}q_3d^mk_3 \frac{1}{[q_3^2 + (k_3^2)^2 + m_{\tau}^{2\tau}] \left\{(q_1 + q_3 + p)^2 + [(k_1 + k_3 + K')^2]^2 + m_{\tau}^{2\tau}  \right\}}  \right] .
\end{eqnarray}

A integral em $q_2$ e $k_2$ é dada por

\begin{eqnarray}
&& \int d^{d-m}q_2 d^mk_2 \frac{1}{[q_2^2 + (k_2^2)^2 + m_{\tau}^{2\tau}]\left\{(q_1 + q_2 + p)^2 + [(k_1 + k_2 + K')^2]^2 + m_{\tau}^{2\tau}  \right\} }= \nonumber \\
&& \frac{1}{\varepsilon_L} [1+([i_2]_m - 1)\varepsilon_L] \int_{0}^{1}dx \left\{[(q_1 + p)^2 + [(k_1 + K')^2]^2]x(1-x) + m_{\tau}^{2\tau}\right\}^{-\varepsilon_L /2} .
\end{eqnarray}

Como a integral em $q_3$ e $k_3$ possui também o mesmo resultado, temos que

\begin{eqnarray}
&& D_{\tau}(p,K')= -\left[\frac{p}{2} \frac{\partial}{\partial p} + \frac{K'}{4} \frac{\partial}{\partial K'}  \right] \frac{1}{\varepsilon_L^2}[1+2([i_2]_m-1)\varepsilon_L]\int_{0}^{1}dx[x(1-x)]^{-\varepsilon_L} \times \nonumber \\
&& \int d^{d-m}q_1 d^mk_1 \frac{1}{[q_1^2 + (k_1^2)^2 + m_{\tau}^{2\tau}] \left\{(q_1 + p)^2 + [(k_1 + K')^2]^2 + \frac{m_{\tau}^{2\tau}}{x(1-x)} \right\}^{\varepsilon_L}} .
\end{eqnarray}

Utilizando agora os parâmetros de Feynman, integrando em $q_1$ e $k_1$, e absorvendo o fator angular (\ref{fatorlif}) em uma redefinição da constante de acoplamento, obtemos

\begin{eqnarray}
D_{\tau}(p,K')&=&\frac{1}{\mu_{\tau}^{3\tau \varepsilon_L}} \frac{p^2 + (K'^2)^2}{6\varepsilon_L^2}[1+2([i_2]_m-1)\varepsilon_L]\left(1-\frac{\varepsilon_L}{2}\psi\left(2-\frac{m}{4}\right) \right) \left(1+\frac{3}{2}\varepsilon_L \psi(1) \right) \times \nonumber \\
&& \left(1+\varepsilon_L \psi(1) \right)^{-1} \Biggl\{1+\frac{\varepsilon_L}{2} -3\varepsilon_L \int_{0}^{1}dxdy(1-y) ln \Biggl\{\frac{[p^2 + (K'^2)^2]y(1-y)}{\mu_{\tau}^{2\tau}} + \nonumber \\
&& \frac{m_{\tau}^{2\tau}}{\mu_{\tau}^{2\tau}} \left[(1-y) + \frac{y}{x(1-x)}  \right] \Biggr\} \Biggr\} .
\end{eqnarray}

Substituindo o resultado $D_{\tau}(p,K')$ na expressão (\ref{intelif15}) obtemos o resultado final para o oitavo diagrama de Feynman, isto é,

\begin{eqnarray}
 \begin{picture}(30,22) (427,-96)
    \SetWidth{1.0}
    \SetColor{Black}
    \Arc(442,-91)(10.05,276,636)
    \Arc(429.125,-77.75)(13.558,-77.758,-18.268)
    \Arc[clock](454,-79)(12.748,-101.31,-168.69)
    \Vertex(442,-81){1.5}
    \Vertex(432,-91){1.5}
    \Vertex(452,-91){1.5}
    \Line(452,-91)(457,-91)
    \Line(432,-91)(427,-91)
  \end{picture}\bigg|_{m_{\tau}^{2\tau}=0} &=& \frac{g_{\tau}^3[p^2 + (K'^2)^2]}{6 \varepsilon_L^2} \Biggl\{1 + (3[i_2]_m-1)\varepsilon_L -3\varepsilon_L \int_{0}^{1}dxdy(1-y)\times  \nonumber \\
&& ln \Biggl\{ \frac{[p^2 + (K'^2)^2]y(1-y)}{\mu_{\tau}^{2\tau}} + \frac{m_{\tau}^{2\tau}}{\mu_{\tau}^{2\tau}} \left[(1-y) + \frac{y}{x(1-x)}  \right] \Biggr\} \Biggr\}.
\end{eqnarray}

Agora iremos calcular o primeiro diagrama de contra-termo dado por

\begin{eqnarray}
\begin{picture}(16,21) (216,-193)
    \SetWidth{1.0}
    \SetColor{Black}
    \Arc(224,-182)(7,270,630)
    \Line(215,-190)(233,-190)
    \Vertex(224,-189.5){1.5}
    \Line(221,-172)(227,-179)
    \Line(227,-172)(221,-179)
  \end{picture}=
  \begin{picture}(30,24) (218,-161)
    \SetWidth{0.8}
    \SetColor{Black}
    \Arc(233,-157)(8,252,612)
    \Line(241.5,-157)(246,-151)
    \Line(241.5,-157)(246,-163)
    \Line(224.5,-157)(220,-151)
    \Line(224.5,-157)(220,-163)
    \Vertex(224.5,-157){1.5}
    \Vertex(241.5,-157){1.5}
  \end{picture}\bigg|_{p^2 + (K'^2)^2=0,-\mu_{\tau}^{\tau \varepsilon_L}g_{\tau} \rightarrow -m_{\tau}^{2\tau}c_{m_{\tau}^{2\tau}}^{1}} .
\end{eqnarray}

Cujo resultado, para o primeiro diagrama de contra-termo, se encontra na forma

\begin{eqnarray}
\begin{picture}(16,21) (216,-193)
    \SetWidth{1.0}
    \SetColor{Black}
    \Arc(224,-182)(7,270,630)
    \Line(215,-190)(233,-190)
    \Vertex(224,-189.5){1.5}
    \Line(221,-172)(227,-179)
    \Line(227,-172)(221,-179)
  \end{picture}=
  \frac{m_{\tau}^{2\tau} g_{\tau}^2}{2\varepsilon_L^2} \left[1+([i_2]_m-1)\varepsilon_L - \frac{\varepsilon_L}{2}ln \left(\frac{m_{\tau}^{2\tau}}{\mu_{\tau}^{2\tau}}  \right)  \right] .
\end{eqnarray}

O segundo diagrama de contra-termo é dado por

\begin{eqnarray}
\begin{picture}(30,17) (212,-153)
    \SetWidth{1.0}
    \SetColor{Black}
    \Arc(227,-144)(7.071,262,622)
    \Line(213,-152)(241,-152)
    \Vertex(227,-151){3}
  \end{picture}=
 \begin{picture}(30,17) (212,-153)
    \SetWidth{0.9}
    \SetColor{Black}
    \Arc(227,-144)(7.071,262,622)
    \Line(213,-152)(241,-152)
    \Vertex(227,-151.5){1.5}
  \end{picture}\bigg|_{-\mu_{\tau}^{\tau \varepsilon_L} g_{\tau}\rightarrow - \mu_{\tau}^{\tau \varepsilon_L}g_{\tau} c_{g_{\tau}}^{1}}  .
\end{eqnarray}

No qual obtemos o resultado para o segundo diagrama de contra-termo na forma

\begin{eqnarray}
\begin{picture}(30,17) (212,-153)
    \SetWidth{1.0}
    \SetColor{Black}
    \Arc(227,-144)(7.071,262,622)
    \Line(213,-152)(241,-152)
    \Vertex(227,-151){3}
  \end{picture}=
  \frac{3m_{\tau}^{2\tau} g_{\tau}^{2}}{2 \varepsilon_L^2} \left[1+\left([i_2]_m-\frac{1}{2}\right)\varepsilon_L -\frac{\varepsilon_L}{2}ln \left(\frac{m_{\tau}^{2\tau}}{\mu_{\tau}^{2\tau}}\right)  \right] .
\end{eqnarray}

O terceiro diagrama de contra-termo que iremos calcular, é dado por

\begin{eqnarray}
\begin{picture}(29,18) (152,-178)
    \SetWidth{1.0}
    \SetColor{Black}
    \Arc(166,-173)(8.246,256,616)
    \Line(152,-173)(180,-173)
    \Vertex(158,-173){1.5}
    \Vertex(174,-173){2.8}
  \end{picture}=
  \begin{picture}(29,18) (152,-178)
    \SetWidth{1.0}
    \SetColor{Black}
    \Arc(166,-173)(8.246,256,616)
    \Line(152,-173)(180,-173)
    \Vertex(158,-173){1.5}
    \Vertex(174,-173){1.5}
  \end{picture}\bigg|_{m_{\tau}^{2\tau}=0, -\mu_{\tau}^{\tau \varepsilon_L}g_{\tau} \rightarrow -\mu_{\tau}^{\tau \varepsilon_L}g_{\tau}c_{g_{\tau}}^{1} } .
\end{eqnarray}

Portanto obtemos o resultado para o terceiro diagrama de contra-termo, isto é, 

\begin{eqnarray}
\begin{picture}(29,18) (152,-178)
    \SetWidth{1.0}
    \SetColor{Black}
    \Arc(166,-173)(8.246,256,616)
    \Line(152,-173)(180,-173)
    \Vertex(158,-173){1.5}
    \Vertex(174,-173){2.8}
  \end{picture}&=&
  -\frac{3g_{\tau}^3[p^2 + (K'^2)^2]}{16\varepsilon_L^2} \Biggl\{1+ \left(2[i_2]_m-\frac{3}{4}\right)\varepsilon_L -2\varepsilon_L \int_{0}^{1}dxdy(1-y)\times \nonumber \\
&& ln \left\{\frac{[p^2 + (K'^2)^2]y(1-y)}{\mu_{\tau}^{2\tau}} + \frac{m_{\tau}^{2\tau}}{\mu_{\tau}^{2\tau}} \left[(1-y)+\frac{y}{x(1-x)}  \right] \right\} \Biggr\}  .
\end{eqnarray}

Calcularemos agora o quarto diagrama de contra-termo dado por

\begin{eqnarray}
\begin{picture}(30,24) (218,-161)
    \SetWidth{1.0}
    \SetColor{Black}
    \Arc(233,-157)(8,252,612)
    \Line(241.5,-157)(246,-151)
    \Line(241.5,-157)(246,-163)
    \Line(224.5,-157)(220,-151)
    \Line(224.5,-157)(220,-163)
    \Vertex(224.5,-157){1.5}
    \Vertex(241.5,-157){3}
  \end{picture}=
  \begin{picture}(30,24) (218,-161)
    \SetWidth{0.8}
    \SetColor{Black}
    \Arc(233,-157)(8,252,612)
    \Line(241.5,-157)(246,-151)
    \Line(241.5,-157)(246,-163)
    \Line(224.5,-157)(220,-151)
    \Line(224.5,-157)(220,-163)
    \Vertex(224.5,-157){1.5}
    \Vertex(241.5,-157){1.5}
  \end{picture}\bigg|_{-\mu_{\tau}^{\tau \varepsilon_L} g_{\tau} \rightarrow -\mu_{\tau}^{\tau \varepsilon_L} g_{\tau}c_{g_{\tau}}^{1}} ,
\end{eqnarray}

Que podemos obter o resultado na forma

\begin{eqnarray}
\begin{picture}(30,24) (218,-161)
    \SetWidth{1.0}
    \SetColor{Black}
    \Arc(233,-157)(8,252,612)
    \Line(241.5,-157)(246,-151)
    \Line(241.5,-157)(246,-163)
    \Line(224.5,-157)(220,-151)
    \Line(224.5,-157)(220,-163)
    \Vertex(224.5,-157){1.5}
    \Vertex(241.5,-157){3}
  \end{picture}=
  \frac{3g_{\tau}^3 \mu_{\tau}^{\tau \varepsilon_L}}{2 \varepsilon_L^2} \left\{1 + ([i_2]_m-1)\varepsilon_L - \frac{\varepsilon_L}{2} \int_{0}^{1} dx ln \left\{\frac{[p^2 + (K'^2)^2]x(1-x) + m_{\tau}^{2\tau}}{\mu_{\tau}^{2\tau}}  \right\}  \right\} .
\end{eqnarray}

O quinto diagrama de contra-termo que iremos calcular é dado por

\begin{eqnarray}
\begin{picture}(30,24) (218,-161)
    \SetWidth{1.0}
    \SetColor{Black}
    \Arc(233,-157)(8,252,612)
    \Line(241.5,-157)(246,-151)
    \Line(241.5,-157)(246,-163)
    \Line(224.5,-157)(220,-151)
    \Line(224.5,-157)(220,-163)
    \Vertex(224.5,-157){1.5}
    \Vertex(241.5,-157){1.5}
    \Line(230,-145.5)(236,-152.5)
    \Line(236,-145.5)(230,-152.5)
  \end{picture}= 
  -\frac{1}{2} \mathcal{K}
  \left(
 \parbox{8mm} {\begin{picture}(30,24) (218,-168)
    \SetWidth{1.0}
    \SetColor{Black}
    \Arc(233,-157)(8,252,612)
    \Arc(233,-145)(4.472,117,477)
    \Line(241,-157)(246,-151)
    \Line(241,-157)(246,-163)
    \Line(224,-157)(220,-151)
    \Line(224,-157)(220,-163)
    \Vertex(233,-149){1.5}
    \Vertex(224.7,-157){1.5}
    \Vertex(241.5,-157){1.5}
  \end{picture}}\hspace*{0.2cm} \right) .
\end{eqnarray}

Consequentemente obtemos o resultado escrito na forma

\begin{eqnarray}
\begin{picture}(30,24) (218,-161)
    \SetWidth{1.0}
    \SetColor{Black}
    \Arc(233,-157)(8,252,612)
    \Line(241.5,-157)(246,-151)
    \Line(241.5,-157)(246,-163)
    \Line(224.5,-157)(220,-151)
    \Line(224.5,-157)(220,-163)
    \Vertex(224.5,-157){1.5}
    \Vertex(241.5,-157){1.5}
    \Line(230,-145.5)(236,-152.5)
    \Line(236,-145.5)(230,-152.5)
  \end{picture}=
  -\frac{g_{\tau}^3 \mu_{\tau}^{\tau \varepsilon_L}}{4\varepsilon_L}\int_{0}^{1}dx \frac{m_{\tau}^{2\tau}(1-x)}{[p^2 + (K'^2)^2]x(1-x) + m_{\tau}^{2\tau}} .
\end{eqnarray}

\section{Diagramas de Feynman}
 
Nessa seção será listado os diagramas utilizados no cáculo das constantes de renormalização abordado no capítulo 1. A resolução detalhada desses diagramas pode ser encontrado na referência  \cite{kleinert:2000}. 

\begin{eqnarray}
\begin{picture}(30,17) (212,-153)
    \SetWidth{0.9}
    \SetColor{Black}
    \Arc(227,-144)(7.071,262,622)
    \Line(213,-152)(241,-152)
    \Vertex(227,-151.5){1.5}
  \end{picture}=
  m^2 \frac{g}{(4\pi)^2} \left[\frac{2}{\varepsilon} + \psi(2) + log\left(\frac{4\pi \mu^2}{m^2}  \right) + O(\varepsilon) \right] , \label{tadklei}
\end{eqnarray}

\begin{eqnarray}
\begin{picture}(30,24) (218,-161)
    \SetWidth{1.0}
    \SetColor{Black}
    \Arc(233,-157)(8,252,612)
    \Line(241.5,-157)(246,-151)
    \Line(241.5,-157)(246,-163)
    \Line(224.5,-157)(220,-151)
    \Line(224.5,-157)(220,-163)
    \Vertex(224.5,-157){1.5}
    \Vertex(241.5,-157){1.5}
  \end{picture}=
  g\mu^{\varepsilon} \frac{g}{(4\pi)^2} \left\{\frac{2}{\varepsilon} + \psi(1) + \int_{0}^{1}dx log \left[\frac{4\pi \mu^2}{\vec{k}x(1-x) + m^2} \right] + O(\varepsilon) \right\} \label{2klei} ,
\end{eqnarray}

\begin{eqnarray}
\begin{picture}(28,30) (204,-138)
    \SetWidth{0.9}
    \SetColor{Black}
    \Arc(218,-129)(6.083,261,621)
    \Line(205,-136)(231,-136)
    \Vertex(218,-135.5){1.5}
    \Arc(218,-116)(6.083,261,621)
    \Vertex(218,-122){1.5}
  \end{picture}=
  -\frac{m^2 g^2}{(4\pi)^4} \left[\frac{4}{\varepsilon^2} + \frac{2}{\varepsilon}(\psi(1) + \psi(2)) - \frac{4}{\varepsilon}log \left(\frac{m^2}{4\pi \mu^2}  \right) + O(\varepsilon^0)  \right] ,
\end{eqnarray}

\begin{eqnarray}
\begin{picture}(29,18) (152,-178)
    \SetWidth{1.0}
    \SetColor{Black}
    \Arc(166,-173)(8.246,256,616)
    \Line(152,-173)(180,-173)
    \Vertex(158,-173){1.5}
    \Vertex(174,-173){1.5}
  \end{picture}=
  -g^2 \frac{m^2}{(4\pi)^4} \left\{\frac{6}{\varepsilon^2} + \frac{6}{\varepsilon} \left[\frac{3}{2} + \psi(1) + log \left(\frac{4\pi \mu^2}{m^2}\right)\right] + \frac{\vec{k}^2}{2m^2\varepsilon} + O(\varepsilon^0)  \right\} ,
\end{eqnarray}

\begin{eqnarray}
\begin{picture}(44,16) (142,-138)
    \SetWidth{1.0}
    \SetColor{Black}
    \Arc(156,-134)(7.28,254,614)
    \Arc(171,-134)(7.28,254,614)
    \Vertex(163.5,-134){1.5}
    \Line(178,-134)(185,-128)
    \Line(178,-134)(185,-141)
    \Line(149,-134)(143,-128)
    \Line(149,-134)(143,-141)
    \Vertex(149,-134){1.5}
    \Vertex(178,-134){1.5}
  \end{picture}=
  -g\mu^{\varepsilon}\frac{g^2}{(4\pi)^4} \left\{\frac{4}{\varepsilon^2} + \frac{4}{\varepsilon}\psi(1) + \frac{4}{\varepsilon}\int_{0}^{1}dx log \left[\frac{4\pi \mu^2}{\vec{k}^2x(1-x)+m^2} \right] + O(\varepsilon^0)     \right\} ,
\end{eqnarray}

\begin{eqnarray}
\begin{picture}(30,24) (218,-161)
    \SetWidth{1.0}
    \SetColor{Black}
    \Arc(233,-157)(8,252,612)
    \Arc(233,-145)(4.472,117,477)
    \Line(241,-157)(246,-151)
    \Line(241,-157)(246,-163)
    \Line(224,-157)(220,-151)
    \Line(224,-157)(220,-163)
    \Vertex(233,-149){1.5}
    \Vertex(224.7,-157){1.5}
    \Vertex(241.5,-157){1.5}
  \end{picture}=
  -g\mu^{\varepsilon}\frac{g^2}{(4\pi)^4}\frac{2}{\varepsilon} \left[\int_{0}^{1}dx \frac{m^2(1-x)}{\vec{k}^2x(1-x) + m^2} + O(\varepsilon) \right] ,
\end{eqnarray}

\begin{eqnarray}
\begin{picture}(29,25) (216,-187)
    \SetWidth{0.9}
    \SetColor{Black}
    \Vertex(231,-172){1.5}
    \Arc(231,-181)(8.544,291,651)
    \Line(231,-172)(236,-167)
    \Line(217,-181)(244,-181)
    \Vertex(222.5,-181){1.5}
    \Vertex(239.5,-181){1.5}
    \Line(231,-172)(226,-167)
  \end{picture}=
  -g\mu^{\varepsilon}\frac{g^2}{(4\pi)^4} \frac{2}{\varepsilon^2} \left\{1 + \frac{\varepsilon}{2} + \varepsilon \psi(1) - \varepsilon \int_{0}^{1}dxlog \left[\frac{\vec{k}^2x(1-x)+m^2}{4\pi \mu^2}  \right] + O(\varepsilon^0)   \right\} ,
\end{eqnarray}

\begin{eqnarray}
\begin{picture}(16,21) (216,-193)
    \SetWidth{1.0}
    \SetColor{Black}
    \Arc(224,-182)(7,270,630)
    \Line(215,-190)(233,-190)
    \Vertex(224,-189.5){1.5}
    \Line(221,-172)(227,-179)
    \Line(227,-172)(221,-179)
  \end{picture}= 
  \frac{m^2 g^2}{(4\pi)^4} \left\{\frac{2}{\varepsilon^2} + \frac{\psi(1)}{\varepsilon} -\frac{1}{\varepsilon}log \left(\frac{m^2}{4\pi \mu^2} \right) + O(\varepsilon^0)  \right\} ,
\end{eqnarray}

\begin{eqnarray}
\begin{picture}(30,17) (212,-153)
    \SetWidth{1.0}
    \SetColor{Black}
    \Arc(227,-144)(7.071,262,622)
    \Line(213,-152)(241,-152)
    \Vertex(227,-151){3}
  \end{picture}=
  \frac{m^2 g^2}{(4\pi)^4} \left[\frac{6}{\varepsilon^2} + \frac{3}{\varepsilon}\psi(2) - \frac{3}{\varepsilon}log \left(\frac{m^2}{4\pi \mu^2}  \right) + O(\varepsilon^0)  \right] ,
\end{eqnarray}

\begin{eqnarray}
\begin{picture}(30,24) (218,-161)
    \SetWidth{1.0}
    \SetColor{Black}
    \Arc(233,-157)(8,252,612)
    \Line(241.5,-157)(246,-151)
    \Line(241.5,-157)(246,-163)
    \Line(224.5,-157)(220,-151)
    \Line(224.5,-157)(220,-163)
    \Vertex(224.5,-157){1.5}
    \Vertex(241.5,-157){3}
  \end{picture}=
  g\mu^{\varepsilon} \frac{g^2}{(4\pi)^4} \left\{\frac{6}{\varepsilon^2} + \frac{3}{\varepsilon}\psi(1) - \frac{3}{\varepsilon}\int_{0}^{1}dx log \left[\frac{\vec{k}^2x(1-x)+ m^2}{4\pi \mu^2}  \right]  + O(\varepsilon^0) \right\} ,
\end{eqnarray}

\begin{eqnarray}
\begin{picture}(30,24) (218,-161)
    \SetWidth{1.0}
    \SetColor{Black}
    \Arc(233,-157)(8,252,612)
    \Line(241.5,-157)(246,-151)
    \Line(241.5,-157)(246,-163)
    \Line(224.5,-157)(220,-151)
    \Line(224.5,-157)(220,-163)
    \Vertex(224.5,-157){1.5}
    \Vertex(241.5,-157){1.5}
    \Line(230,-145.5)(236,-152.5)
    \Line(236,-145.5)(230,-152.5)
  \end{picture}=
  g\mu^{\varepsilon}\frac{g^2}{(4\pi)^4}\frac{1}{\varepsilon} \left[\int_{0}^{1}dx \frac{m^2(1-x)}{\vec{k}^2x(1-x)+m^2} + O(\varepsilon) \right] .
\end{eqnarray}

\section{Fatores de Simetria para Simetria O(N)}

Nesta seção será listado os fatores de simetria para uma simetria $O(N)$ dos diagramas de Feynman usados no cálculo das constantes de renormalização discutido no capítulo 1. Uma análise mais detalhada desses fatores pode ser encontrado na referência \cite{kleinert:2000}.

\begin{equation}
S_{\scalebox{0.3}{ \begin{picture}(30,17) (212,-153) 
    \SetWidth{0.9}
    \SetColor{Black}
    \Arc(227,-144)(7.071,262,622)
    \Line(213,-152)(241,-152)
    \Vertex(227,-151.5){1.5}
  \end{picture}}} = \frac{N+2}{3}
\end{equation}

\begin{equation}
S_{\scalebox{0.3}{ \begin{picture}(30,24) (218,-165)
    \SetWidth{1.0}
    \SetColor{Black}
    \Arc(233,-157)(8,252,612)
    \Line(241.5,-157)(246,-151)
    \Line(241.5,-157)(246,-163)
    \Line(224.5,-157)(220,-151)
    \Line(224.5,-157)(220,-163)
    \Vertex(224.5,-157){1.5}
    \Vertex(241.5,-157){1.5}
  \end{picture}}}=\frac{N+8}{9}
\end{equation}

\begin{equation}
S_{\scalebox{0.3}{ \begin{picture}(28,30) (204,-138)
    \SetWidth{0.9}
    \SetColor{Black}
    \Arc(218,-129)(6.083,261,621)
    \Line(205,-136)(231,-136)
    \Vertex(218,-135.5){1.5}
    \Arc(218,-116)(6.083,261,621)
    \Vertex(218,-122){1.5}
  \end{picture}}} = \left( \frac{N+2}{3} \right)^2
\end{equation}

\begin{equation}
S_{\scalebox{0.3}{ \begin{picture}(29,18) (152,-182)
    \SetWidth{1.0}
    \SetColor{Black}
    \Arc(166,-173)(8.246,256,616)
    \Line(152,-173)(180,-173)
    \Vertex(158,-173){1.5}
    \Vertex(174,-173){1.5}
  \end{picture}}} = \frac{N+2}{3}
\end{equation}

\begin{eqnarray}
S_{\scalebox{0.3}{ \begin{picture}(44,16) (142,-142)
    \SetWidth{1.0}
    \SetColor{Black}
    \Arc(156,-134)(7.28,254,614)
    \Arc(171,-134)(7.28,254,614)
    \Vertex(163.5,-134){1.5}
    \Line(178,-134)(185,-128)
    \Line(178,-134)(185,-141)
    \Line(149,-134)(143,-128)
    \Line(149,-134)(143,-141)
    \Vertex(149,-134){1.5}
    \Vertex(178,-134){1.5}
  \end{picture}}}= \frac{N^2+6N+20}{27}
\end{eqnarray}

\begin{eqnarray}
S_{\scalebox{0.3}{ \begin{picture}(30,24) (218,-165)
    \SetWidth{1.0}
    \SetColor{Black}
    \Arc(233,-157)(8,252,612)
    \Arc(233,-145)(4.472,117,477)
    \Line(241,-157)(246,-151)
    \Line(241,-157)(246,-163)
    \Line(224,-157)(220,-151)
    \Line(224,-157)(220,-163)
    \Vertex(233,-149){1.5}
    \Vertex(224.7,-157){1.5}
    \Vertex(241.5,-157){1.5}
  \end{picture}}}= \frac{N+2}{3}\frac{N+8}{9}
\end{eqnarray}

\begin{eqnarray}
S_{\scalebox{0.3}{ \begin{picture}(29,25) (216,-191)
    \SetWidth{0.9}
    \SetColor{Black}
    \Vertex(231,-172){1.5}
    \Arc(231,-181)(8.544,291,651)
    \Line(231,-172)(236,-167)
    \Line(217,-181)(244,-181)
    \Vertex(222.5,-181){1.5}
    \Vertex(239.5,-181){1.5}
    \Line(231,-172)(226,-167)
  \end{picture}}}=
   S_{ \scalebox{0.3}{\begin{picture}(39,29) (156,-201)
    \SetWidth{0.8}
    \SetColor{Black}
    \Arc(179,-186)(11,270,630)
    \Line(168,-186)(157,-175)
    \Line(168,-186)(157,-197)
    \Arc(193.5,-186.5)(13.509,87.879,272.121)
    \Vertex(168,-186){1.5}
    \Vertex(184,-177){1.5}
    \Vertex(184,-196){1.5}
  \end{picture}}}  =\frac{5N+22}{27}
\end{eqnarray}

\begin{eqnarray}
S_{\scalebox{0.3}{ \begin{picture}(16,21) (216,-193)
    \SetWidth{1.0}
    \SetColor{Black}
    \Arc(224,-182)(7,270,630)
    \Line(215,-190)(233,-190)
    \Vertex(224,-189.5){1.5}
    \Line(221,-172)(227,-179)
    \Line(227,-172)(221,-179)
  \end{picture}}}= \left( \frac{N+2}{3} \right)^2 
\end{eqnarray}

\begin{eqnarray}
S_{\scalebox{0.3}{ \begin{picture}(30,17) (212,-153)
    \SetWidth{1.0}
    \SetColor{Black}
    \Arc(227,-144)(7.071,262,622)
    \Line(213,-152)(241,-152)
    \Vertex(227,-151){3}
  \end{picture}}}= \frac{N+2}{3} \frac{N+8}{9}
\end{eqnarray}

\begin{eqnarray}
S_{\scalebox{0.3}{ \begin{picture}(30,24) (218,-165)
    \SetWidth{1.0}
    \SetColor{Black}
    \Arc(233,-157)(8,252,612)
    \Line(241.5,-157)(246,-151)
    \Line(241.5,-157)(246,-163)
    \Line(224.5,-157)(220,-151)
    \Line(224.5,-157)(220,-163)
    \Vertex(224.5,-157){1.5}
    \Vertex(241.5,-157){3}
  \end{picture}}}= \left( \frac{N+8}{9} \right)^2
\end{eqnarray}

\begin{eqnarray}
S_{\scalebox{0.3}{ \begin{picture}(30,24) (218,-165)
    \SetWidth{1.0}
    \SetColor{Black}
    \Arc(233,-157)(8,252,612)
    \Line(241.5,-157)(246,-151)
    \Line(241.5,-157)(246,-163)
    \Line(224.5,-157)(220,-151)
    \Line(224.5,-157)(220,-163)
    \Vertex(224.5,-157){1.5}
    \Vertex(241.5,-157){1.5}
    \Line(230,-145.5)(236,-152.5)
    \Line(236,-145.5)(230,-152.5)
  \end{picture}}}= \frac{N+2}{3} \frac{N+8}{9}
\end{eqnarray}

\begin{eqnarray}
S_{\scalebox{0.3}{ \begin{picture}(30,22) (427,-102)
    \SetWidth{1.0}
    \SetColor{Black}
    \Arc(442,-91)(10.05,276,636)
    \Arc(429.125,-77.75)(13.558,-77.758,-18.268)
    \Arc[clock](454,-79)(12.748,-101.31,-168.69)
    \Vertex(442,-81){1.5}
    \Vertex(432,-91){1.5}
    \Vertex(452,-91){1.5}
    \Line(452,-91)(457,-91)
    \Line(432,-91)(427,-91)
  \end{picture}}}=\frac{(N+2)(N+8)}{27}
\end{eqnarray}

\begin{eqnarray}
S_{\scalebox{0.3}{ \begin{picture}(29,18) (152,-182)
    \SetWidth{1.0}
    \SetColor{Black}
    \Arc(166,-173)(8.246,256,616)
    \Line(152,-173)(180,-173)
    \Vertex(158,-173){1.5}
    \Vertex(174,-173){2.8}
  \end{picture}}}= \left( \frac{N+2}{3} \right) \left(  \frac{N+8}{9} \right)
\end{eqnarray}

\bibliography{emanuel}

\end{document}